\documentclass[12pt, a4paper]{article}

\usepackage[height=25cm,width=16cm]{geometry}
\usepackage{feynmp}
\usepackage{amsmath}
\usepackage{ascmac}
\usepackage{dcolumn}
\usepackage{bm,here}
\usepackage[FIGTOPCAP]{subfigure}
\usepackage{comment}
\usepackage{ifpdf}
\usepackage{slashed}
\usepackage{colortbl}
\usepackage{color}
\usepackage{ulem}
\usepackage{comment}
\usepackage{braket}
\usepackage[mathscr]{eucal}
\usepackage[sort&compress, numbers, merge]{natbib}
\usepackage{cancel}
\usepackage{tikz}
\usepackage[compat=1.1.0]{tikz-feynhand}

\ifpdf
\usepackage{graphicx}     
\usepackage[bookmarksopen,colorlinks=true,linkcolor=bblue,citecolor=ppink,urlcolor=ppink]{hyperref}
\else     
\usepackage[dvipdfmx]{graphicx}     
\usepackage[dvipdfmx,bookmarksopen, colorlinks=true, linkcolor=bblue, citecolor=ppink, urlcolor=ppink]{hyperref}
\fi
\usepackage{subcaption}

\usepackage{multicol}
\definecolor{red}{rgb}{1,0,0}
\definecolor{ppink}{rgb}{0.921545,0.440586,0.687243}
\definecolor{bblue}{rgb}{0.400000,0.400000,1.000000}
\usepackage[charter]{mathdesign}
\usepackage{soul}
\usepackage{wrapfig}

\begin{document}

\begin{titlepage}
\begin{center}

~\vskip 1.5cm
{\large\bf Minimal Majoron Dark Matter from a Discrete $Z_N$ Gauge Symmetry}

\vskip 1.5cm
{\large
Subaru Fujisawa$^{a}$,
Qiuyue Liang$^{a}$,
Shigeki Matsumoto$^{a}$, \\ [.3em]
Michiru Uwabo-Niibo$^{b}$,
and
Tsutomu T. Yanagida$^{a,c}$}

\vskip 1.5cm
$^{a}${\sl Kavli IPMU (WPI), UTIAS, University of Tokyo, Kashiwa 277-8583, Japan} \\ [.3em]
$^{b}${\sl Cosmology, Gravity, and Astroparticle Physics Group, \\ Center for Theoretical Physics of the Universe, \\ Institute for Basic Science (IBS), Daejeon, Korea} \\ [.3em]
$^{c}${\sl Tsung-Dao Lee Institute, School of Physics and Astronomy, \\ Shanghai Jiao Tong University, Shanghai 200240, China} \\ [.1em]

\vskip 3.0cm
\begin{abstract}
    We investigate majoron dark matter in a minimal setup, where the Standard Model is extended by three right-handed neutrinos and a complex scalar field. The theory is defined by an exact discrete gauge symmetry, $Z_N\subset U(1)_{3(B-L)}$, while the global $U(1)_{B-L}$ symmetry emerges only as an accidental symmetry at low energies. For nontrivial choices of the discrete symmetry $Z_N$, such as $Z_5$, $Z_7$, $Z_{11}$, and $Z_{13}$, Planck-suppressed operators explicitly break this accidental symmetry and generate a small majoron mass, making the resulting pseudo-Nambu--Goldstone boson a well-motivated dark matter candidate. We study its production via the misalignment mechanism after inflation, considering both radiation-dominated and early matter-dominated cosmological histories, and confront the viable parameter space with isocurvature bounds, cosmological constraints, and indirect dark matter searches. We find that the $Z_5$ model is excluded by limits on the dominant dark matter decay into neutrinos, whereas the other models remain viable. In particular, the $Z_7$ scenario predicts a majoron mass in the $1$--$10\,{\rm MeV}$ range and can be sensitively probed by future MeV gamma-ray observations, especially with COSI, through the 511\,keV line from the majoron decay into an electron--positron pair and the monochromatic gamma-ray line from its decay into two photons.
\end{abstract}

\end{center}
\end{titlepage}

\tableofcontents
\newpage
\setcounter{page}{1}

\section{Introduction}

Dark matter provides one of the most compelling indications of physics beyond the Standard Model, and identifying its microscopic nature remains a central challenge in particle physics and cosmology~\cite{Bertone:2004pz}. Over the past decades, many well-motivated dark matter candidates have been proposed in connection with other fundamental problems in particle physics. Prominent examples include weakly interacting massive particles, which are closely tied to the electroweak scale and the hierarchy problem, as exemplified by neutralino dark matter in supersymmetric extensions of the Standard Model~\cite{Jungman:1995df}, and axions~\cite{Kim:1979if,Shifman:1979if,Zhitnitsky:1980tq,Dine:1981rt}, which arise from the Peccei--Quinn solution to the strong CP problem~\cite{Peccei:1977hh,Peccei:1977ur,Weinberg:1977ma,Wilczek:1977pj}. In this work, we focus instead on a dark matter candidate associated with the origin of neutrino masses and mixings, as revealed by neutrino oscillations. In particular, we consider a scenario based on the seesaw mechanism, the simplest explanation for the smallness of neutrino masses~\cite{Yanagida:1979as,Yanagida:1979gs,Gell-Mann:1979vob,Minkowski:1977sc}.\footnote{
    The term ``seesaw mechanism'' was first coined by one of the present authors, T.~T.~Yanagida, at the INS Symposium held in Tokyo in 1981~\cite{INS:1981qlp}. He subsequently used the term at the 1981 International Symposium in Bonn, where it was recorded in the proceedings; see his comment on p.~865 of the proceedings~\cite{Pfeil:1981vb}.}

It is well known that two right-handed neutrinos are sufficient to account for the observed neutrino oscillation data, and that they can also generate the observed baryon asymmetry of the Universe through leptogenesis~\cite{Frampton:2002qc,Ibarra:2003up,Antusch:2011nz}. A simple further extension is to introduce a third right-handed neutrino, in analogy with the three generations of quarks and leptons, and identify it with dark matter~\cite{Dodelson:1993je,Shi:1998km,Asaka:2005an,Asaka:2005pn,Laine:2008pg,Kusenko:2010ik}. This scenario is highly constrained, since producing the right-handed-neutrino dark matter in the early Universe requires a sizable primordial lepton asymmetry, and its mass must lie around the keV scale~\cite{Boyarsky:2008xj,Boyarsky:2009ix,Boyarsky:2018tvu,Kasai:2025xaw, Akita:2025txo}. Nevertheless, it remains a viable dark matter candidate. Another well-motivated possibility is to promote the global $U(1)_{B-L}$ symmetry of the Standard Model to a local gauge symmetry. In this framework, three right-handed neutrinos are naturally introduced to make the gauged $U(1)_{B-L}$ symmetry anomaly-free. Two of them account for neutrino masses and leptogenesis, while the remaining one can play the role of dark matter. Unlike the right-handed-neutrino dark matter scenario without this new gauge interaction, the $U(1)_{B-L}$ gauge interaction opens up a wider range of viable dark matter masses~\cite{Kaneta:2016vkq,Okada:2016gsh,Sheng:2023dix,Fujisawa:2025yqi}. Moreover, when the corresponding gauge coupling is extremely small, the $U(1)_{B-L}$ gauge boson itself can be a dark matter candidate, often referred to as F\'eeton dark matter in literature~\cite{Lin:2022xbu,Lin:2022mqe,Sheng:2023iup,Cheng:2024vqb,Hayashi:2024not}.

Another dark matter candidate closely connected to the origin of neutrino masses and mixings is the majoron, which we focus in this article~\cite{Berezinsky:1993fm,Lattanzi:2007ux,Bazzocchi:2008fh,Garcia-Cely:2017oco,Akita:2023qiz,Manna:2022gwn,King:2024idj,Obata:2026qwx,Akita:2026gzk,deGiorgi:2026jqn,Batell:2026avi}. In the majoron framework, the masses of right-handed neutrinos arise from the spontaneous breaking of a global $U(1)_{B-L}$ symmetry, thereby relating the seesaw scale to the scale of $B-L$ breaking~\cite{Chikashige:1980ui,Gelmini:1980re}.\footnote{
    Originally, the majoron was proposed as the Nambu--Goldstone boson associated with the spontaneous breaking of $U(1)_L$. However, one may equally well consider a corresponding global $U(1)_{B-L}$ symmetry instead.}
The associated Nambu--Goldstone boson can become a pseudo-Nambu--Goldstone boson once this symmetry is explicitly broken, and may serve as dark matter. Indeed, global symmetries are generally expected to be violated by quantum-gravity effects~\cite{Kallosh:1995hi,Harlow:2018jwu}; however, the resulting symmetry breaking, here the $U(1)_{B-L}$ breaking, is not necessarily sufficiently suppressed. If symmetry-breaking operators are sizable, the would-be majoron may become too heavy or too unstable to serve as dark matter. Moreover, such uncontrolled breaking may undermine the original motivation for explaining the right-handed-neutrino mass scale in terms of the $B-L$ breaking scale. Motivated by this issue, we consider a setup in which an exact discrete gauge symmetry, $Z_N\subset U(1)_{3(B-L)}$, is imposed instead of a fundamental global $U(1)_{B-L}$ symmetry~\cite{Krauss:1988zc,Ibanez:1991pr,Sheng:2025sou}.\footnote{
    The notation $Z_N\subset U(1)_{3(B-L)}$ is used only as a convenient way to specify the $Z_N$ charge assignments. We do not assume that $U(1)_{B-L}$ is itself a fundamental symmetry or part of the UV completion of our setup.}
With three right-handed neutrinos, the $U(1)_{B-L}$ symmetry is made anomaly-free even if gauged, including mixed gravitational anomalies and the Dai--Freed anomaly~\cite{Dai:1994kq,Yonekura:2016wuc,Garcia-Etxebarria:2018ajm,Kawasaki:2023mjm}, and hence its discrete subgroup can consistently be used as a gauge symmetry. For suitable choices of $N$, operators that violate $U(1)_{B-L}$ but respect $Z_N$ first appear only at sufficiently high dimension. As a result, the global $U(1)_{B-L}$ symmetry emerges accidentally at low energies, while its breaking generates a light pseudo-Nambu--Goldstone boson. This provides a predictive framework for majoron dark matter.

Among the possible discrete subgroups $Z_N\subset U(1)_{3(B-L)}$, we identify several phenomenologically relevant choices, including $Z_5$, $Z_7$, $Z_{11}$, $Z_{13}$, etc., for which the global $U(1)_{B-L}$ symmetry survives as an accidental low-energy symmetry. Assuming that majoron dark matter is produced through the misalignment mechanism in the pre-inflationary symmetry-breaking scenario, we determine the mass range predicted in each $Z_N$ model and examine its consistency with isocurvature bounds, cosmological constraints, and indirect searches for majoron decay. We find that the $Z_5$ model is already excluded by constraints on the dominant decay into neutrinos, whereas the other models remain viable. In particular, the $Z_7$ model predicts a majoron mass in the $1$--$10\,{\rm MeV}$ range. This mass range can be efficiently tested with high sensitivity by upcoming MeV gamma-ray observations, especially COSI, because majoron decay produces characteristic line signals: in particular, the 511\,keV line from the electron--positron channel and a monochromatic gamma-ray line from the two-photon channel.

This article is organized as follows. In Sec.~\ref{sec: model}, we introduce the construction of the minimal majoron model based on a discrete gauge symmetry $Z_N\subset U(1)_{3(B-L)}$ and discuss how the majoron mass is generated by higher-dimensional $U(1)_{B-L}$-breaking operators. In Sec.~\ref{sec: phenomenology}, we study the cosmological and phenomenological aspects of majoron dark matter, including its production through the misalignment mechanism and constraints from isocurvature perturbations, cosmology, and indirect searches. In Sec.~\ref{sec: COSI}, we investigate the prospects for testing the MeV-scale region, especially the $Z_7$ scenario, with MeV gamma-ray observations through the 511\,keV line and the monochromatic gamma-ray line. Sec.~\ref{sec: conclusion} is devoted to our conclusions. Technical details of the Sommerfeld effect and the theoretical estimates of the relevant majoron couplings are summarized in Appendices~\ref{app: Sommerfeld effect} and~\ref{app: scannings}, respectively.

\section{The Minimal Majoron Model}
\label{sec: model}

The majoron is a (pseudo-)Nambu--Goldstone boson associated with the spontaneous breaking of a global $U(1)_{B-L}$ symmetry. Higher-dimensional operators allowed by an appropriate discrete gauge symmetry can explicitly break this global symmetry, giving the majoron a nonzero mass and making it a potential dark matter candidate. We consider a minimal realization of this scenario, in which only a discrete $Z_N$ gauge symmetry, taken to be a subgroup of $U(1)_{3(B-L)}$, is imposed. Throughout this paper, for clarity and notational convenience, we specify the charge assignments using the conventional $B-L$ charges, with the corresponding $Z_N$ charges given by $3(B-L)$ modulo $N$. The particle content of the Standard Model (SM) is extended by a complex scalar field $\Phi$, whose vacuum expectation value spontaneously breaks an accidental global $U(1)_{B-L}$ symmetry, and three right-handed neutrinos $\mathcal{N}_I$. The latter also play an essential role in rendering the discrete gauge symmetry anomaly-free, including the Dai--Freed anomaly~\cite{Dai:1994kq,Yonekura:2016wuc,Garcia-Etxebarria:2018ajm,Kawasaki:2023mjm}. The corresponding Lagrangian is
\begin{align}
    \mathcal{L} =
    &
    \mathcal{L}_{\rm SM}
    +|\partial^\mu \Phi|^2
    -\mu_\Phi^2 |\Phi|^2
    -\frac{\lambda_\Phi}{4} |\Phi|^4
    -\lambda_{\Phi H} |\Phi|^2 |H|^2
    +\sum_I \bar{\mathcal{N}}_I i \slashed{\partial} \mathcal{N}_I
    \nonumber \\
    &
    -\sum_I \frac{y^{(\mathcal{N})}_I}{2} (\bar{\mathcal{N}}_I^c \mathcal{N}_I \Phi + h.c.)
    -\sum_{i, I} (y^{(\nu)}_{iI} \bar{L}_i H^c \mathcal{N}_I + h.c.)
    +{\cal L}_{\slashed{L}}\ ,
    \label{eq: original lagrangian}
\end{align}
where $\mathcal{L}_{\rm SM}$ denotes the SM Lagrangian. The fields $H$ and $L_i$ are the SM Higgs and lepton doublets, respectively. The complex scalar field $\Phi$ carries $B-L$ charge $2$, while the right-handed neutrinos $\mathcal{N}_I$ carry $B-L$ charge $-1$. The Yukawa couplings between the right-handed neutrinos and $\Phi$, denoted by $y_I^{(\mathcal{N})}$, can be chosen real and positive without loss of generality by field redefinitions of $\mathcal{N}_I$, whereas the Yukawa couplings involving the right-handed neutrinos, the lepton doublets, and the SM Higgs, $y_{iI}^{(\nu)}$, are in general complex. Apart from the last term, the Lagrangian possesses an accidental global $U(1)_{B-L}$ symmetry. The last term is invariant under the imposed $Z_N$ gauge symmetry but explicitly breaks the accidental global symmetry, thereby inducing a majoron mass, as discussed below.

For $\mu_\Phi^2<0$, the scalar field $\Phi$ acquires a vacuum expectation value (VEV) $v_\Phi$, and the global $U(1)_{B-L}$ symmetry is spontaneously broken down. We expand $\Phi$ around the VEV in the nonlinear representation as $\Phi=(v_\Phi+\rho)e^{iJ/v_\Phi}/\sqrt{2}$, where $\rho$ and $J$ denote the radial and Nambu--Goldstone modes, respectively. Redefining the right-handed neutrinos as $\mathcal{N}_I \to e^{-iJ/(2v_\Phi)}\mathcal{N}_I$, together with appropriate field redefinitions of the SM fields, we obtain\footnote{
    Since U(1)$_{B-L}$ is anomaly-free in this model, no anomaly-induced term appears in the Lagrangian.}
\begin{align}
    {\cal L} =
    &
    {\cal L}_{\rm SM}
    +\frac{1}{2} (\partial J)^2
    +\sum_I \bar{\mathcal{N}}_I i \slashed{\partial} \mathcal{N}_I
    -\sum_I \frac{m_{\mathcal{N}_I}}{2}(\bar{\mathcal{N}}_I^c \mathcal{N}_I + h.c.)
    \nonumber \\
    &
    -\sum_{i, I} (y^{(\nu)}_{iI} \bar{L}_i H^c \mathcal{N}_I + h.c.)
    -\frac{1}{2v_\Phi}(\partial_\mu J)\,J^\mu_{B-L}
    +{\cal L}_{\slashed{L}}
    +\cdots,
\end{align}
where $m_{\mathcal{N}_I}=y_I^{(\mathcal{N})} v_\Phi/\sqrt{2}$, and $J^\mu_{B-L}$ is the $B-L$ current constructed from the SM and $\mathcal{N}_I$ fields. With $Q_i$, $U_i$, $D_i$, and $E_i$ denoting the quark doublets, up-type quark singlets, down-type quark singlets, and charged-lepton singlets, respectively, its explicit form is given by
\begin{align}
    J^\mu_{B-L} =
    -\sum_I \bar{\mathcal{N}}_I \gamma^\mu \mathcal{N}_I
    +\sum_i \left( \frac{1}{3}\bar{Q}_i \gamma^\mu Q_i + \frac{1}{3} \bar{U}_i \gamma^\mu U_i + \frac{1}{3} \bar{D}_i \gamma^\mu D_i-\bar L_i \gamma^\mu L_i - \bar E_i \gamma^\mu E_i \right).
\end{align}
We have omitted terms involving $\rho$ from substituting $\Phi$ into the Lagrangian in Eq.~(\ref{eq: original lagrangian}), since they are irrelevant at low energies. After electroweak symmetry breaking, the Higgs field acquires the vacuum expectation value $v_{\rm EW} \simeq 246\,\mathrm{GeV}$ and is written as $H=(0,v_{\rm EW}+h)^T/\sqrt{2}$ in the unitary gauge, where $h$ denotes the Higgs boson. The Yukawa interactions between the right-handed neutrinos and the SM lepton doublets induce the Dirac mass term $\sum_{i,I}[(m_D)_{iI} \bar{\nu}_i \mathcal{N}_I+{\rm h.c.}]$, with $(m_D)_{iI} \equiv y_{iI}v_{\rm EW}/\sqrt{2}$. The resulting neutrino mass matrix is
\begin{align}
    \mathcal{L} \supset -\frac{1}{2}
    \begin{pmatrix}
        \bar{\nu} & \overline{\mathcal{N}^c} \\
    \end{pmatrix}
    \begin{pmatrix}
        0 & m_D \\
        m_D^T & m_\mathcal{N} \\
    \end{pmatrix}
    \begin{pmatrix}
        \nu^c \\
        \mathcal{N} \\
    \end{pmatrix}
    + h.c.,
    \label{eq: nu-mass matrix}
\end{align}
where flavor indices are suppressed, and $m_\mathcal{N} \equiv {\rm diag}(m_{\mathcal{N}_1}, m_{\mathcal{N}_2}, m_{\mathcal{N}_3})$. Since this $6\times 6$ mass matrix is complex symmetric, it can be diagonalized by a unitary matrix $V$ as~\cite{Minkowski:1977sc, Yanagida:1979as, Yanagida:1979gs, Gell-Mann:1979vob}
\begin{align}
    V^T
    \begin{pmatrix}
    0 & m_D \\
    m_D^T & m_\mathcal{N} \\
    \end{pmatrix}
    V
    =
    {\rm diag}(m_1, m_2, m_3, M_1, M_2, M_3).
\end{align}
Since we focus on the parameter region with $m_D < m_\mathcal{N}$, where the mass scale governing $m_D$ is much smaller than that governing $m_\mathcal{N}$, the seesaw mechanism leads to a strong hierarchy between the light active neutrino masses and the heavy sterile right-handed neutrino masses, $m_{1,2,3} \ll M_{1,2,3}$. Here, $m_{1,2,3}$ denote the active ``SM'' neutrino masses, while $M_{1,2,3}$ denote the sterile ``right-handed'' neutrino masses\,\cite{Yanagida:1979as,Yanagida:1979gs,Gell-Mann:1979vob}. For the discussion below, it is convenient to parametrize $m_D$, or equivalently the Yukawa coupling matrix, in terms of these masses and the Pontecorvo--Maki--Nakagawa--Sakata (PMNS) matrix $U_{\rm PMNS}$\,\cite{Pontecorvo:1957qd,Maki:1962mu},
\begin{align}
    m_D \equiv y^{(\nu)} v_{\rm EW}/\sqrt{2} = i U_{\rm PMNS}^* D_\nu^{1/2} R D_\mathcal{N}^{1/2},
\end{align}
where $D_\nu \equiv {\rm diag}(m_1,m_2,m_3)$ and $D_\mathcal{N} \equiv {\rm diag}(M_1,M_2,M_3)$, while $R$ is a complex $3 \times 3$ matrix satisfying $R^T R = \mathbb{I}_{3 \times 3}$, with $\mathbb{I}_{3 \times 3}$ denoting the unit matrix. The PMNS matrix contains three mixing angles, one Dirac phase, and two Majorana phases. In addition, the matrix $R$ is parameterized by six independent real parameters beyond low-energy observables~\cite{Casas:2001sr}.

We now discuss the origin of the explicit $U(1)_{B-L}$-breaking terms in ${\cal L}_{\slashed{L}}$, which generate the majoron mass. A concrete mechanism for controlling these terms is desirable~\cite{Ghosh:2025cxp}, since introducing them by hand would reduce predictivity. As introduced above, we impose a discrete gauge symmetry $Z_N$, defined as a subgroup of $U(1)_{3(B-L)}$\,\cite{Krauss:1988zc,Ibanez:1991pr,Sheng:2025sou}. With three right-handed neutrinos, the continuous $U(1)_{3(B-L)}$ symmetry is free of both gauge and gravitational anomalies. Therefore, its discrete subgroups can be realized as anomaly-free gauge symmetries. The model can be defined by imposing a discrete gauge symmetry, $Z_N\subset U(1)_{3(B-L)}$, together with the SM gauge groups, without imposing a global $U(1)_{B-L}$ symmetry.\footnote{
    A possible ultraviolet realization is obtained by gauging $U(1)_{3(B-L)}$ and subsequently breaking it to $Z_N$ through the vacuum expectation value of an additional scalar $S$ with charge $Q_S=3(B-L)=N$ at a sufficiently high scale. If this scale is much larger than $v_\Phi$, the physical Nambu--Goldstone mode orthogonal to the one eaten by the massive gauge boson is predominantly the phase of $\Phi$, reproducing Eq.~(\ref{eq: original lagrangian}) at low energies. Such a construction also provides a framework for satisfying the Dai--Freed anomaly cancellation condition~\cite{Dai:1994kq}.}
In this setup, the global $U(1)_{B-L}$ symmetry emerges only as an accidental symmetry of the low-energy theory. It then remains to determine which discrete $Z_N$ subgroup should be chosen. Working in the regime $v_\Phi \ll m_{\rm pl}$ and assuming that the cutoff scale of the model is the Planck scale, $m_{\rm pl} \simeq 2.4 \times 10^{18}$~GeV, the leading contribution to the majoron mass is expected to arise from the lowest-dimensional operator of the form $\Phi^n/m_{\rm pl}^{n-4}$ allowed by the discrete symmetry. To keep the majoron mass sufficiently small compared with $v_\Phi$, the discrete symmetry $Z_N$ should forbid all operators of the form $\Phi^n$ with $n\leq 4$. Since $\Phi$ carries the integer-normalized charge $Q_\Phi=3(B-L)=6$, such an operator is allowed only when $6n=0$ modulo $N$. Table~\ref{tab: Phi^n} summarizes the lowest-dimensional operator of this form allowed by each $Z_N$. When different choices of $Z_N$ allow the same leading operator $\Phi^n$, we focus on the smallest such $N$, since it gives the same leading majoron potential. With this convention, the phenomenologically relevant choices are $Z_5$, $Z_7$, $Z_{11}$, and $Z_{13}$, for which the leading operator is $\Phi^N$. The majoron mass induced by the leading operator is obtained as
\begin{align}
    {\cal L}_\slashed{L} \supset \frac{\kappa}{n!} \frac{\Phi^n}{m_{\rm pl}^{n-4}} + h.c.
    \qquad \to \qquad
    m_J = \left[\frac{|\kappa|\,n}{(n-1)!\,2^{n/2-1}}\right]^{1/2} \frac{v_\Phi^{n/2-1}}{m_{\rm pl}^{n/2-2}}.
    \label{eq: massdiscrete}
\end{align}
We summarize the masses for different values of $n$ in Table~\ref{tab: Phi^n} for $\kappa = 1$ and $v_\Phi = 10^{10}$\,GeV.\footnote{
    In an SU(5) embedding of the minimal majoron model, the relevant discrete gauge symmetry should instead be a $Z_N$ subgroup of $\mathrm{U}(1)_\chi = \mathrm{U}(1)_{5(B-L)-4Y}$, rather than of $\mathrm{U}(1)_{3(B-L)}$\,\cite{Borzumati:2000fe}. Interestingly, $Z_7$ is then the smallest such symmetry yielding a suppressed majoron mass term, namely an operator $\Phi^n$ with $n>4$.}
It is also worth emphasizing that a $Z_N$ symmetry that suppresses the majoron mass sufficiently, as exemplified by $Z_5$, $Z_7$, and $Z_{11}$, suppresses other operators by powers of $v_\Phi/m_{\rm pl}$ as well.

\begin{table}[t]
    \begin{center}
    \begin{tabular}{c|ccccccccccc}
        & $Z_2$ & $Z_3$ & $Z_4$ & $Z_5$ & $Z_6$ & $Z_7$ & $Z_8$ & $Z_9$ & $Z_{10}$ & $Z_{11}$ & $Z_{12}$ \\
        \hline
        Lowest $\Phi^n$ & $\Phi$ & $\Phi$ & $\Phi^2$ & $\Phi^5$ & $\Phi$ & $\Phi^7$ & $\Phi^4$ & $\Phi^3$ & $\Phi^5$ & $\Phi^{11}$ & $\Phi^2$ \\
        $m_J$ & -- & --  & -- & 0.2\,PeV & -- & 0.1\,MeV & -- & -- & 0.2\,PeV & 20\,feV & -- \\
        \hline
    \end{tabular}
    \caption{\small \sl The lowest $\Phi^n$ operator allowed by the discrete gauged $Z_N$ symmetry. The predicted majoron mass for $\kappa = 1$ and $v_\Phi = 10^{10}$\,GeV, equivalently $m_J/[\kappa^{1/2}(v_\Phi/10^{10}\,{\rm GeV})^{n/2-1}]$, is also shown.}
    \label{tab: Phi^n}        
    \end{center}
\end{table}

We now focus on the interactions of the majoron field $J$. In what follows, we adopt the non-linear representation of $\Phi$ introduced above, namely $\Phi=(v_\Phi+\rho)e^{iJ/v_\Phi}/\sqrt{2}$. In this representation, the terms in the Lagrangian involving the majoron field are given by
\begin{align}
    \mathcal{L}
    \supset
    -\frac{1}{2} J\,(\Box + m_J^2)\,J
    -\frac{1}{2v_\Phi}(\partial_\mu J)\,J^\mu_{B-L}
    +\cdots,
    \label{eq: majoron lagrangian}
\end{align}
where $J_{B-L}^\mu$ denotes the $B-L$ current and governs the majoron decay into SM particles. We do not explicitly display the interactions between $J$ and $\rho$, the radial mode of $\Phi$, or those induced by the explicit $U(1)_{B-L}$-breaking terms in ${\cal L}_{\slashed{L}}$, since they are irrelevant at the low energies considered in the following discussion of majoron decay phenomenology.

The interaction above shows that the dominant decay channel of majoron dark matter is the tree-level decay into two active SM neutrinos. The partial decay width is given by
\begin{align}
    \Gamma\,[J \to \nu\nu] \simeq \frac{m_J}{16 \pi v_\Phi^2} \sum_{i = 1}^3 m_i^2,
    \label{eq: decays@tree}
\end{align}
with $m_i$ denoting the active SM neutrino masses, as defined above. Other tree-level decay modes into two SM particles are absent, even though $J$ couples to other SM fermions through the Lagrangian~(\ref{eq: majoron lagrangian}). This is because these couplings are vector-like; for example, $J^\mu_{B-L} \supset -\bar{e}_{L,i}\gamma^\mu e_{L,i}-\bar{E}_i\gamma^\mu E_i = -\bar{e}_i\gamma^\mu e_i$, with $L_i=(\nu_i,e_{L,i})^T$. The corresponding amplitudes vanish by the on-shell equations of motion for the final-state fermions. Therefore, this neutrino channel is the dominant process as long as $m_J$ is sufficiently below the electroweak scale.

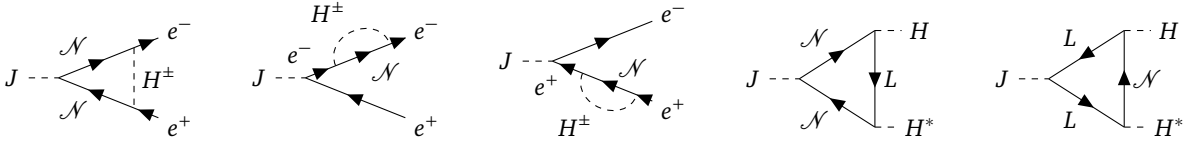
\begin{figure}
    \centering
    \begin{subfigure}
    \centering
    \scalebox{0.8}{
    \begin{tikzpicture}
    \begin{feynhand}
    \vertex [particle] (J) at (0,0) {\(J\)};
    \vertex (v1) at (0.75,0);
    \vertex (v2) at (2.0,0.5);
    \vertex (v3) at (2.0,-0.5);
    \vertex [particle] (e1) at (2.75,0.8) {\(e^-\)};
    \vertex [particle] (e2) at (2.75,-0.8) {\(e^+\)};
    \propag [scalar] (J) to (v1);
    \propag [fermion] (v1) to [edge label = \(\mathcal{N}\)] (v2);
    \propag [anti fermion] (v1) to [edge label' = \(\mathcal{N}\)] (v3);
    \propag [scalar] (v2) to [edge label = \(H^\pm\)] (v3);
    \propag [fermion] (v2) to (e1);
    \propag [anti fermion] (v3) to (e2);
    \end{feynhand}
    \end{tikzpicture}}
    \end{subfigure}
    \hfill
    \begin{subfigure}
    \centering
    \scalebox{0.8}{
    \begin{tikzpicture}
    \begin{feynhand}
    \vertex [particle] (J) at (0,0) {\(J\)};
    \vertex (v1) at (0.75,0);
    \vertex (v2) at (1.25,0.2);
    \vertex (v3) at (2.15,0.56);
    \vertex [particle] (e1) at (2.75,0.8) {\(e^-\)};
    \vertex [particle] (e2) at (2.75,-0.8) {\(e^+\)};
    \propag [scalar] (J) to (v1);
    \propag [fermion] (v1) to [edge label = \(e^-\)] (v2);
    \propag [fermion] (v2) to [edge label' = \(\mathcal{N}\)] (v3);
    \propag [fermion] (v3) to (e1);
    \propag [scalar] (v2) to [out=111.8, in=111.8, looseness=1.5, edge label = \(H^\pm\)] (v3);
    \propag [anti fermion] (v1) to (e2);
    \end{feynhand}
    \end{tikzpicture}}
    \end{subfigure}
    \hfill
    \begin{subfigure}
    \centering
    \scalebox{0.8}{
    \begin{tikzpicture}
    \begin{feynhand}
    \vertex [particle] (J) at (0,0) {\(J\)};
    \vertex (v1) at (0.75,0);
    \vertex (v2) at (1.25,-0.2);
    \vertex (v3) at (2.15,-0.56);
    \vertex [particle] (e1) at (2.75,0.8) {\(e^-\)};
    \vertex [particle] (e2) at (2.75,-0.8) {\(e^+\)};
    \propag [scalar] (J) to (v1);
    \propag [anti fermion] (v1) to [edge label' = \(e^+\)] (v2);
    \propag [anti fermion] (v2) to [edge label = \(\mathcal{N}\)] (v3);
    \propag [anti fermion] (v3) to (e2);
    \propag [scalar] (v2) to [out=248.2, in=248.2, looseness=1.5, edge label' = \(H^\pm\)] (v3);
    \propag [fermion] (v1) to (e1);
    \end{feynhand}
    \end{tikzpicture}}
    \end{subfigure}
    \hfill
    \begin{subfigure}
    \centering
    \scalebox{0.8}{
    \begin{tikzpicture}
    \begin{feynhand}
    \vertex [particle] (J) at (0,0) {\(J\)};
    \vertex [particle] (H1) at (2.75,0.8) {\(H\)};
    \vertex [particle] (H2) at (2.75,-0.8) {\(H^*\)};
    \vertex (v1) at (0.75,0);
    \vertex (v2) at (2.0,0.8);
    \vertex (v3) at (2.0,-0.8);
    \propag [scalar] (J) to (v1);
    \propag [fermion] (v1) to [edge label=\(\mathcal{N}\)] (v2);
    \propag [fermion] (v2) to [edge label=\(L\)] (v3);
    \propag [fermion] (v3) to [edge label=\(\mathcal{N}\)] (v1);
    \propag [scalar] (v2) to (H1);
    \propag [scalar] (v3) to (H2);
    \end{feynhand}
    \end{tikzpicture}}
    \end{subfigure}
    \hfill
    \begin{subfigure}
    \centering
    \scalebox{0.8}{
    \begin{tikzpicture}
    \begin{feynhand}
    \vertex [particle] (J) at (0,0) {\(J\)};
    \vertex [particle] (H1) at (2.75,0.8) {\(H\)};
    \vertex [particle] (H2) at (2.75,-0.8) {\(H^*\)};
    \vertex (v1) at (0.75,0);
    \vertex (v2) at (2.0,0.8);
    \vertex (v3) at (2.0,-0.8);
    \propag [scalar] (J) to (v1);
    \propag [anti fermion] (v1) to [edge label=\(L\)] (v2);
    \propag [anti fermion] (v2) to [edge label=\(\mathcal{N}\)] (v3);
    \propag [anti fermion] (v3) to [edge label=\(L\)] (v1);
    \propag [scalar] (v2) to (H1);
    \propag [scalar] (v3) to (H2);
    \end{feynhand}
    \end{tikzpicture}}
    \end{subfigure}
    \caption{\small\sl Feynman diagrams for the one-loop effective Lagrangian of the minimal majoron model.}
    \label{fig: diagrams}
\end{figure}

At next-to-leading order, majoron dark matter can decay into SM particles through one-loop diagrams. For example, the electron--positron channel has the decay width~\cite{Chikashige:1980ui, Pilaftsis:1993af, Garcia-Cely:2017oco},
\begin{align}
    \Gamma_0[J \to e^- e^+] \simeq
    \frac{m_J m_e^2}{2048 \pi^5 v_{\rm EW}^2}
    (K_{11} - K_{22} - K_{33})^2
    \sqrt{1 - \frac{4 m_e^2}{m_J^2}},
    \quad {\rm with} \quad
    K_{ij} =
    \frac{(m_D m_D^\dagger)_{ij}}{v_{\rm EW}\,v_\Phi},
    \label{eq: J to ee}
\end{align}
where $m_D$ is the Dirac mass matrix defined in Eq.~(\ref{eq: nu-mass matrix}). This result follows from the one-loop effective Lagrangian obtained by integrating out the right-handed neutrinos from the Lagrangian in Eq.~(\ref{eq: majoron lagrangian}), as illustrated in Fig.~\ref{fig: diagrams}. The effective Lagrangian takes the form,
\begin{align}
    \mathcal{L}_{\rm 1-loop} =
    \frac{(m_D m_D^\dagger)_{ij}}{8 \pi^2 v_\Phi v_{\rm EW}^2} (\partial_\mu J) (\bar{L}_i \gamma^\mu L_j)
    -\frac{(m_D m_D^\dagger)_{ii}}{16\pi^2 v_\Phi v_{\rm EW}^2}(\partial_\mu J)(H^\dagger i\overleftrightarrow{\partial^\mu} H)
    +\cdots.
    \label{eq: Leff}
\end{align}
Each diagram in the figure is divergent by itself, whereas the sum of all diagrams contributing to each interaction term is finite. In the diagrammatic calculation, the derivative coupling of the external majoron probes the divergence of the $U(1)_{B-L}$ current. By the Ward--Takahashi identity, the contribution from the conserved part of the current cancels in the complete sum over diagrams, leaving an amplitude proportional to the right-handed-neutrino Majorana mass insertions, which originate from the spontaneous breaking of $U(1)_{B-L}$. Therefore, if these $U(1)_{B-L}$-violating mass insertions are removed, the current becomes conserved and the complete summed amplitude vanishes, as required by the Ward--Takahashi identity. The partial decay width of majoron dark matter into an electron--positron pair then follows directly from the relevant operator coefficients in Eq.~(\ref{eq: Leff}).

Furthermore, at next-to-next-to-leading order, namely at the two-loop level, the majoron can also decay into two photons, with the corresponding partial decay width given by
\begin{align}
    &
    \Gamma[J \to \gamma \gamma] \simeq
    \frac{\alpha^2 m_J^3}{4096 \pi^7 v_{\rm EW}^2}
    \left|
        {\rm tr}[K] \sum_\mathfrak{f} N^c_\mathfrak{f} Q_\mathfrak{f}^2 T^3_\mathfrak{f}\,h[m_J^2/(4 m_\mathfrak{f}^2)]
    + \sum_i K_{ii}\,h[m_J^2/(4 m_{\ell_i}^2)]
    \right|^2,
    \label{eq: J to gammagamma}
\end{align}
where $\alpha$ denotes the fine-structure constant, and $\mathfrak{f}$ denotes an SM fermion, namely the six quarks $u_i$ and $d_i$ and the three charged leptons $\ell_i$, with $i$ the flavor index~\cite{Heeck:2019guh}. Here, $N^c_\mathfrak{f}$, $Q_\mathfrak{f}$, and $m_\mathfrak{f}$ represent the number of colors, electric charge, and mass of $\mathfrak{f}$, respectively, while the weak isospin $T^3_\mathfrak{f}$ is given by $T^3_{u_i} = -T^3_{d_i,\ell_i} = 1/2$. The loop function $h[x]$ is defined as
\begin{align}
    h[x]
    =
    -\frac{1}{4 x} \left[ \log\left(1 - 2x + 2\sqrt{x (x - 1)} \right) \right]^2 - 1.
\end{align}
It is important to note that this decay is not induced by the anomalous electromagnetic term, $J F_{\mu\nu}\tilde F^{\mu\nu}$. Such a term is forbidden by the Adler--Bardeen theorem, since the U(1)$_{B-L}$ symmetry is anomaly-free. Indeed, one finds that $h[x] \to x/3$ as $x\to 0$, and hence $\Gamma[J\to\gamma\gamma]\propto m_J^7$ as $m_J\to 0$. This confirms the above statement, since a nonzero contribution from the $J F_{\mu\nu}\tilde F^{\mu\nu}$ term would give a decay width proportional to $m_J^3$ in the same limit.

\section{Phenomenology of Majoron Dark Matter}
\label{sec: phenomenology}

In this section, we summarize the phenomenology of minimal majoron dark matter. We first discuss its production through misalignment and derive the associated constraint from isocurvature perturbations. We subsequently examine constraints from cosmological observations, including CMB, BBN, and Ly-$\alpha$ data, as well as from indirect dark matter searches.

\subsection{Misalignment mechanism}
\label{subsec: misalignment}

We focus in particular on the misalignment mechanism in the pre-inflationary scenario, where the U(1)$_{B-L}$ symmetry is already spontaneously broken before the onset of inflation~\cite{Preskill:1982cy, Abbott:1982af, Dine:1982ah, Turner:1983he}. This is because, in the post-inflationary scenario, the gauged $Z_N$ remnant can give rise to nontrivial defect configurations, whose interpretation and stability may depend sensitively on the global structure of the theory and on its UV completion \cite{OHare:2024nmr,Suzuki:2026xvf}. We therefore restrict our attention throughout this article to the pre-inflationary scenario.

\begin{figure}[t]
    \centering
    \includegraphics[width=0.92 \linewidth]{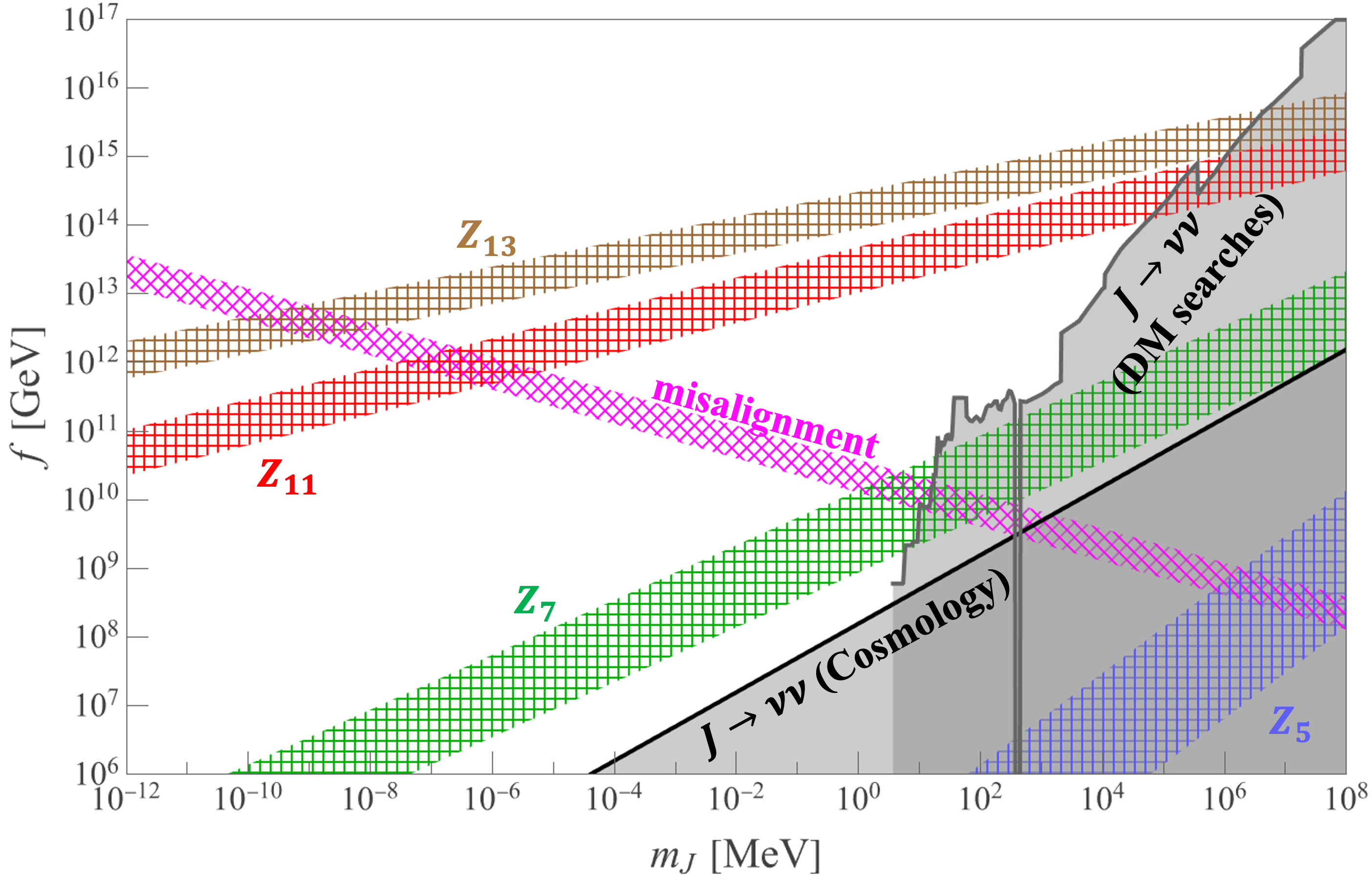} 
    \caption{\small\sl The magenta hatched region shows where the observed dark matter abundance is reproduced through the misalignment mechanism in Eq.~\eqref{eq: misalignment}, with the initial angle varied over $0.5 \leq \theta_0 \leq 1.5$ and the Standard Model values of $g_*(T)$ adopted. Also shown are the mass predictions for the different $Z_N$ gauge symmetry models discussed in Table~\ref{tab: Phi^n} and Eq.~\eqref{eq: massdiscrete}. The corresponding colored hatched regions are obtained by varying $10^{-3} \leq \kappa \leq 10^3$. The grey-shaded regions indicate the cosmological and dark matter search constraints from majoron decay into neutrino pairs for the $Z_5$ case; the constraints become progressively weaker for larger $N$. We assume normal ordering of the active neutrino masses with the lightest neutrino mass set to zero, as discussed in Section~\ref{subsubsec: constraints from J to nunu}.}
    \label{fig: RD result}
\end{figure}

The equation of motion for the homogeneous majoron mode is given, in the harmonic approximation, by $\ddot{J}+3H\dot{J}+m_J^2J\simeq0$. Solving it in a specified cosmological era yields the scalar-field energy density, $\rho_J\simeq\dot{J}^2/2+m_J^2J^2/2$. In the standard misalignment mechanism, this energy density is determined by the initial displacement at the onset of scalar-field oscillations in the expanding Universe. In standard cosmology, the oscillation begins in the radiation-dominated era, and the resulting dark matter abundance is expressed as~\cite{Blinov:2019rhb}
\begin{align}
    \Omega_J h^2 \simeq
    0.12\,\bigg[\frac{f\,\theta_0}{1.9 \times 10^{10}\,\mathrm{GeV}}\bigg]^2
    \bigg[\frac{m_J}{1\,\mathrm{MeV}}\bigg]^{1/2}
    \bigg[\frac{106.75}{g_*(T_{\mathrm{osc}})}\bigg]^{1/4},
    \label{eq: misalignment}
\end{align}
where $f\equiv v_\Phi/N$, with $N$ labeling the $Z_N$ symmetry, and $\theta_0$ denotes the initial displacement from the nearest minimum, which can be restricted to $0\leq\theta_0\leq\pi$ by the periodicity of the majoron potential.\footnote{
    If the lowest-dimensional operator allowed by the $Z_N$ symmetry has the form $\Phi^n$ with $n\neq N$ (see Table~\ref{tab: Phi^n}), then the decay constant is given by $f=v_\Phi/n$. For the $Z_5$, $Z_7$, $Z_{11}$, and $Z_{13}$ cases shown in the figure, $n=N$.}
Here, $g_*(T_{\rm osc})$ is the effective number of relativistic degrees of freedom when the majoron starts to oscillate, and $T_{\rm osc}$ is determined by solving $m_J\simeq q_0H(T_{\rm osc})$, where $H(T_{\rm osc})$ is the Hubble parameter at $T=T_{\rm osc}$ and $q_0=1.6$~\cite{Blinov:2019rhb}.\footnote{
    For simplicity, we take $g_{\ast}(T)=g_{\ast\,S}(T)$ in all of the numerical estimates throughout this section.}
In Fig.~\ref{fig: RD result}, the magenta hatched region provides an order-of-magnitude estimate of the parameter space where the dark matter abundance agrees with the observed value, $\Omega_{\rm DM}h^2\simeq 0.12$~\cite{Planck:2018vyg}. This region is estimated using Eq.~(\ref{eq: misalignment}) by varying $\theta_0$ between 0.5 and 1.5 and evaluating $g_*(T)$ using the SM relativistic degrees of freedom.\footnote{
    The exact periodic majoron potential is $V(J)\propto1-\cos(J/f)$, with $\ddot{\theta}+3H\dot{\theta}+m_J^2\sin\theta=0$ for $\theta=J/f$. For $0.5\leq\theta_0\leq1.5$, anharmonic corrections to the relic abundance typically remain at the $\mathcal O(10\%)$ level.}
The other hatched bands show the corresponding theoretical predictions from explicit $\mathrm{U}(1)_{B-L}$ breaking, assuming a discrete $Z_5$, $Z_7$, $Z_{11}$, or $Z_{13}$ gauge symmetry and varying $\kappa$ between $10^{-3}$ and $10^3$ in Eq.~(\ref{eq: massdiscrete}).\footnote{
    This range of $\kappa$ is motivated by representative ultraviolet estimates, taking $N=7$ as a benchmark. With the normalization in Eq.~(\ref{eq: massdiscrete}), an ${\cal O}(1)$ Wilson coefficient corresponds to $\kappa\sim N!$, giving $\kappa\sim10^3$ for $N=7$. Conversely, in the gauged $U(1)_{B-L}$ completion discussed above, the operator $c\,S^6\Phi^N/m_{\rm Pl}^{N+2}$ gives a sixth-power suppression after symmetry breaking, allowing $\kappa\sim10^{-3}$ for $N=7$ even if the $B-L$-breaking scale is only about an order of magnitude below $m_{\rm Pl}$. For larger $N$, an even broader range of $\kappa$ can be expected.}
We see that the $Z_7$ model predicts a majoron mass of 1--10~MeV, which is the parameter region we consider in the following discussion. For the $Z_7$ model, with $m_J\sim 1\,\mathrm{MeV}$, the oscillation temperature is $T_{\rm osc}\sim 10^7\,\mathrm{GeV}$, so the radiation-dominated misalignment scenario requires $T_{\rm RH} \gtrsim T_{\rm osc}$.

At the same time, in the pre-inflationary scenario considered here, the $U(1)_{B-L}$ symmetry must not be thermally restored after inflation, requiring $T_{\rm RH} \lesssim v_\Phi=Nf$. The condition $T_{\rm osc} \lesssim T_{\rm RH} \lesssim Nf$ has a qualitatively different implication depending on $N$. Since $m_J\propto f^{N/2-1}$, the oscillation temperature scales as $T_{\rm osc}\propto f^{(N-2)/4}$ during radiation domination. Then, $T_{\rm osc} \lesssim Nf$ gives a lower bound on $f$ for $N<6$, whereas it gives an upper bound for $N>6$; it requires $f\gtrsim O(10^{15})\,\kappa\,{\rm GeV}$ for $Z_5$, for which the misalignment abundance is far above the observed dark matter abundance. Thus, the $Z_5$ model cannot realize the radiation-dominated pre-inflationary misalignment scenario considered here. For $Z_7$, $Z_{11}$ and $Z_{13}$, the corresponding upper bounds on $f$ are well above the parameter region of interest.

We note that, in addition to misalignment, the majoron can also be produced thermally after reheating through interactions involving the right-handed neutrinos in the early Universe~\cite{Manna:2022gwn, Akita:2026gzk, deGiorgi:2026jqn}. This additional thermal contribution is strongly suppressed for $T_{\rm RH}\ll M_I$, while it can become relevant for $T_{\rm RH}\gtrsim M_I$. To estimate it, we first note that right-handed neutrinos are almost thermalized around $T\sim M_I$ for representative seesaw parameters compatible with neutrino oscillation data, as expected in the strong-washout regime of thermal leptogenesis~\cite{Buchmuller:2003gz, Engelhard:2007kf}. Their interaction rate at $T\sim M_I$ satisfies $\Gamma_{\mathcal N_I}/H={\cal O}(1)\,[(y^{(\nu)\dagger}y^{(\nu)})_{\mathcal N_I\mathcal N_I}^{1/2}/10^{-3}]^2(10^{10}\,{\rm GeV}/M_I)(106.75/g_*)^{1/2}$, typically exceeding unity when $M_I\lesssim10^{10}$\,GeV. The production processes arise from the $y_I^{(\mathcal N)}\Phi\mathcal N_I\mathcal N_I$ and $y_{\alpha I}^{(\nu)}L_\alpha H\mathcal N_I$ interactions; at leading order, they are $\mathcal N_I\mathcal N_I\to JJ$, $LH\to\mathcal N_IJ$, $H\mathcal N_I\to LJ$, $L\mathcal N_I\to HJ$, and $\mathcal N_I\to LHJ$. Assuming $T_{\rm RH}\sim M_I$ and an approximately thermal $\mathcal N_I$ abundance, we find
{\small
\begin{align}
    \Omega_J^{\rm FI} h^2 \simeq
    \left(\frac{m_J}{1\,{\rm MeV}}\right)
    \left(\frac{10^9\,{\rm GeV}}{M_I}\right)
    \left[
        0.38\left(\frac{y_I^{(\mathcal N)}}{10^{-2}}\right)^4
        +0.058\left(\frac{y_I^{(\mathcal N)}}{10^{-2}}\right)^2
        \left(\frac{\left(y^{(\nu)\dagger}y^{(\nu)}\right)_{\mathcal N_I\mathcal N_I}^{1/2}}{10^{-3}}\right)^2
    \right].
    \label{eq: abundance from FI}
\end{align}
}The first term arises from $\mathcal N_I\mathcal N_I\to JJ$, while the second includes the three scattering processes involving $y^{(\nu)}$ as well as the three-body decay $\mathcal N_I\to LHJ$. Details of the derivation are given in Appendix~\ref{app: freeze-in production}. Using $M_I=y_I^{(\mathcal N)}Nf/\sqrt{2}$, we find that, for given $N$, $m_J$, $f$, and $y^{(\nu)}$, the freeze-in contribution can be sufficiently suppressed by taking $y_I^{(\mathcal N)}$ small. For example, for $N=7$, $m_J=1\,{\rm MeV}$, $f=2\times10^{10}\,{\rm GeV}$, and $(y^{(\nu)\dagger}y^{(\nu)})_{\mathcal N_I\mathcal N_I}^{1/2}=10^{-3}$, the radiation-dominated misalignment mechanism can produce a majoron abundance of the same order as the observed dark matter abundance, as shown in Eq.~(\ref{eq: misalignment}). In this case, Eq.~(\ref{eq: abundance from FI}) shows that the freeze-in abundance is at most comparable to the misalignment abundance for $y_I^{(\mathcal N)}\lesssim6\times10^{-3}$ and decreases further for smaller couplings. Exploring the full parameter space including this freeze-in contribution would be interesting, but is beyond the scope of the present work. Our purpose here is to demonstrate that viable parameter regions exist in gauged $Z_N$ majoron models and that they can be tested by near-future observations. We therefore leave a systematic quantitative analysis including freeze-in production for future work. Accordingly, for cosmological scenarios in which the conventional misalignment mechanism is effective, we restrict our discussion to parameter regions where the thermal contribution is negligible; $T_{\rm RH}\ll M_I$, or $T_{\rm RH}\sim M_I$ with sufficiently suppressed $y_I^{(\mathcal N)}$.

On the other hand, if the inflaton couples only very weakly to other particles and reheating after inflation is prolonged, the inflaton energy density dominates the Universe for a certain period, leading to a nontrivial cosmological epoch known as early matter domination. In this case, the majoron dark matter abundance produced through the misalignment mechanism is modified from the standard radiation-dominated result as follows~\cite{Blinov:2019rhb, Nelson:2018via}:
\begin{equation}
    \Omega_J h^2 \simeq
    0.12 \times \bigg[\frac{f \theta_0}{ 10^{10}\,\mathrm{GeV}}\bigg]^2 
    \bigg[\frac{T_{\mathrm{RH}}}{8 \times 10^7\,\mathrm{GeV}}\bigg].
    \label{eq: misalgnmentinEMD}
\end{equation}
The absence of an explicit $m_J$ dependence can be understood intuitively as follows. During an early matter-dominated era, once the majoron field starts oscillating at $H_{\rm osc} \sim m_J$, it behaves as non-relativistic matter. Since the background inflaton energy density also redshifts as matter, the fractional energy density of the majoron remains approximately constant throughout the matter-dominated era, in contrast to the standard radiation-dominated scenario. At the onset of oscillation, both the majoron energy density and the total energy density scale as $m_J^2$, namely $\rho_J\sim m_J^2 f^2\theta_0^2$ and $\rho_{\rm tot}\sim 3m_{\rm pl}^2 m_J^2$. Their ratio is therefore independent of $m_J$, being set only by $(f\theta_0/m_{\rm pl})^2$. The final abundance is then determined by this fractional energy density together with the entropy production at reheating, and hence scales as $\Omega_J h^2\propto (f\theta_0)^2 T_{\rm RH}$, with no residual dependence on the majoron mass.

\begin{figure}[t]
    \centering
    \includegraphics[width=0.92\linewidth]{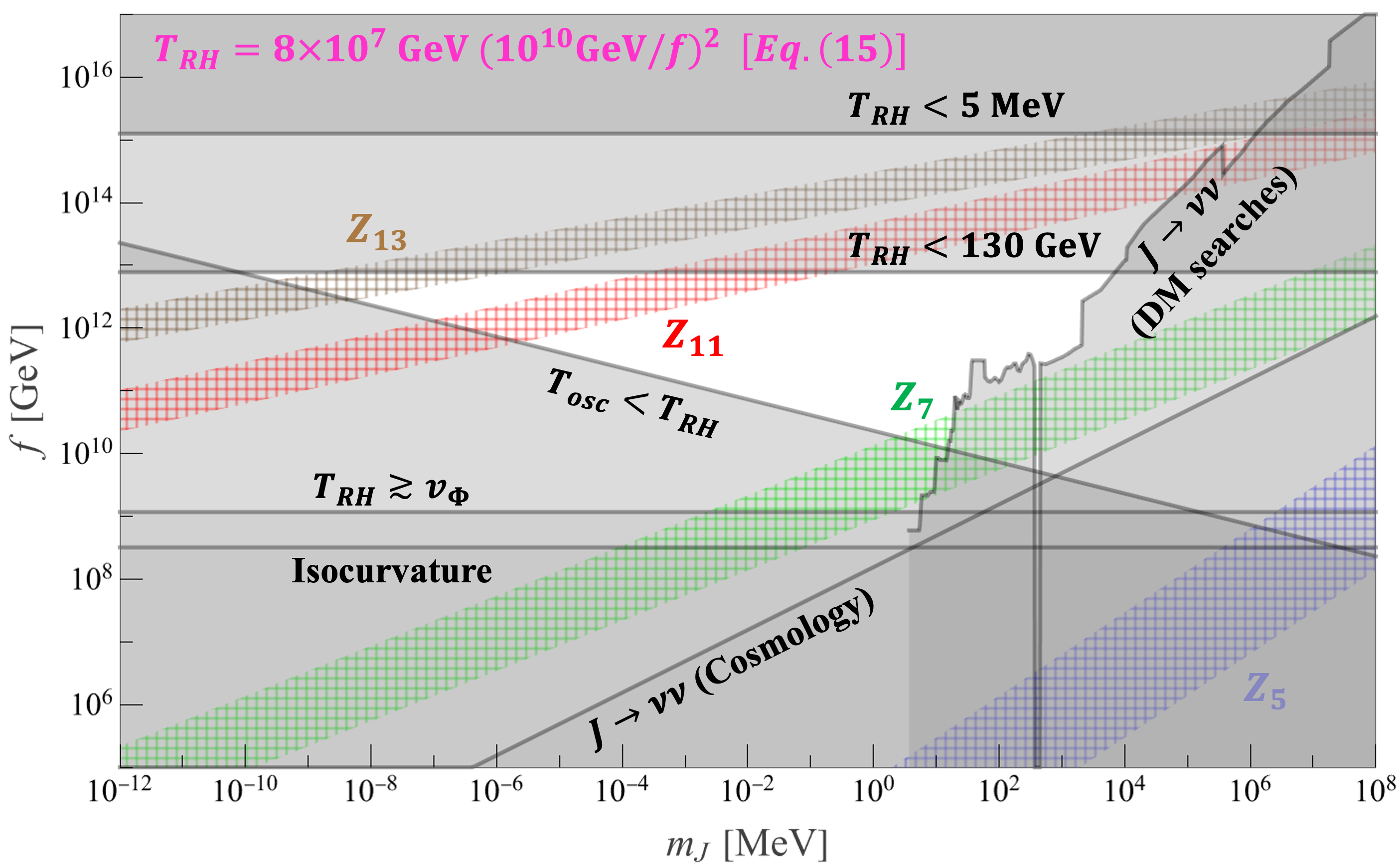} 
    \caption{\small\sl Constraints on the $(m_J,f)$ parameter space for majoron dark matter produced via the misalignment mechanism during an early matter-dominated era, assuming the observed dark matter abundance $\Omega_J h^2 = 0.12$ and taking $\theta_0=1$ in Eq.~\eqref{eq: misalgnmentinEMD}. Shown are the region with $T_{\rm osc}<T_{\rm RH}$, where the early matter-dominated misalignment formula is not applicable; the region excluded by successful BBN, $T_{\rm RH}<5\,{\rm MeV}$; the region violating the pre-inflationary condition, $T_{\rm RH}\gtrsim v_\Phi=Nf$, shown for $N=5$ (larger $N$ gives a weaker constraint); the isocurvature constraint in Eq.~\eqref{eq: isoconstraint}; the region incompatible with leptogenesis based on sphaleron conversion, $T_{\rm RH}<130\,{\rm GeV}$; and the cosmological and dark matter search constraints from majoron decay into neutrinos, also shown in Fig.~\ref{fig: RD result}. The hatched bands show the mass predictions for $Z_N$ gauge symmetry models, as in Fig.~\ref{fig: RD result}.}
    \label{fig: EMD result}
\end{figure}

For the early matter-dominated expression to apply, the majoron must start oscillating before reheating is completed, namely ``$T_{\rm osc}$''$>T_{\rm RH}$.\footnote{
    Here ``$T_{\rm osc}$'' denotes the formal temperature from the radiation-dominated expression given above. In the early matter-dominated case, this temperature is not realized in the actual thermal history, but merely serves as a convenient label for the condition $H \sim m_J$, which is why we explicitly enclose it in quotation marks.}
In addition, the reheating temperature is subject to the lower bound from successful BBN, $T_{\rm RH}\gtrsim 5\,{\rm MeV}$\,\cite{deSalas:2015glj, Barbieri:2025moq}. Moreover, the pre-inflationary scenario considered here requires $T_{\rm RH}<v_\Phi=Nf$, so that the $U(1)_{B-L}$ symmetry is not thermally restored after inflation.\footnote{
    If inflaton decay is not instantaneous, the temperature of the radiation bath can first rise to a maximum value $T_{\rm max}>T_{\rm RH}$ and then decrease, with $T_{\rm RH}$ conventionally associated with the epoch when the radiation and inflaton energy densities become comparable. The precise condition for avoiding thermal restoration of the $U(1)_{B-L}$ symmetry therefore depends on the reheating and thermalization dynamics and may be stronger than $T_{\rm RH}<v_\Phi$. The $Z_5$ model is most sensitive to such a stronger condition, but it is already excluded by the majoron decay into neutrinos discussed in the next subsection; we therefore do not pursue this issue further.}
Combining $\Omega_Jh^2\simeq0.12$ in Eq.~(\ref{eq: misalgnmentinEMD}) with $T_{\rm RH}<Nf$, we obtain $f\gtrsim 2\times10^9\,{\rm GeV}\,N^{-1/3}\theta_0^{-2/3}$. The condition ``$T_{\rm osc}$''$>T_{\rm RH}$ is more easily satisfied for a heavier majoron, since a larger $m_J$ implies an earlier onset of oscillations. In this regime, the absence of the mass dependence discussed above means that, for fixed $T_{\rm RH}$ and $\theta_0$, reproducing the observed dark matter abundance requires a larger decay constant $f$ than in the standard radiation-dominated case. In Fig.~\ref{fig: EMD result}, we show these constraints, assuming that the majoron accounts for the observed dark matter abundance and taking $\theta_0=1$, i.e., $T_{\rm RH}\simeq8\times10^7\,{\rm GeV}\,(10^{10}\,{\rm GeV}/f)^2$. The gray-shaded regions indicate where the early matter-dominated expression is not applicable (``$T_{\rm osc}$''$<T_{\rm RH}$), where the reheating temperature is below the BBN bound ($T_{\rm RH}\lesssim5\,{\rm MeV}$), or where the pre-inflationary condition is violated ($T_{\rm RH}\gtrsim Nf$), taking $N=5$ for the last condition (with larger $N$ giving a weaker constraint). As in Fig.~\ref{fig: RD result}, we show the parameter regions predicted by the different $Z_N$ gauge symmetry models. We find that the $Z_7$ case predicts a dark matter mass larger than a few MeV, somewhat larger than in the standard radiation-dominated misalignment mechanism shown in Fig.~\ref{fig: RD result}, reflecting the larger value of $f$ required in the early matter-dominated case.

Unlike in the radiation-dominated misalignment scenario, thermal production is negligible in the early matter-dominated case considered here. This is because $T_{\rm RH}$ is typically well below $M_I$, strongly suppressing production involving $\mathcal N_I$. Even when $T_{\rm RH}\sim M_I$, Eq.~(\ref{eq: misalgnmentinEMD}) together with $M_I=y_I^{(\mathcal N)}Nf/\sqrt{2}$ gives $y_I^{(\mathcal N)}\simeq1.6\times10^{-3}(7/N)(10^{10}\,{\rm GeV}/f)^3\theta_0^{-2}$. For the viable models with $N\geq7$ considered in our analysis, the relatively large values of $f$ required by the misalignment abundance can thus suppress $y_I^{(\mathcal N)}$ sufficiently for the freeze-in contribution to be negligible. The $Z_5$ case, for which this suppression can be weaker, is independently excluded by the majoron decay into neutrinos discussed in the next subsection.

\subsection{Constraints on majoron dark matter}

Production of majoron dark matter via misalignment is subject to isocurvature constraints, which bound the inflationary scale and reheating temperature. Through leptogenesis, these bounds further imply constraints on the right-handed neutrino masses, or equivalently on the $\mathrm{U}(1)_{B-L}$ breaking scale. Majoron dark matter is also constrained by its decays through cosmological observations and indirect searches. We summarize these constraints below.

\subsubsection{Isocurvature constraints}
\label{subsec: isocurvature}

In the pre-inflationary scenario, where the $\mathrm{U}(1)_{B-L}$ symmetry is broken during inflation, the initial misalignment angle $\theta_0$ acquires quantum fluctuations of magnitude $\delta\theta = H_I/(2\pi f)$, with $H_I$ denoting the Hubble scale during inflation. As is generic for a light scalar field with a mass much smaller than $H_I$ during this epoch, these fluctuations give rise to isocurvature perturbations~\cite{OHare:2024nmr, Hertzberg:2008wr,Kobayashi:2013nva}, which are constrained by CMB observations~\cite{Planck:2018jri} as follows:
{\small
\begin{align}
    A_{\rm iso} =
    \left(\frac{\delta \rho_J}{\rho_J}\right)^2 =
    \left(\frac{2 \delta \theta}{\theta_0}\right)^2 =
    \left(\frac{H_I}{\pi f \theta_0}\right)^2 \lesssim
    8.4 \times 10^{-11}
    \rightarrow
    H_I \lesssim 2.9 \times 10^5\,\mathrm{GeV} ~\left(\frac{f \theta_0}{10^{10}\,\mathrm{GeV}}\right),
    \label{eq: iso constraint}
\end{align}
}using $\rho_J \propto \theta_0^2$, as in Eqs.~(\ref{eq: misalignment}, \ref{eq: misalgnmentinEMD}).\footnote{
    Anharmonicity modifies the factor $2/\theta_0$ in the isocurvature fluctuation formula to $\partial\ln\Omega_J/\partial\theta_0$. For $0.5\leq\theta_0\leq1.5$ adopted here, the resulting correction remains at the $\mathcal O(10\%)$ level and does not alter our conclusions.}
This gives an upper limit on the reheating temperature,
\begin{align}
    T_{\rm RH} \lesssim
    \left(\frac{90}{\pi^2 g_\star(T_{\rm RH})}\right)^{1/4} \sqrt{H_I m_{\rm pl}} \lesssim
    4.5 \times 10^{11}\,\text{GeV}\,\left(\frac{f \theta_0  }{10^{10}\,\mathrm{GeV}}\right)^{1/2}.
    \label{eq: reheating constraint}
\end{align}
In addition, the non-observation of primordial tensor modes in CMB measurements, as constrained by Planck, places an upper bound on the tensor-to-scalar ratio $r$, implying $H_I \lesssim 10^{13}\,\mathrm{GeV}$~\cite{Planck:2018jri}. Combining this observational bound with the isocurvature and reheating constraints in Eqs.~(\ref{eq: iso constraint}) and (\ref{eq: reheating constraint}), we obtain upper limits on $H_I$ and $T_{\rm RH}$ as
\begin{align}
    \,H_I \lesssim &
    {\rm min}\,\bigg[2.9 \times 10^5\,\mathrm{GeV} \left(\frac{f \theta_0}{10^{10}\,\mathrm{GeV}}\right),
    \quad
    10^{13}\,\mathrm{GeV}\bigg],
    \nonumber\\
    T_{\rm RH} \lesssim &
    {\rm min}\,\bigg[4.5 \times 10^{11}\,\text{GeV}\,\left(\frac{f \theta_0}{10^{10}\,\mathrm{GeV}}\right)^{1/2},
    \quad
    2.7 \times 10^{15}\,\mathrm{GeV} \bigg].
    \label{eq: iso constrint 2}
\end{align}
In the following, we discuss how these constraints shape the allowed model parameter space in the radiation-dominated and early matter-dominated misalignment scenarios, and what implications they have for leptogenesis~\cite{Fukugita:1986hr} as the origin of the observed baryon asymmetry.

\paragraph{Radiation-dominated misalignment scenario:}

The conventional isocurvature bound on $H_I$ implies a low inflation scale. For example, the $Z_7$ model, which predicts $m_J$ in the MeV range, requires $H_I\lesssim\mathcal{O}(10^5)\,\mathrm{GeV}$, while the $Z_{13}$ model allows a higher scale, $H_I\lesssim\mathcal{O}(10^8)\,\mathrm{GeV}$. The pre-inflationary misalignment scenario thus favors low-scale inflation. 

An interesting possibility is that the isocurvature constraint on $H_I$ can be relaxed if the radial component of the $U(1)_{B-L}$-breaking field $\Phi$ takes a value much larger than $v_\Phi$ during inflation, as in Linde's mechanism for the QCD axion~\cite{Linde:1991km}. In this case, the majoron misalignment-angle fluctuation is suppressed as $\delta\theta_J \simeq H_I/(2\pi f_I)$ with $f_I \gg f$, rather than being determined by the low-energy decay constant $f = v_\Phi/N$. As emphasized in subsequent analyses~\cite{Kawasaki:2018qwp}, however, the post-inflationary dynamics of the radial mode, such as possible parametric resonance and defect formation, must be under control. If these issues are avoided, $f_I$ can be as large as $\mathcal{O}(10^4)\,f$, leading to a substantial suppression of isocurvature perturbations.\footnote{
    The enhancement is bounded by the requirement that the field value of the radial component during inflation, $|\Phi_{\rm inf}|$, remain below $m_{\rm pl}$, unless trans-Planckian field values are justified within the UV completion.}
Under these conditions, the majoron misalignment mechanism can accommodate an inflation scale well above the conventional isocurvature bound.

On the other hand, the bounds on the reheating temperature are less restrictive. The consistency conditions $T_{\rm osc}\lesssim T_{\rm RH}\lesssim v_\Phi=Nf$ determine the reheating window for the radiation-dominated misalignment scenario. Using $f\sim10^{10}\,\mathrm{GeV}$ for the $Z_7$ model and $f\sim10^{12.5}\,\mathrm{GeV}$ for the $Z_{13}$ model, we obtain $10^7\,\mathrm{GeV}\lesssim T_{\rm RH}\lesssim10^{11}\,\mathrm{GeV}$ and $10^3\,\mathrm{GeV}\lesssim T_{\rm RH}\lesssim10^{14}\,\mathrm{GeV}$, respectively. The isocurvature bound given above should then be imposed on these windows; for $\theta_0\sim1$, it does not further restrict the $Z_7$ window, whereas for the $Z_{13}$ model it lowers the upper end of the allowed range to approximately $10^{13}\,\mathrm{GeV}$.

These considerations have implications for leptogenesis. In conventional thermal leptogenesis with a hierarchical right-handed-neutrino spectrum, successful baryogenesis typically requires $T_{\rm RH}\gtrsim M_1\gtrsim10^9\,{\rm GeV}$~\cite{Davidson:2002qv,Buchmuller:2004nz}. This lower bound can be relaxed to $M_1\sim10^8$--$10^9\,{\rm GeV}$ when flavor effects and a mildly non-hierarchical right-handed-neutrino spectrum are taken into account~\cite{Nardi:2006fx,Hambye:2003rt,Blanchet:2008pw}. As discussed in Sec.~\ref{subsec: misalignment}, such scenarios can be compatible with a misalignment-dominated majoron abundance when $y_1^{(\mathcal N)}$ is sufficiently suppressed, particularly in models with larger $f$. Alternatively, leptogenesis can proceed nonthermally through inflaton decays into $\mathcal N_I$ with $T_{\rm RH}\ll M_I$~\cite{Asaka:1999yd,Asaka:1999jb}, in which case thermal production involving $\mathcal N_I$ is strongly suppressed.\footnote{
    Majoron production from these $\mathcal N_I$ is also suppressed, leaving the abundance dominated by misalignment.}
At even lower scales, resonant leptogenesis with quasi-degenerate right-handed neutrinos can generate the observed baryon asymmetry~\cite{Pilaftsis:2003gt}. In particular, it can operate for $T_{\rm RH}\sim M_I\ll f$, corresponding to suppressed $y_I^{(\mathcal N)}$, where the thermal majoron contribution remains negligible. Thus, viable leptogenesis scenarios are also compatible with the misalignment-dominated regions considered here.

\paragraph{Early matter-dominated misalignment scenario:}

The isocurvature bound on $H_I$ has a different implication from the radiation-dominated case discussed above. Assuming that the majoron accounts for the observed dark matter abundance, $\Omega_J h^2=0.12$, the misalignment relic-abundance condition in Eq.~(\ref{eq: misalgnmentinEMD}) gives $f\theta_0\propto T_{\rm RH}^{-1/2}$. Thus, a lower reheating temperature requires a larger $f\theta_0$ and relaxes the bound on $H_I$. While a high reheating temperature favors low-scale inflation, a lower $T_{\rm RH}$ can allow a substantially higher inflation scale.

The upper limit on $T_{\rm RH}$ in Eq.~(\ref{eq: iso constrint 2}) implies a lower limit on $f\theta_0$ using the misalignment relic-abundance condition in the early matter-dominated scenario, namely $f\theta_0\propto T_{\rm RH}^{-1/2}$:
\begin{align}
    f \theta_0 \gtrsim
    3.2 \times 10^8\,{\rm GeV}.
    \label{eq: isoconstraint}
\end{align}
This bound is shown as the corresponding gray-shaded region in Fig.~\ref{fig: EMD result}. Using the relic-abundance condition with $\theta_0=1$, it can equivalently be expressed as $T_{\rm RH} \lesssim 8.0 \times10^{10}\,\mathrm{GeV}$. Indeed, as seen in Fig.~\ref{fig: EMD result}, comparison with the predictions of the specific $Z_N$ models shows that reproducing the observed dark matter abundance generally requires even lower reheating temperatures. Consequently, all the $Z_N$ models except $Z_5$ fail to allow a reheating temperature high enough for conventional thermal leptogenesis to be successful.\footnote{
    The $Z_5$ model is excluded by cosmological and dark matter search constraints, as shown later. Hence, none of the models in the early matter-dominated scenario is compatible with conventional thermal leptogenesis.}

Nevertheless, the lower reheating temperatures required in the early matter-dominated scenario can still be compatible with leptogenesis. As discussed above, nonthermal or low-scale mechanisms, such as resonant leptogenesis, can generate the observed baryon asymmetry even for $T_{\rm RH}\ll10^9\,{\rm GeV}$. Note that thermal production of the majoron is typically negligible in this scenario: $T_{\rm RH}$ is generally below $M_I$, while even for $T_{\rm RH}\sim M_I$, the values of $f$ required by the misalignment abundance imply a suppressed $y_I^{(\mathcal N)}$ for the viable models with $N\geq7$. The remaining requirement is that the lepton asymmetry be generated while electroweak sphalerons are active and converted into a baryon asymmetry. Since sphalerons freeze out below about $130\,{\rm GeV}$~\cite{DOnofrio:2014rug}, we require $T_{\rm RH}\gtrsim130\,{\rm GeV}$. Accordingly, in Fig.~\ref{fig: EMD result}, we shade the region with $T_{\rm RH}<130\,{\rm GeV}$, while the $Z_7$, $Z_{11}$, and $Z_{13}$ models remain viable.

\subsubsection{Constraints on majoron decay into neutrinos}
\label{subsubsec: constraints from J to nunu}

Here we discuss majoron dark matter decay into active SM neutrinos, its effects on the early Universe, and the constraints from cosmology, as well as those from indirect dark matter searches. As shown in Eq.~(\ref{eq: decays@tree}), the decay into neutrinos dominates the decay rate and may modify cosmic expansion if its rate is sufficiently large. If an appreciable fraction of dark matter is converted into relativistic neutrinos, the matter and radiation energy densities are redistributed, changing the Hubble rate and affecting early-Universe observables. Precision CMB and baryon acoustic oscillation data strongly constrain such departures from standard cosmological evolution, leading to an upper bound on the decay width in Eq.~(\ref{eq: decays@tree}):
\begin{align}
    \Gamma[J \to \nu\nu] \simeq \frac{m_J}{16 \pi N^2 f^2} \sum_{i = 1}^3 m_i^2 <
    1.3 \times 10^{-19}\,\mathrm{s}^{-1},
    \label{eq: nunu}
\end{align}
at 95\% C.L.\,\cite{Alvi:2022aam}, where $N$ labels the $Z_N$ symmetry. This limit is shown by the gray shaded regions in Figs.~\ref{fig: RD result} and \ref{fig: EMD result} for the $Z_5$ case (larger N gives a weaker constraint), assuming normal ordering of the active neutrino masses with the lightest neutrino mass set to zero.

In addition to the cosmological bound discussed above, this decay mode is constrained by indirect dark matter searches. We use the lifetime limits on dark matter decaying into neutrino final states from various experiments, as compiled in Refs.~\cite{Arguelles:2022nbl, Akita:2023qiz}. The majoron dark matter decays into a neutrino pair, $J\to \nu_i\nu_i$, in the neutrino mass-eigenstate basis, where $i$ labels the mass eigenstates, with a rate proportional to $m_i^2$. Thus, for example, for normal ordering with the lightest neutrino mass set to zero, the decay is dominated by the heaviest mass eigenstate, leading to a flavor composition approximately given by $\nu_e:\nu_\mu:\nu_\tau \simeq 0.02:0.5:0.5$. Taking this flavor dependence into account is particularly important in the low-$m_J$ region, where the relevant searches rely on $\bar{\nu}_e$ detection through inverse beta decay. We therefore include the flavor dependence of the neutrino signal when deriving the limits for $m_J\lesssim 100~{\rm MeV}$. In the higher-mass region, on the other hand, we use the results of Ref.~\cite{Arguelles:2022nbl}, where the limits were derived assuming universal couplings to all neutrino flavors. The difference between this flavor-dependent treatment and the flavor-universal approximation in the higher-mass region is expected to be at the level of several tens of percent, comparable to the astrophysical uncertainty associated with the $D$-factor, namely that from the assumed Galactic dark matter profile. In Figs.~\ref{fig: RD result} and \ref{fig: EMD result}, using Eq.~(\ref{eq: decays@tree}), we show for comparison the constraint for the $Z_5$ discrete symmetry as the gray-shaded region, assuming normal ordering of the active SM neutrino masses with the lightest neutrino mass set to zero. The constraints become progressively weaker for larger $N$.

As seen in the figures, the cosmological and astrophysical constraints from majoron dark matter decay into neutrinos exclude the $Z_5$ model. On the other hand, the $Z_7$, $Z_{11}$, and $Z_{13}$ models remain viable because the decay width is more strongly suppressed for larger $N$, and models with larger $N$ predict larger values of $f$ for a given $m_J$, as shown in Eq.~(\ref{eq: nunu}). In particular, the favored majoron mass in the $Z_7$ model is around the MeV scale in both standard and early matter-dominated misalignment scenarios. This region will be a main target of future gamma-ray observations, as discussed in the following sections.

\subsubsection{Constraints on majoron decay into an electron--positron pair}
\label{subsubsec: constraint on J to e e}

Majoron dark matter can decay into an electron--positron pair through a one-loop process when its mass exceeds the kinematic threshold, $m_J>2m_e$. Among the viable $Z_n$ models, the $Z_7$ model is the only one that predicts a majoron mass above this threshold and hence allows the decay $J\to e^-e^+$. The resulting phenomenology in the MeV mass range can, in particular, be extensively probed by MeV gamma-ray observatories such as COSI~\cite{Tomsick:2023aue}.

When the majoron dark matter mass lies near the $e^-e^+$ threshold, as suggested by the $Z_7$ model, the decay into an electron--positron pair occurs close to threshold. Photon exchange between the final-state particles induces a long-range force and gives rise to a threshold singularity. This effect can modify the decay rate and must be included. We incorporate the threshold, namely Sommerfeld, effect in estimating the decay width using the potential non-relativistic Lagrangian method\,\cite{Pineda:1998kn, Brambilla:1999xf}. Details are given in Appendix\,\ref{app: Sommerfeld effect}. Accordingly, the decay width is modified from its leading-order expression in Eq.~(\ref{eq: J to ee}) as\,\cite{Hayashi:2024not}
\begin{align}
    \Gamma[J \to e^- e^+] \simeq
    \frac{\pi \alpha /\sqrt{m_J/m_e - 2}}{1 - \exp[-\pi \alpha /\sqrt{m_J/m_e - 2}]}
    \Gamma_0[J \to e^-e^+],
    \label{eq: J to ee kai}
\end{align}
where $\alpha$ denotes the fine-structure constant associated with the Coulomb potential between the final-state particles. In the absence of the Coulomb interaction, corresponding to the limit $\alpha \to 0$, the decay width reduces to the leading-order result, $\Gamma_0[J \to e^-e^+]$ in Eq.~(\ref{eq: J to ee}), as expected. By contrast, in the threshold region $m_J \simeq 2m_e$, the Sommerfeld effect becomes appreciably significant, and the decay width remains finite even at the threshold $m_J = 2m_e$.

\begin{figure}[t]
    \centering
    \includegraphics[keepaspectratio, scale=0.43]{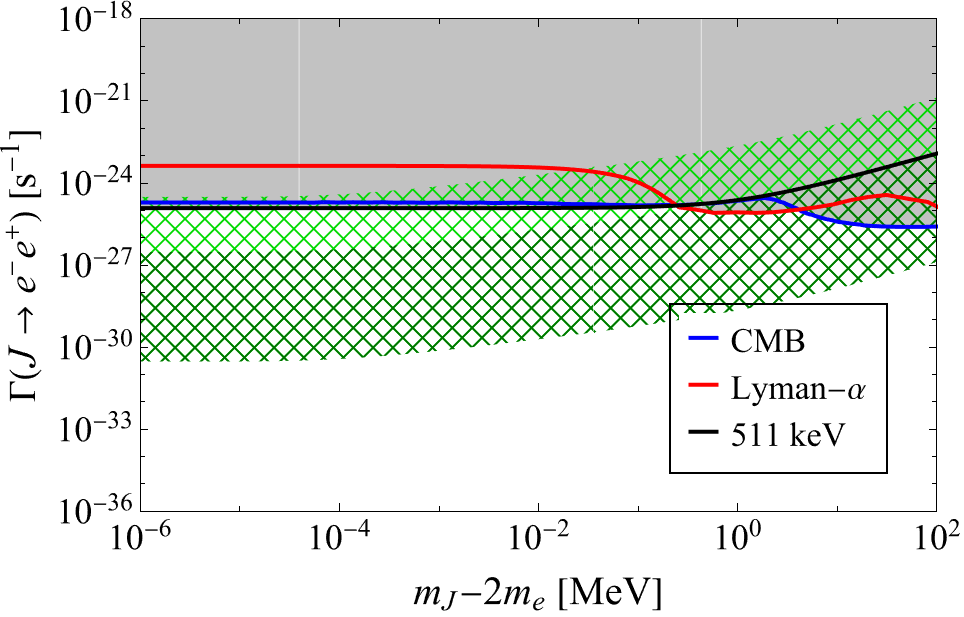}
    \qquad
    \includegraphics[keepaspectratio, scale=0.43]{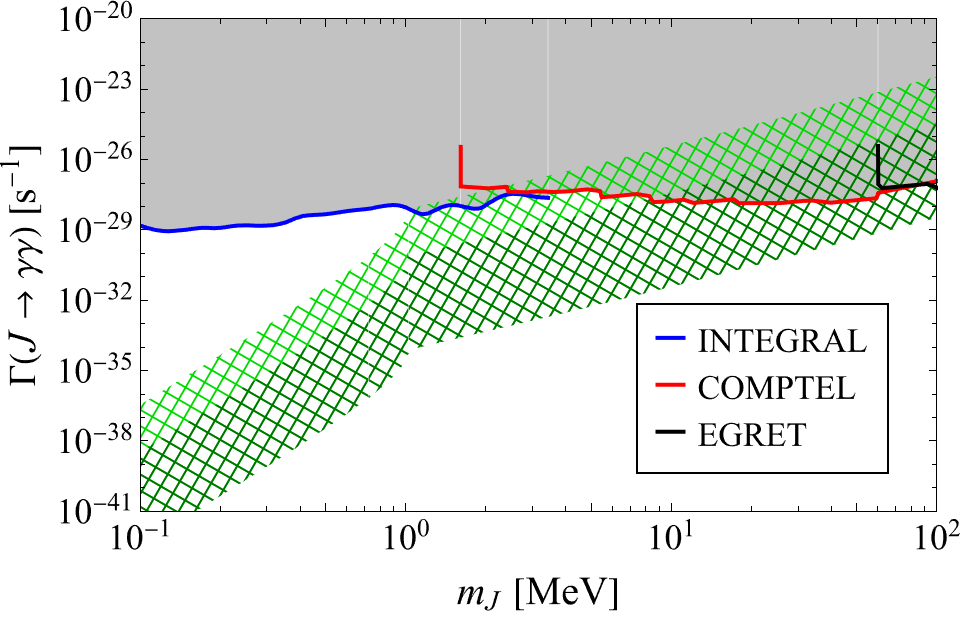}
    \caption{\small\sl {\bf Left panel:} Range of $\Gamma[J \to e^-e^+]$ in the $Z_7$ symmetry model, including the Sommerfeld enhancement. The green hatched region, including light and dark green, shows the central 99.8\,\% interval obtained in our parameter scan as a function of $m_J-2m_e$, while the dark green region corresponds to the subset with $y_I^{(\mathcal N)}<0.01$. The blue and red lines show the upper limits from the CMB and Lyman-$\alpha$ forest, while the black line shows the 511\,keV line-search limit. {\bf Right panel:} Same as the left panel, but for $\Gamma[J \to \gamma\gamma]$ as a function of $m_J$. The blue, red, and black lines show the upper limits from $\gamma$-ray line searches with INTEGRAL, COMPTEL, and EGRET, respectively.}
    \label{fig: ee gammagamma}
\end{figure}

When majoron dark matter decays into $e^-e^+$ in the early Universe, the resulting electromagnetic energy injection can modify the thermal history of the Universe, affecting Big Bang nucleosynthesis (BBN), the ionization history around recombination\,\cite{Xu:2024vdn}, and the thermal evolution of the intergalactic medium (IGM) probed by Ly-$\alpha$ forest observations\,\cite{Liu:2020wqz}. Observationally, the consistency of these probes with the standard cosmological history constrains the corresponding partial decay width. In the mass range of interest, the strongest cosmological limits are provided by CMB measurements and Ly-$\alpha$ forest data, whereas BBN bounds are weaker~\cite{Forestell:2018txr,Depta:2020zbh}. The corresponding 95\,\% C.L. upper bounds on the width are shown in the left panel of Fig.~\ref{fig: ee gammagamma} as the gray-shaded regions with blue and red lines.

In the present Universe, the same decay into an electron--positron pair can instead give rise to signals in indirect dark matter searches. Direct observation of these electrons and positrons near Earth is, however, inefficient for MeV-scale majoron masses, since such low-energy charged particles are strongly affected by solar modulation and cannot efficiently penetrate the heliosphere. Although Voyager~1, which is located beyond the heliopause, could in principle detect such low-energy particles, its sensitivity is not sufficient to probe MeV-scale electrons and positrons from dark matter decay~\cite{Boudaud:2016mos}.\footnote{
    The sensitivity depends on poorly understood low-energy electron and positron propagation in the Galaxy. We neglect possible reacceleration-induced signal enhancement~\cite{DelaTorreLuque:2023olp} to derive conservative constraints.}
Indirect detection is nevertheless possible through MeV-scale positrons produced in majoron dark matter decay. After capturing ambient electrons in the Galaxy, these positrons form positronium, whose two-photon decay, with a branching fraction of about 25\%, produces the characteristic 511~keV line, referred to as a \textit{tertiary photon signal}. The 511~keV line emission has been observed from the Galactic bulge, with a flux of $F_{511}^{\rm obs} \simeq (4.8$--$9.6) \times 10^{-4}$\,ph\,cm$^{-2}$\,s$^{-1}$~\cite{Siegert:2015knp}, although its origin remains unclear.\footnote{
    Here, the bulge region is defined as a circular region of radius $10.3^\circ$ centered on the Galactic center.}
Possible astrophysical sources include microquasars, supernovae, and massive stars~\cite{Siegert:2017thesis, Siegert:2023wus}, while the observed flux morphology in the Galactic bulge disfavors a dominant dark matter decay origin. We therefore use the observed 511~keV flux conservatively to set an upper bound on the partial decay width as
\begin{align}
     F_{511}^{\rm DM} =
     2 \times
     \frac{1}{4} \times
     f_{Ps} \times
     {\cal E}_{ff}
     \times
     \frac{\Gamma[J \to e^-e^+]}{4\pi m_J}\,D
     < F_{511}^{\rm Obs},
     \label{eq: const from 511}
 \end{align}
where the factors of 2 and $1/4$ account for the photons emitted per positronium decay and the two-photon branching fraction of positronium, respectively. The parameter $f_{\rm Ps}$ denotes the fraction of $e^-e^+$ annihilations producing the 511\,keV flux through positronium formation. Since it is observationally consistent with 100\%\,\cite{Siegert:2015knp}, we set it to unity. The $D$-factor is defined by the line-of-sight integral of the dark matter density over the bulge, $D \equiv \int_{\Omega_{\rm Bulge}} d\Omega \int_{\rm l.o.s.} d\ell\,\rho_{\rm DM}[r(\ell,\Omega)]$, where we assume a spherically symmetric profile, $\rho_{\rm DM}[\vec r]=\rho_{\rm DM}[r]$, with $r=|\vec{r}|$ measured from the Galactic center. Following Ref.~\cite{Hayashi:2024not}, we adopt the conservative value $D=3.4 \times 10^{21}\,\mathrm{GeV/cm^2}$. We include the efficiency factor ${\cal E}_{\rm ff}$ to account for positrons produced in the bulge that may escape without annihilating. Since estimates range from 6\% to 100\%~\cite{Siegert:2017thesis, Siegert:2023wus}, we conservatively take the lowest value. The upper bound on $\Gamma(J\to e^-e^+)$ from Eq.~(\ref{eq: const from 511}) is shown in the left panel of Fig.~\ref{fig: ee gammagamma}.\footnote{
     Although Refs.~\cite{DelaTorreLuque:2023cef, Nguyen:2025tkl} show that spatial-morphology data strengthens the constraint, we instead adopt a conservative bound without using this information, owing to its non-negligible uncertainties, as noted there.}

On the theoretical side, Eq.~\eqref{eq: J to ee} shows that the new-physics dependence of $\Gamma[J\to e^-e^+]$ enters through $m_J$ and $|K_{11}-K_{22}-K_{33}|^2$, where $K_{ij}=(m_Dm_D^\dagger)_{ij}/(v_{\rm EW}v_\Phi)$. Although $K_{ij}$ is of order ${\cal O}(m_i y^{(\mathcal{N})}_I/v_{\rm EW})$, with $m_i$ denoting an active neutrino mass and $y^{(\mathcal{N})}_I$ the Yukawa coupling of the right-handed neutrinos to $\Phi$, its precise value also depends on the PMNS matrix $U_{\rm PMNS}$ and the Casas--Ibarra matrix $R$. Consequently, the parameter dependence is more complicated than in the neutrino-decay channel; indeed, $K_{11}-K_{22}-K_{33}$, and hence $\Gamma[J\to e^-e^+]$, can be tuned to vanish. We therefore estimate the typical decay width by scanning over the relevant parameter space, as detailed in Appendix~\ref{app: scannings}. We vary $0\leq m_1\leq 0.1~{\rm eV}$ and $0\leq \alpha_i\leq 2\pi$ $(i=1,2)$, where $\alpha_i$ are the Majorana phases, while fixing the oscillation parameters to their latest normal-ordering best-fit values: $\Delta m_{21}^2=7.49\times10^{-5}~{\rm eV}^2$, $\Delta m_{31}^2=2.513\times10^{-3}~{\rm eV}^2$, $\theta_{12}=33.68^\circ$, $\theta_{23}=43.3^\circ$, $\theta_{13}=8.56^\circ$, and $\delta_{\rm CP}=212^\circ$\,\cite{Esteban:2024eli}. We scan the Casas--Ibarra parameters over $0\leq {\rm Re}[z_i]\leq 2\pi$ with ${\rm Im}[z_i]=0$, and vary the right-handed-neutrino Yukawa couplings over $10^{-3} \leq y^{(\mathcal{N})}_I\leq 10^{-1}$, focusing on relatively heavy right-handed neutrinos, as discussed in Sec.~\ref{subsec: isocurvature}. Assuming flat priors within our scan ranges, we find that the central 99.8\% interval of $|K_{11}-K_{22}-K_{33}|$ is $3.5\times10^{-17}$--$3.6\times10^{-14}$, yielding the range of $\Gamma[J\to e^-e^+]$ shown by the green hatched region in the left panel of Fig.~\ref{fig: ee gammagamma}. Within this band, the dark green region highlights the subset with $y_I^{(\mathcal N)}<0.01$, using $0.01$ as a representative reference value below which freeze-in remains subdominant even for $T_{\rm RH}\sim M_I$, as relevant for thermal leptogenesis discussed in that section. As seen in the figure, part of this parameter space is excluded by current observations, while the region with smaller decay widths remains phenomenologically viable. It is worth emphasizing that $\Gamma[J\to e^-e^+]$ remains finite near the threshold, $m_J\simeq 2m_e$, thanks to the Sommerfeld effect, making this region an interesting target for future indirect dark matter searches.

\subsubsection{Constraints on majoron decay into a photon pair}
\label{subsubsec: constraint on J to gamma gamma}

Majoron dark matter can decay into two photons through the two-loop process given in Eq.~\eqref{eq: J to gammagamma}. We again focus on the $Z_7$ model, for reasons similar to those for the $e^-e^+$ decay.

As seen in Eq.~(\ref{eq: J to gammagamma}), majoron dark matter decays into two photons through diagrams in which charged SM particles run in loops. As discussed there, the contribution to the $J\to\gamma\gamma$ decay amplitude is suppressed when the charged particle running in the loop is much heavier than the majoron, because the Adler--Bardeen theorem ensures the absence of an unsuppressed local $J F_{\mu\nu}\tilde F^{\mu\nu}$ term. Thus, in the MeV range relevant for the $Z_7$ model, contributions from charged SM particles beyond the first generation are negligible. For first-generation quarks, the perturbative light-quark-loop description should be replaced by a hadronic one, in which the relevant contributions are described by neutral pseudoscalar-meson poles and chiral loops. In the main region of interest, $m_J=O(1$--$10)\,{\rm MeV}$, one has $m_J \ll m_\pi$, and these hadronic contributions are expected to be suppressed by $m_J^2/m_{\rm had}^2$, with $m_{\rm had} \gtrsim m_\pi$. We therefore expect them to be numerically subdominant compared with the leading electron-loop contribution. In this estimate we therefore retain only the electron-loop contribution. For masses approaching the hadronic scale, however, this estimate should be regarded as indicative, since a dedicated hadronic treatment would be required.

When this decay occurs in the early Universe, it injects electromagnetic energy into the cosmological plasma and is therefore subject to constraints from CMB measurements and Ly-$\alpha$ forest observations, as in the $e^-e^+$ channel. However, since the two-photon mode arises only at the two-loop level, the predicted partial decay width is much more suppressed than that of the one-loop $e^-e^+$ mode, and the resulting cosmological constraints are correspondingly less severe. We therefore do not explicitly impose these constraints in our analysis.

On the other hand, the decay into two photons in the present Universe gives a distinctive indirect-detection signature. In general, majoron decay can produce photon signals in two different ways: final-state radiation associated with the three-body decay $J \to e^- e^+ \gamma$, and the monochromatic line from the two-photon decay $J\to\gamma\gamma$, whose partial width is given in Eq.~(\ref{eq: J to gammagamma}). The former contribution produces a continuum spectrum and is suppressed by the loop-induced $J\to e^-e^+$ coupling as well as the three-body phase space. It therefore does not lead to a stronger constraint than that from the $e^-e^+$ channel itself. The latter contribution is two-loop suppressed, but it produces a sharp monochromatic photon line at $E_\gamma=m_J/2$, which is a particularly clean target for indirect searches against astrophysical backgrounds. Since no such line has been observed by current X-ray and $\gamma$-ray telescopes, such as INTEGRAL, COMPTEL, and EGRET~\cite{Calore:2022pks, Essig:2013goa}, the resulting null results place an upper bound on the partial width $\Gamma(J\to\gamma\gamma)$, as shown in the right panel of Fig.~\ref{fig: ee gammagamma}.

As in the case of the decay into $e^-e^+$, we also estimate the partial decay width into two photons using the full parameter scan detailed in Appendix~\ref{app: scannings}. The predicted range of $\Gamma[J\to \gamma\gamma]$ is shown in the right panel of Fig.~\ref{fig: ee gammagamma} as the full green hatched region, with the dark green subset defined in the same way as in the left panel. As seen from the figure, within the electron-loop approximation adopted above, some of the parameter region with $m_J \gtrsim$ a few MeV appears to be excluded by current indirect-detection observations, despite the two-loop suppression of the decay rate. Meanwhile, the region with $m_J \lesssim$ a few MeV remains only rather weakly constrained, primarily because of the very strong mass dependence of the decay width itself, $\Gamma[J\to\gamma\gamma]\propto m_J^7/(m_e^4\,v_{\rm EW}^2)$, as directly follows from Eq.~(\ref{eq: J to gammagamma}).

\section{Testing Majoron Dark Matter with MeV Gamma Rays}
\label{sec: COSI}

As discussed in the previous section, the minimal majoron model based on a gauged $Z_N$ subgroup of $U(1)_{3(B-L)}$ provides a promising dark matter candidate. In particular, the model with a gauged $Z_7$ symmetry predicts a majoron dark matter mass in the MeV range, making a broad region of its parameter space testable through searches for signals from majoron decays into an $e^-e^+$ pair and two photons in the present Universe. In this section, we investigate the prospects for testing this scenario with forthcoming MeV gamma-ray observations, with particular emphasis on \textit{COSI}, a NASA SMEX mission scheduled for launch in 2027\,\cite{Tomsick:2023aue}. The excellent energy resolution of COSI in the 200\,keV--5\,MeV energy range provides a powerful handle on narrow gamma-ray features, including the 511\,keV line from the $e^-e^+$ decay channel and the monochromatic line at $E_\gamma=m_J/2$ from the two-photon decay channel.

\subsection{Search for the 511~keV line signal}

To assess the sensitivity of COSI to the 511\,keV line from majoron dark matter decay, we take the region of interest (RoI) to be the sky outside the Galactic bulge throughout this analysis. This choice is guided by the spatial distribution of the observed 511\,keV emission. As discussed in the previous section, its morphology is difficult to reconcile with a dominant dark matter origin and instead points to astrophysical sources as the main contributors. Since this astrophysical component is strongly concentrated toward the bulge, whereas the decay signal from dark matter is expected to be more extended over the sky, excluding the bulge can reduce background contamination while preserving sensitivity to a possible dark matter contribution. This strategy is particularly well suited to Compton telescopes such as COSI. The signal would appear as a broad diffuse component in the 511\,keV sky, which can be efficiently probed with COSI's Compton-imaging capability, whereas previous searches relying on coded-mask instruments are less suited to such extended emission\,\cite{Tomsick:2021wed, Aramaki:2022zpw, Tomsick:2023aue}.

\begin{figure}[t]
    \centering
    \includegraphics[keepaspectratio, scale=0.52]{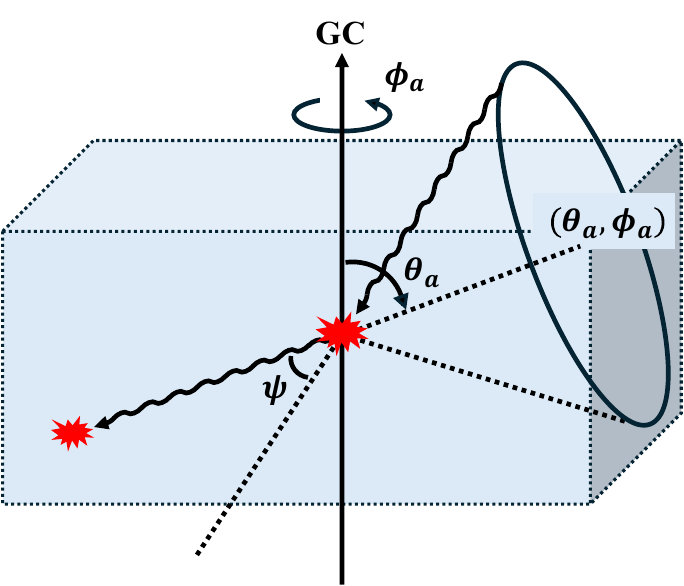}
    \qquad\qquad\quad
    \includegraphics[keepaspectratio, scale=0.32]{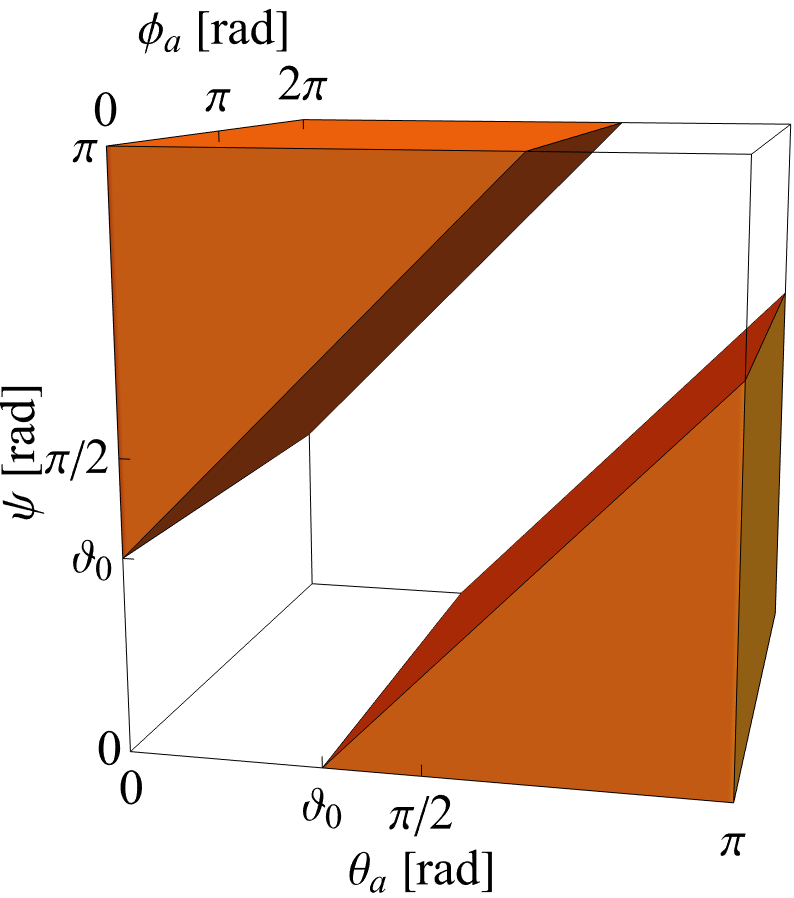}
    \caption{\small \sl {\bf Left panel:} Schematic illustration of the Compton-event variables used in this work for the 511\,keV line analysis. The angle $\psi$ denotes the Compton scattering angle, while $(\theta_a,\phi_a)$ specifies the apparent event direction as reconstructed by the detector. {\bf Right panel:} Representation of the corresponding Compton data space, parameterized by $(\psi,\theta_a,\phi_a)$. Each detected event is mapped to a point in this space, and the red shaded region indicates the RoI used in our analysis.}
    \label{fig: detector}
 \end{figure}

To specify the RoI quantitatively, we define the bulge region as an annulus satisfying $\theta_{\rm RoI,1}(\psi) < \theta_a < \theta_{\rm RoI,2}(\psi)$, where $\psi$ is the event-by-event Compton scattering angle reconstructed by COSI. The $\psi$ dependence arises because, in Compton imaging, each event is represented by a point in the three-dimensional Compton data space $(\theta_a,\phi_a,\psi)$, rather than by a unique incident direction, as illustrated in the left panel of Fig.~\ref{fig: detector}. We determine this annulus from a reference angular radius $\vartheta_0$ defined in the sky-coordinate slice at $\psi=0$.

We choose this angular radius $\vartheta_0$ using the following benchmark prescription. First, $\vartheta_0$ is motivated by the angular position at which the astrophysical component of the observed 511\,keV flux equals the dark matter contribution, $d{\cal F}_{511}^{\rm Obs}(\vartheta_0)/d\Omega-d{\cal F}^{\rm DM}_{511}(\vartheta_0)/d\Omega=d{\cal F}^{\rm DM}_{511}(\vartheta_0)/d\Omega$. Here, the two relevant surface brightnesses are estimated as follows:
\begin{align}
    \frac{d{\cal F}_{511}^{\rm Obs}(\theta)}{d\Omega} =
    \frac{1}{4\pi} \int_{\rm l.o.s} d\ell\,n_{511}(r),
    \quad
    \frac{d{\cal F}_{511}^{\rm DM}(\theta)}{d\Omega} \simeq
    \frac{\Gamma(J \to e^- e^+)}{8 \pi m_J}
    \int_{\rm l.o.s} d\ell\,\rho_{\rm DM}[r].
    \label{eq:equation for reference angular radius}
\end{align}
The observed 511\,keV bulge profile is modeled as $n_{511}(r)=n_{\rm BB}(r)+n_{\rm NB}(r)+n_{\rm c}(r)$, where $n_{\rm BB}(r)$, $n_{\rm NB}(r)$, and $n_{\rm c}(r)$ denote the broad, narrow, and central bulge components, respectively, from Ref.~\cite{Skinner:2015}. We take $n_{\rm BB}(r)=f_{\rm BB}e^{-r^2/(2\sigma_{\rm BB}^2)}$, $n_{\rm NB}(r)=f_{\rm NB}e^{-r^2/(2\sigma_{\rm NB}^2)}$, and $n_{\rm c}(r)=f_{\rm c}\delta(r)$, with $(f_{\rm BB},\sigma_{\rm BB})=(3.80\times10^{-2}\,\mathrm{cm^{-2}\,s^{-1}\,kpc^{-1}},1.26\,\mathrm{kpc})$, $(f_{\rm NB},\sigma_{\rm NB})=(3.08\times10^{-1}\,\mathrm{cm^{-2}\,s^{-1}\,kpc^{-1}},0.363\,\mathrm{kpc})$, and $f_{\rm c}=8.0\times10^{-5}\,\mathrm{cm^{-2}\,s^{-1}}$. For the dark matter contribution, we take its morphology to trace the dark matter density profile, adopting the spherically symmetric cored profile as in Eq.~(\ref{eq: const from 511}), $\rho_{\rm DM}[r]=\rho_s/{(1+r/r_s)[1+(r/r_s)^2]}$, with $\rho_s\simeq0.71\,{\rm GeV/cm^3}$ and $r_s=12.7\,{\rm kpc}$. In evaluating the signal flux, we neglect the spatial smearing of the 511\,keV emission caused by positron propagation before annihilation. Such propagation could enhance the 511\,keV signal outside the bulge, since positrons produced in the bulge may travel outward before forming positronium. However, the propagation of low-energy positrons in the Galaxy remains uncertain~\cite{Jean:2009zj, Prantzos:2010wi, Siegert:2021upv}; we do not include this effect and leave its estimate for future work. We also adopt $f_{\rm Ps}=1$ as noted in Sec.~\ref{subsubsec: constraint on J to e e}. In addition, the possible reduction of the 511 keV emission due to positron escape is not included, since the equation is used only as an approximate surface-brightness estimate for defining $\vartheta_0$. Since $d{\cal F}^{\rm DM}_{511}/d\Omega\propto \Gamma(J\to e^{-}e^{+})/m_J$, the surface-brightness condition alone does not uniquely determine $\vartheta_0$. It instead relates $\vartheta_0$ to the assumed signal normalization. 

Next, we introduce the corresponding $\psi$-dependent angular boundaries $\theta_{\rm RoI,1}(\psi)$ and $\theta_{\rm RoI,2}(\psi)$, as schematically illustrated in the right panel of Fig.~\ref{fig: detector}. While the bulge region is defined at $\psi=0$ by $0 \leq \theta_a \leq \vartheta_0$, for each nonzero $\psi$ the angular range contaminated by photons emitted from the bulge, as effectively viewed from the detector, is shifted to $\theta_{\rm RoI,1}(\psi) \leq \theta_a \leq \theta_{\rm RoI,2}(\psi)$, where $\theta_{\rm RoI,1}(\psi)\equiv \max[0,\psi-\vartheta_0]$ and $\theta_{\rm RoI,2}(\psi)\equiv \vartheta_0+\psi$. Thus, at fixed $\psi$, the complementary angular region, $0 \leq \theta_a < \theta_{\rm RoI,1}(\psi)$ and $\theta_{\rm RoI,2}(\psi) < \theta_a \leq \pi$, is therefore free from contamination by photons from the bulge region. We therefore adopt this complementary region as the RoI for the dark matter signal search in our analysis.

Accordingly, the total signal flux integrated over the RoI can be written in the form:
\begin{align}
    &
    {\cal F}^{\rm DM}_{511}|_{\rm RoI} =
    (2\pi)\int_0^\pi 
    P(\psi)\,d\psi
    \int_0^\pi \sin\theta\,d\theta
    \int_{\theta_a(\theta, \psi, \varphi)\,\subset\,{\rm RoI}} \frac{d\varphi}{2\pi}\,
    \widetilde{\mathcal E}_{ff}(\theta)\,\frac{d{\cal F}_{\rm DM}^{511}(\theta)}{d\Omega},
    \nonumber \\
    &
    P(\psi) =
    \frac{9 \sin\psi}{40 - 27\log 3}
    \frac{2 + (1-\cos\psi)^3}{(2-\cos\psi)^3}.
\end{align}
Here $P(\psi)$ denotes the normalized Klein--Nishina angular distribution, defined by $P(\psi)\equiv \sigma_{\rm KN}^{-1}\,d\sigma_{\rm KN}/d\psi$, with $\sigma_{\rm KN}$ being the total Klein--Nishina cross section. In evaluating this function, we fix the incoming photon energy to $E_\gamma=m_e=511\,\mathrm{keV}$. The variable $\varphi$ denotes the azimuthal angle of the Compton scattering. We integrate over this angle only for events whose apparent direction, determined by $\theta$, $\psi$ and $\varphi$, lies within the RoI shown in Fig.~\ref{fig: detector}. 

We also introduce an efficiency factor {\small $\widetilde{\mathcal E}_{ff}(\theta)$} to account for the possibility that positrons produced by majoron decay escape from the Galaxy before forming positronium. Positrons injected inside the Galactic magnetic diffusion zone are expected to lose energy and form positronium before escaping. We therefore define $\widetilde{\mathcal E}_{ff}(\theta)$ as follows: for each line of sight specified by $(\theta,\phi)$, we first compute the fraction of dark-matter-induced positrons injected within the diffusion zone, and then average this fraction over the azimuthal angle $\phi$:
\begin{align}
    \widetilde{\mathcal E}_{ff}(\theta) =
    \frac{1}{2\pi} \int_0^{2\pi} d\phi\,
    \frac{\int_{{\rm l.o.s} \subset {\rm DZ}}\,d\ell\,\rho_{\rm DM}[r(\theta,\phi)]}{\int_{\rm l.o.s}\,d\ell\,\rho_{\rm DM}[r(\theta,\phi)]},
\end{align}
where ``DZ'' denotes the diffusion zone. We model the diffusion zone as a cylindrical region with a sufficiently large radial extent, taken to be effectively infinite, and a vertical half-height $L$. We then compute the corresponding dark matter signal flux for several representative values of $L$. The formal limit $L\to\infty$ corresponds to the no-escape limit, $\widetilde{\cal E}_{ ff}\to1$.

To estimate the 511\,keV line sensitivity of COSI, we start from the publicly released narrow-line sensitivity in Ref.~\cite{Tomsick:2023aue}, which gives a $3\sigma$ point-source flux threshold of ${\cal F}^{\rm PS}_{511}=1.2\times10^{-5}\,\mathrm{cm}^{-2}\,\mathrm{s}^{-1}$ for a 24-month observation. This sensitivity characterizes the expected background-limited performance of COSI for a point-like source. At 511\,keV, the instrumental annihilation-line background is expected to receive its largest contribution from atmospheric albedo positrons~\cite{Gallego:2025xxx}. This background is approximately smooth in the sky-coordinate dimensions of the Compton data space, although it has a nontrivial dependence on the Compton scattering angle. Neglecting additional systematic uncertainties and following the extended-source scaling used in Ref.~\cite{Caputo:2022dkz}, we estimate the corresponding sensitivity for our diffuse dark-matter signal by accounting for the larger region of the Compton data space occupied by the signal. For untracked Compton events, the source extension broadens the signal region in the Compton data space approximately in proportion to its angular radius. Since the sensitivity in the background-dominated regime scales with the square root of the background counts, the resulting degradation scales approximately as the fourth root of the source angular area relative to the angular-resolution area. Using our dark-matter profile and RoI together with the 511\,keV Klein--Nishina angular distribution, and including the Compton-angle dependence of the instrumental background, we find that the flux sensitivity is degraded by approximately a factor of five relative to the point-source value for the RoI geometry adopted here. Although the RoI changes with the diffusion-zone half-height $L$, the weak fourth-root dependence implies that moderate changes in its geometry have only a small effect on this factor. We therefore use the same factor of five for different values of $L$. Accordingly, we adopt ${\cal F}^{\rm limit}_{511}\simeq6\times10^{-5}\,\mathrm{cm}^{-2}\,\mathrm{s}^{-1}$ as an indicative $3\sigma$ sensitivity to the extended dark-matter signal. The benchmark signal normalization is then fixed by requiring ${\cal F}^{\rm DM}_{511}|_{\rm RoI}={\cal F}^{\rm limit}_{511}$. Together with the surface-brightness condition used to determine $\vartheta_0$, this fixes both the representative RoI boundary and the sensitivity to $\Gamma(J\to e^-e^+)/m_J$. In the no-escape limit, $L\to\infty$, we find a representative value $\vartheta_0\simeq30^\circ$. For finite $L$, the efficiency factor $\widetilde{\cal E}_{ff}$ reduces the RoI flux, and for each choice of $L$ we solve the surface-brightness and RoI-flux conditions simultaneously to determine the corresponding $\vartheta_0$ and sensitivity.

\begin{figure}[t]
    \centering
    \includegraphics[keepaspectratio, scale=0.43]{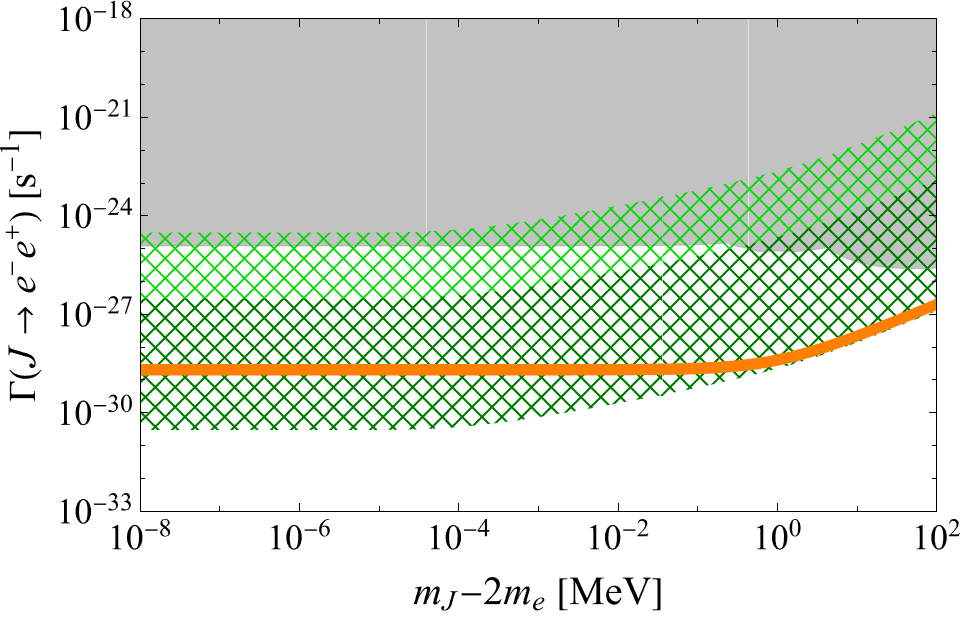}
    \qquad
    \includegraphics[keepaspectratio, scale=0.43]{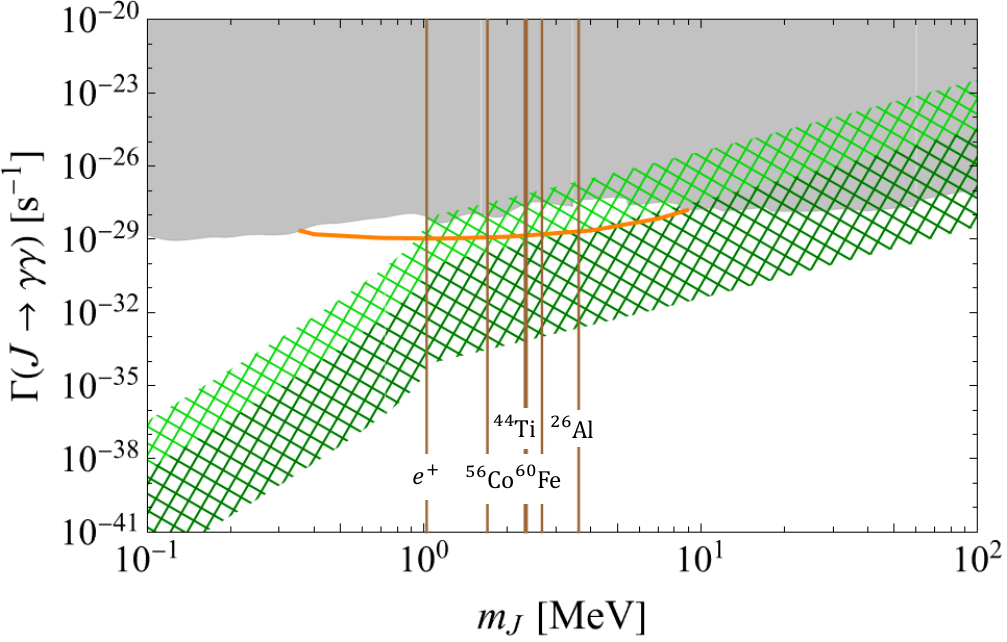}
    \caption{\small\sl {\bf Left panel:} Range of $\Gamma[J\to e^-e^+]$ in the $Z_7$ symmetry model, including the Sommerfeld enhancement. The full green hatched region, including light and dark green, shows the central 99.8\,\% interval obtained from our parameter scan, while the dark green region corresponds to the subset with $y_I^{(\mathcal N)}<0.01$. The orange band indicates the estimated 24-month COSI sensitivity to the 511\,keV line, with its upper and lower edges corresponding to diffusion-zone half-heights of $L=3\,{\rm kpc}$ and $L=\infty$, respectively. {\bf Right panel:} Same as the left panel, but for $\Gamma[J\to\gamma\gamma]$ as a function of $m_J$. The corresponding green hatched regions are obtained from the same parameter scan, while the orange line shows the estimated COSI sensitivity to a monochromatic line at $E_\gamma=m_J/2$. In both panels, the gray-shaded regions show the current observational constraints summarized in Fig.~\ref{fig: ee gammagamma}.}
    \label{fig: ee gammagamma F}
\end{figure}

With this prescription, the contour ${\cal F}^{\rm DM}_{511}|_{\rm RoI} = {\cal F}^{\rm limit}_{511}$ is shown as an orange solid line in the left panel of Fig.\,\ref{fig: ee gammagamma F}. The finite width of the line reflects the variation of the diffusion-zone half-height from its minimum value allowed by cosmic-ray observations, $L = 3$\,kpc~\cite{Lavalle:2014kca, Weinrich:2020ftb, Maurin:2022gfm}, corresponding to the upper edge, to $L = \infty$, corresponding to the lower edge. In this indicative estimate, the projected COSI sensitivity covers a large fraction of the full green hatched region, which shows the central 99.8\,\% interval obtained from our parameter scan, extending also into part of the dark-green region. We emphasize again that the signal strength is not significantly suppressed near the threshold, $m_J \simeq 2m_e$, thanks to the Sommerfeld effect, helping to maintain strong sensitivity over much of this region. Moreover, the $J \to e^-e^+$ channel provides sensitivity to the dark matter mass region above $2m_e$, precisely the region favored by the misalignment mechanisms shown in Figs.~\ref{fig: RD result} and \ref{fig: EMD result}.

A more robust sensitivity assessment would require several improvements beyond the indicative pre-flight estimate adopted here. First, $e^+$ propagation should be included, since positrons produced near the Galactic center may propagate into our RoI before annihilating, potentially increasing the dark-matter signal and improving the sensitivity. Second, residual astrophysical 511\,keV emission from the Galactic disk and bright sources should either be suppressed through an optimized RoI in the Compton data space or incorporated explicitly as background components in the likelihood, which would generally reduce the sensitivity. Though the baseline point-source sensitivity is taken to include the statistical impact of the expected instrumental and diffuse continuum backgrounds, their systematic uncertainties are not included in our estimate. In particular, the normalization and Compton-data-space shape of the 511\,keV background, whose largest individual contribution is expected to arise from atmospheric albedo positrons~\cite{Gallego:2025xxx}, as well as the normalizations and spectral and spatial distributions of diffuse continuum backgrounds, should be constrained from data and treated as nuisance parameters in a realistic likelihood analysis. Finally, both the signal and these background components should be forward modeled through the full COSI response, including its time-dependent exposure, energy and angular responses, and event reconstruction and selection. Determining the optimal analysis scheme for extracting a 511\,keV signal from dark matter decay requires both a detailed response model of the Compton detector, which is currently under development, and real flight data. We therefore regard such an analysis as beyond the scope of the present study and leave it for future work.

\subsection{Search for the line signal at \texorpdfstring{$E_\gamma = m_J/2$}{Egamma = mJ/2}}

We next consider the projected COSI sensitivity to monochromatic photons from the two-photon decay of dark matter, $J\to\gamma\gamma$, for which the photon energy is fixed at $E_\gamma=m_J/2$. For this search, we take the Galactic Center as the region of interest (RoI), following Refs.~\cite{Essig:2013goa, Laha:2020ivk, Fischer:2022pse}. Since astrophysical gamma-ray backgrounds are smooth in energy, a narrow line at an energy set by the dark matter mass would provide a distinctive signature, except near known nuclear gamma-ray lines. To estimate the reach of COSI, we follow the method of Ref.~\cite{Caputo:2022dkz} and use the publicly available COSI line-sensitivity curve reported in Ref.~\cite{Tomsick:2023aue}. We evaluate the sensitivity for emission from a circular region of radius $10^\circ$ centered on the Galactic Center, assuming an observation time of 24 months. The resulting projected sensitivity is shown by the orange solid line in the right panel of Fig.~\ref{fig: ee gammagamma F}. The brown vertical lines, labeled by $^{56}\mathrm{Co}$, $^{44}\mathrm{Ti}$, $^{60}\mathrm{Fe}$, and $^{26}\mathrm{Al}$, indicate photon energies at which the line sensitivity is degraded by known astrophysical gamma-ray lines associated with these isotopes.

As shown in the right panel, COSI can probe a large fraction of the full green hatched region, primarily the light-green part, which represents the central 99.8\,\% interval obtained from our parameter scan for the majoron model with the $Z_7$ discrete symmetry, for $m_J=1$--$10$\,MeV. This mass range closely coincides with the theoretically favored region for production through the misalignment mechanism in the early Universe, as shown in Figs.~\ref{fig: RD result} and \ref{fig: EMD result}. Although the corresponding signal is two-loop suppressed, COSI can still access much of this theoretically motivated parameter region thanks to its excellent energy resolution. Notably, the two decay channels, $J \to e^-e^+$ and $J \to \gamma\gamma$, can in principle be probed simultaneously in a single observation within the same misalignment-favored mass range. Since the strengths of these signals are correlated within the electron-loop approximation adopted in Sec.~\ref{subsubsec: constraint on J to gamma gamma}, the concurrent observation of 511\,keV emission outside the bulge and a sharp gamma-ray line at $E_\gamma=m_J/2$ from the bulge would provide strong evidence for this scenario, offering a unique opportunity to directly test it with upcoming MeV gamma-ray missions.

There is, however, an astrophysical uncertainty associated with the gamma-ray line signal from the Galactic bulge region. Since this signal is proportional to the relevant $D$-factor, its normalization depends on the dark matter density profile in the inner Galaxy. The projected reach of COSI should be interpreted with this profile dependence in mind. For our baseline result, we have used the most conservative value of the $D$-factor discussed above. Other profiles can lead to appreciably larger fluxes. For example, a halo model fitted to an alternative set of kinematic data~\cite{McMillan:2016jtx} enhances the expected signal by roughly a factor of three, and a similar enhancement is obtained for the profile motivated by Galactic bar dynamics~\cite{Portail:2016vei}. The increase can be as large as a factor of five for a more centrally concentrated profile, such as the contracted NFW profile constrained by \textit{Gaia} data~\cite{Cautun:2019eaf}. Thus, although the decay rates into $e^-e^+$ and $\gamma\gamma$ are correlated at the particle-physics level, the observed fluxes also depend on astrophysical factors. In particular, the two searches probe different regions of the Galactic halo, and the 511 keV signal is affected by positron propagation and annihilation efficiency. Hence the observable correlation between the two line signals need not coincide directly with the correlation between the decay widths.

\section{Conclusion}
\label{sec: conclusion}

We have studied majoron dark matter in the minimal majoron model with three right-handed neutrinos, in which an exact discrete gauge symmetry, $Z_N\subset U(1)_{3(B-L)}$, is imposed instead of a fundamental global $U(1)_{B-L}$ symmetry. In this setup, the global $U(1)_{B-L}$ symmetry appears only as an accidental symmetry of the low-energy theory, while its explicit breaking is controlled by Planck-suppressed operators. We focused on the phenomenologically nontrivial choices $Z_5$, $Z_7$, $Z_{11}$, and $Z_{13}$, for which $U(1)_{B-L}$-violating operators of dimension four or lower are forbidden and the majoron mass is generated only by higher-dimensional Planck-suppressed operators. The resulting pseudo-Nambu--Goldstone boson has a mass whose scale is predicted by the discrete gauge $Z_N$ symmetry and can be a detectable dark matter candidate.\footnote{
    The same mechanism for predicting the mass of a pseudo-Nambu--Goldstone boson through a discrete gauge symmetry can also be applied more generally to various other candidates, such as axion-like particles.}
We examined the parameter space in which the observed dark matter abundance is produced through the misalignment mechanism after inflation.

We considered radiation-dominated and early-matter-dominated scenarios for misalignment production and confronted the resulting parameter space with isocurvature bounds, cosmological constraints, and indirect-search limits. We found that the $Z_5$ model is not viable: in the radiation-dominated pre-inflationary misalignment scenario, it is incompatible with the thermal-restoration condition and, irrespective of the cosmological history, is excluded by constraints on its dominant decay into neutrinos. The $Z_7$, $Z_{11}$, and $Z_{13}$ models, by contrast, remain phenomenologically viable. We also briefly discussed the resulting implications for leptogenesis. Thermal leptogenesis can remain compatible with the radiation-dominated scenario, in particular when flavor effects or mild right-handed-neutrino mass degeneracies are taken into account. In the early-matter-dominated scenario, however, the restricted reheating temperature makes thermal leptogenesis considerably more difficult to realize. This tension can be further alleviated by resonant or nonthermal leptogenesis.

Among them, the $Z_7$ scenario is especially interesting, since misalignment production naturally places its predicted majoron mass in the MeV range. In this mass range, the majoron can decay into an electron--positron pair and into two photons, leading to characteristic MeV gamma-ray signatures. The electron--positron channel produces a 511\,keV line through positronium formation, and its rate is significantly enhanced by the Sommerfeld effect when the majoron mass lies near the threshold. Thanks to this enhancement, COSI observations are expected to probe a large part of the parameter region obtained in our scan through the 511\,keV line search. The two-photon decay provides an additional monochromatic gamma-ray line at half the majoron mass, offering a complementary probe. A correlated search for both line signals at COSI would provide a particularly distinctive test of the $Z_7$ majoron dark matter scenario, although a refined treatment of positron propagation, Galactic backgrounds, and the Compton-detector response is left for future work.

\section*{Acknowledgments}

S.~F. is supported by JST SPRING, Grant Number JPMJSP2108. S.~M. is supported by the Grant-in-Aid for Scientific Research from the Ministry of Education, Culture, Sports, Science and Technology, Japan (MEXT), under Grant No.~24H00244. M.~U. is supported by IBS under the project code IBS-R018-D3. T.~T.~Y. is supported by the Natural Science Foundation of China (NSFC) under Grant No.~12175134. S.~M. and T.~T.~Y. are also supported by MEXT Grant No.~24H02244. Finally, S.~F., Q.~L., S.~M., and T.~T.~Y. are supported by the World Premier International Research Center Initiative (WPI), MEXT, Japan (Kavli IPMU).

\appendix

\section{Thermal production of the Majoron}
\label{app: freeze-in production}

In addition to the misalignment mechanism, the majoron can be produced through freeze-in processes in the early Universe via interactions involving the right-handed neutrinos. The leading majoron production processes are $\mathcal{N}_I\mathcal{N}_I \to JJ$, $L_\alpha H \to \mathcal{N}_I J$, $H\mathcal{N}_I \to L_\alpha J$, $\mathcal{N}_I L_\alpha \to HJ$, and $\mathcal{N}_I \to L_\alpha HJ$, whose corresponding cross sections and decay width are
\begin{align}
    \sigma (\mathcal{N}_I\mathcal{N}_I \to JJ)
    &
    =
    \frac{\left(y^{(\mathcal N)}_I\right)^4}{256\pi M_I^ 2}\left[\frac{1}{\beta}-\frac{2}{x\beta^2}\ln\left(\frac{1+\beta}{1-\beta}\right)\right], \\
    \sigma (L_\alpha H \to \mathcal{N}_I J)
    &
    =
    \frac{\left(y_I^{(\mathcal N)}\right)^2\left|y_{\alpha I}^{(\nu)}\right|^2}{64\pi M_I^2}\frac{x-1}{x^2}, \\
    \sigma (H\mathcal{N}_I \to L_\alpha J)
    &
    =
    \frac{\left(y_I^{(\mathcal N)}\right)^2\left|y_{\alpha I}^{(\nu)}\right|^2}{64\pi M_I^2}\frac{x}{(x-1)^2}\ln\left(x\right), \\
    \sigma (\mathcal{N}_I L_\alpha \to HJ)
    &
    =
    \frac{\left(y_I^{(\mathcal N)}\right)^2\left|y_{\alpha I}^{(\nu)}\right|^2}{64\pi M_I^2}\frac{x\ln\left(x\right)-(x-1)}{(x-1)^2}, \\
    \Gamma (\mathcal{N}_I \to L_\alpha HJ)
    &
    =
    \frac{\left(y_I^{(\mathcal N)}\right)^2\left|y_{\alpha I}^{(\nu)}\right|^2}{4096\pi^3}M_I,
\end{align}
respectively. The cross sections and decay width are given for each lepton-Higgs component channel, while the sums over the $SU(2)$ components, CP-conjugate channels, and lepton flavors are included when evaluating the total production rate. We define $x\equiv s/M_I^2$ and $\beta\equiv\sqrt{1-4/x}$, where $s$ is the center-of-mass energy squared. In the early Universe, the right-handed neutrinos are generally thermalized, since the equilibrium condition
\begin{align}
    \left.
    \frac{\Gamma_{N_I}}{H}\right|_{T=M_I}
    \simeq 28
    \left[\frac{\left(y^{(\nu)\dagger}y^{(\nu)}\right)_{\mathcal N_I\mathcal N_I}^{1/2}}{10^{-3}}\right]^2
    \left(\frac{10^9\,{\rm GeV}}{M_I}\right)
    \left(\frac{106.75}{g_*}\right)^{1/2}
    \gtrsim 1
\end{align}
is typically satisfied. Then, the majoron number density can be obtained by solving the Boltzmann equation that includes all the relevant production processes mentioned above. Assuming that $\mathcal N_1$ provides the dominant contribution to the final abundance, we have
{\small
\begin{align}
    \frac{dn_J}{dt}+3Hn_J
    =
    \left\langle \sigma v(\mathcal N_1\mathcal N_1\to JJ)\right\rangle n_{\mathcal N_1}^2
    &
    +
    \sum_\alpha
    \left[
        \left\langle \sigma v(L_\alpha H\to\mathcal N_1J)\right\rangle n_{L_\alpha}n_H
        +
        \left\langle \sigma v(H\mathcal N_1\to L_\alpha J)\right\rangle n_Hn_{\mathcal N_1}
    \right.
    \nonumber \\
    &
    \left.
        +
        \left\langle \sigma v(\mathcal N_1L_\alpha\to HJ)\right\rangle n_{\mathcal N_1}n_{L_\alpha}
        +
        \left\langle \Gamma(\mathcal N_1\to L_\alpha HJ)\right\rangle n_{\mathcal N_1}
    \right].
\end{align}
}Here, the sums over the $SU(2)$ components and CP-conjugate channels are implicit. Assuming a thermal abundance of $\mathcal N_1$ and $T_{\rm RH}\simeq M_1$, and taking the SM particles to be in thermal equilibrium, solving the Boltzmann equation gives the freeze-in relic abundance in Eq.~(\ref{eq: abundance from FI}).

\section{Sommerfeld effect on the decay width into \texorpdfstring{$e^-e^+$}{e-e+}}
\label{app: Sommerfeld effect}

Near the electron--positron threshold, the final-state interaction mediated by photon exchange becomes long-ranged and gives rise to a threshold singularity. We incorporate this effect by evaluating the Sommerfeld enhancement of the decay width within the potential non-relativistic (NR) effective theory approach~\cite{Pineda:1998kn, Brambilla:1999xf}. In this framework, the fields are decomposed into several modes characterizing the energy scales according to their scalings with the velocity $\beta$. The hard modes, $\ell^0\sim |\vec{\ell}|\sim m_e$, and the soft modes, $\ell^0\sim |\vec{\ell}|\sim \beta m_e$, together with the potential photon mode, are integrated out by matching onto the low-energy theory. The potential electron and positron modes, $\ell^0\sim \beta^2 m_e$ and $|\vec{\ell}|\sim \beta m_e$, and the ultrasoft photon modes, $\ell^0\sim |\vec{\ell}|\sim \beta^2 m_e$, remain as degrees of freedom.\footnote{
    For the electron field, $\ell^0$ denotes the kinetic energy; namely, $p^0=m_e+\ell^0$, where $p^0$ is the total energy.}

We apply this construction to majoron decay into an electron--positron pair near threshold. In this region, the photons exchanged between the NR electron and positron have potential scaling, $\ell^0\sim \beta^2 m_e$ and $|\vec{\ell}|\sim \beta m_e$, while the intermediate electron and positron remain nearly on shell. For the leading contribution to $\Gamma(J\to e^-e^+)$, it is sufficient to retain the potential electron and positron fields together with the majoron field. The potential photon modes are integrated out into the interaction potential, while ultrasoft photons do not contribute at this order. The potential electron field is then expanded as follows:
\begin{align}
    \psi_{\rm Pot}(x) =
    \begin{pmatrix}
        e^{-i m_e x^0} \eta(x)
        + i e^{i m_e x^0}
        \vec{\nabla}\cdot\vec{\sigma}\,\xi(x)/(2m_e)
        + \cdots \\
        e^{i m_e x^0} \xi(x)
        - i e^{-i m_e x^0}
        \vec{\nabla}\cdot\vec{\sigma}\,\eta(x)/(2m_e)
        + \cdots
    \end{pmatrix},
    \label{eq: NR expansion}
\end{align}
where $\vec{\sigma}$ are the Pauli matrices. Here, $\eta$ $(\eta^\dagger)$ annihilates (creates) a NR electron, while $\xi$ $(\xi^\dagger)$ creates (annihilates) a NR positron. Substituting this field expansion into the matched low-energy theory and retaining the leading NR terms, we obtain the NR Lagrangian
\begin{align}
    &
    \mathcal{L}_{\rm NR}
    \simeq
    -\frac{1}{2} J\,(\Box + m_J^2)\,J
    +\eta^\dagger \left( i\partial_{x^0} + \frac{\nabla_x^2}{2m_e} \right) \eta
    +\xi^\dagger \left( i\partial_{x^0} - \frac{\nabla_x^2}{2m_e} \right) \xi
    \label{eq: NR Lagrangian}
    \\
    & \quad\quad
    +\int d^4y \frac{\alpha\,\delta(x^0 - y^0)}{2|\vec{x}-\vec{y}|}
    \left[ \eta^\dagger(x)\,\xi(y) \right]
    \cdot
    \left[ \xi^\dagger(y)\,\eta(x) \right]
    -ic\,J\,
    \left[
    e^{2 i m_e x^0} \eta^\dagger\,\xi
    +
    e^{-2 i m_e x^0} \xi^\dagger \eta
    \right].
    \nonumber
\end{align}
Here the coefficient is defined as $c \equiv m_e (K_{11}-K_{22}-K_{33})/(16 \pi^2 v_{\rm EW})$. In writing the above expression, we have retained only the two-body channel that is relevant for the majoron coupling. Other configurations that do not directly interact with the majoron, for example $\eta^\dagger(x)\vec{\sigma}\xi(y)$, are not written here. We then describe the corresponding spin-singlet $e^-e^+$ pair by an auxiliary two-body field $\Phi(\vec{r},x)$, which provides a convenient low-energy degree of freedom coupled to the majoron field. Eliminating the original NR fields $\eta$ and $\xi$ in favor of this two-body description leads to the following potential NR (pNR) Lagrangian:
\begin{align}
    \mathcal{L}_{\rm pNR} =
    &
    -\frac{1}{2} J\,(\Box + m_J^2)\,J
    -\sqrt{2} ic\,J \cdot
    \left[ e^{2i m_e x^0} \Phi^\dagger (\vec{0}, x) - e^{-2i m_e x^0} \Phi(\vec{0},x) \right]
    \nonumber \\
    &\qquad\qquad\qquad
    +\int d^3r\,\Phi^\dagger(\vec{r},x) \left[i\partial_{x^0} + \frac{\nabla_x^2}{4m_e} + \frac{\nabla_r^2}{m_e} + \frac{\alpha}{r} \right]  \Phi(\vec{r},x),
    \label{eq: pNR Lagrangian}
\end{align}
where $\Phi(\vec r,x)$ and $\Phi^\dagger(\vec r,x)$ denote the fields that annihilate and create the total-spin-zero (i.e., spin-singlet) $e^-e^+$ state, respectively. Here, $\vec r$ and $x$ are the relative and center-of-mass coordinates of the two-body system. See Ref.~\cite{Matsumoto:2022ojl} for further details of the derivation.

Then, it is useful to expand the field ${\Phi}(\vec r,x)$ in terms of the solutions to the Schr\"odinger equation describing the relative motion between the electron and the positron, as follows:
\begin{align}
    &
    \Phi(\vec{r},x) =
    \sum_{\ell,\,m} \int_0^\infty \frac{dk}{2\pi}\,C_{k \ell m}(x)\,\psi_{k \ell m}(\vec{r})
    + \cdots,
    \label{eq: expansion}
    \\
    &
    \psi_{k \ell m}(\vec{r}) =
    \frac{\Gamma[1 + \ell + i\alpha m_e/(2k)]}{(2\ell + 1)!r}
    \exp\left(\frac{\pi \alpha m_e}{4k}\right)
    M(i\alpha m_e/k, \ell + 1/2, -2ikr)
    Y_{\ell m}(\theta,\varphi),
    \nonumber
\end{align}
where $k=\sqrt{m_e E}$ denotes the wave number associated with the internal kinetic energy $E\geq 0$, and $\ell$ and $m$ label the angular-momentum eigenstates. We use $Y_{\ell m}(\theta,\varphi)$ for the spherical harmonics, $M(a,b,c)$ for the Whittaker function of the first kind, and $\Gamma(x)$ for the Gamma function. The operator $C_{k\ell m}(x)$ annihilates the corresponding spin-singlet continuum state. The continuum wave functions are normalized as $\int d^3r\,\psi^\dagger_{k'\ell'm'}(\vec r)\,\psi_{k\ell m}(\vec r)=(2\pi)\delta(k-k')\delta_{\ell\ell'}\delta_{mm'}$, so that the kinetic term for $C_{k\ell m}(x)$ is canonically normalized. The ellipsis represents bound-state modes with $E<0$, which are not written explicitly above. Inserting this mode decomposition into the pNR Lagrangian in Eq.\,(\ref{eq: pNR Lagrangian}), we obtain
\begin{align}
    \mathcal{L}_{\rm pNR} =
    \label{eq: pNR_Phi1}
    &
    -\frac{1}{2} J\,(\Box + m_J^2)\,J
    +\int \frac{dk}{2\pi} C^\dagger_{k00} \left[ i\partial_{x^0} + \frac{\nabla^2}{4m_e} - \frac{k^2}{m_e} \right] C_{k00}
    \nonumber \\
    &
    -\frac{c}{\sqrt{\pi}} J
    \left[
        i e^{2im_ex^0} \int \frac{dk}{2\pi} \left(\frac{2 \pi m_e\alpha k}{1 - e^{-\pi \alpha m_e/k}}\right)^{1/2} C^\dagger_{k00} + h.c.
    \right]
    + \cdots,
\end{align}
where we have omitted all fields with angular momentum quantum number $\ell \neq 0$, since their corresponding wave functions vanish at the origin, $\vec{r}=0$. Consequently, these modes do not couple directly to the majoron field at leading order in the non-relativistic expansion.

With the above pNR Lagrangian, the partial decay width for $J \to e^-e^+$ near the $e^-e^+$ threshold region is obtained using the LSZ reduction formula. The corresponding asymptotic field $C_{k00}^{(\rm as)}(x)$, which describes the spin-singlet $e^-e^+$ two-body state with $\ell = 0$, satisfies the free equation of motion, $\left[i\partial_{x^0} + \nabla^2/(4m_e) - k^2/m_e\right] C^{(\rm as)}_{k00}(x) = 0$, whose solution is
\begin{align}
    C_{k00}^{({\rm as})}(x) =
    -\int \frac{d^3p}{\sqrt{(2\pi)^3 2E_{p,k}}}
    \frac{1}{2\pi}\,A^{({\rm as})}_k(\vec{p})\,
    e^{-iE_{p,k} x^0 + i\vec{p}\cdot\vec{x}},
    \label{eq: LSZ1}
\end{align}
with $E_{p,k} = \vec{p}^{\,2}/(4m_e) + k^2/m_e$. From the equal-time commutation relation for the canonical variable $C^{({\rm as})}_{k00}(x)$, $[C^{({\rm as})}_{k00}(t,\vec{x}),\,\Pi^{({\rm as})}_{k'00}(t,\vec{y})] = i\delta(\vec{x}-\vec{y})\delta(k-k')$, with the canonical conjugate $\Pi^{({\rm as})}_{k00} = iC_{k00}^{({\rm as})\,\dagger}/(2\pi)$, the operator $A^{({\rm as})}_k(\vec{p})$ is found to satisfy the corresponding relation, $[A^{({\rm as})}_k(\vec{p}),\,A^{({\rm as})\,\dagger}_{k'}(\vec{p}')] = (2\pi)^3(2E_{p,k})\delta(\vec{p}-\vec{p}')\delta(k-k')$. The operators $A^{({\rm as})}_k(\vec{p})$ and $A^{({\rm as})\,\dagger}_k(\vec{p})$ can then be expressed in terms of the annihilation and creation operators as follows:
\begin{align}
    A_k^{({\rm as})\,(\dagger)}(\vec{p}) 
    = \int d^3x\,
    f_{\vec{p},k}^{(*)}(x)\,
    C^{({\rm as})\,(\dagger)}_{k00}(x)
    \quad {\rm with} \quad
    f_{\vec{p}, k}(x) =
    -\sqrt{\frac{E_{p,k}}{\pi}} e^{-iE_{p,k} x^0 + i\vec{p}\cdot\vec{x}}.
\end{align}
By defining the $e^-e^+$ two-body state as $\ket{e^-e^+(\vec p,k),{\rm as}} = A_k^{({\rm as})\dagger}(\vec p)\ket{0}$, with the normalization $\braket{e^-e^+(\vec p,k),{\rm as}|e^-e^+(\vec p\,',k'),{\rm as}} = (2\pi)^3 2E_{p,k}\,\delta^3(\vec p-\vec p\,')\delta(k-k')$ following from the commutation relation given above, the transition amplitude from the initial majoron state $\ket{J(p'),{\rm in}}$ to the two-body final state $\ket{e^-e^+(\vec p,k),{\rm out}}$ is obtained via the LSZ reduction formula as
\begin{align}
    \braket{e^-e^+(\vec{p}, k), {\rm out}|J(p'), {\rm in}} =
    &
    -i \int d^4x\,
    f_{\vec{p},k}^*(x)
    \left(
        i \partial_0 + \frac{\vec{\nabla}^2}{4m_e} - \frac{k^2}{m_e}
    \right)
    \nonumber \\
    &
    \times i\int d^4y\,f_{\vec{p}'}(y) (\Box_y + m^2)
    \bra{0} T[C_{k00}(x)\,J(y)] \ket{0},
\end{align}
where $p'=(E_{p'},\vec p\,')^T$ is the four-momentum of $J$, with $E_{p'}=(|\vec p\,'|^2+m_J^2)^{1/2}$, while $f_{\vec p\,'}(y)=e^{-ip'\cdot y}$ denotes the one-majoron wave function. The invariant amplitude is defined through the matrix element as $\braket{e^-e^+(\vec p,k),{\rm out}|J(p'),{\rm in}} \equiv i(2\pi)^4\delta^{(4)}(p'-p)\mathcal{M}[J\to e^-e^+]$, with $p=(E_{p,k},\vec p)^T$. Therefore, at leading order in the interaction of Eq.\,(\ref{eq: pNR_Phi1}), one obtains
\begin{align}
    {\cal M}[J \rightarrow e^-e^+] =
    -\frac{c\,[(2\pi)^3 2E_{p, k}]^{1/2}}{4 \pi^{5/2}} \left(\frac{2 \pi m_e \alpha k}{1 - e^{-\pi \alpha m_e / k}} \right)^{1/2}.
     \label{eq: Sommerfeld amplitude eenu0}
 \end{align}
Then, the partial decay width of $J \rightarrow e^-e^+$ is obtained via the decay formula as follows:
{\small
\begin{align}
    \Gamma[J \rightarrow e^-e^+]
    &=\frac{1}{2 m_J}
    \int d\Phi \left|{\cal M}[J \rightarrow e^-e^+]\right|^2\,
    =\frac{1}{2 m_J}
    \int \frac{dk\,d^3 p}{(2\pi)^3\,2E_{p,k}} (2 \pi)^4 \delta(p' - p)
    |{\cal M}|^2,
    \\
    &=\frac{\alpha c^2 m_J}{8\,[1 - \exp(-\pi \alpha/\sqrt{m_J/m_e - 2})]}
    \simeq \frac{\pi \alpha /\sqrt{m_J/m_e - 2}}{1 - \exp[-\pi \alpha /\sqrt{m_J/m_e - 2}]}
    \Gamma_0[J \to e^-e^+]\,,
    \nonumber
\end{align}
}for $m_J\geq  2 m_e$. The last expression is motivated by the fact that the Sommerfeld effect becomes sizable only near the threshold region. The Sommerfeld factor, namely the factor multiplying $\Gamma_0[J\to e^-e^+]$, is given by the ratio of the wave functions at the origin, $|\psi(\vec 0)|^2/|\psi_0(\vec 0)|^2$, where $\psi(\vec r)$ and $\psi_0(\vec r)$ describe the relative motion of the electron--positron pair with and without the long-range force, respectively. This factor rapidly approaches unity when $m_J$ is far above the threshold, $m_J=2m_e$. We therefore combine the result obtained by the pNR Lagrangian method with the perturbative decay width to obtain an expression applicable over the entire mass range of $m_J$, as shown in the last expression.

\section{Theoretical expectations for the majoron couplings}
\label{app: scannings}

The decay widths of majoron dark matter into an electron-positron pair and two photons depend on the so-called $K_{ij}$ factors, defined as $K_{ij}=(m_Dm_D^\dagger)_{ij}/(v_{\rm EW}v_\Phi)$, as discussed in the main text. Here, we estimate the typical size and distribution of these factors obtained in the minimal majoron model, with particular focus on these two decay modes.

The $K_{ij}$ factors depend on several new-physics parameters: the active-neutrino masses $m_i$; the right-handed-neutrino Yukawa couplings $y^{(\mathcal{N})}_I$; the PMNS matrix $U_{\rm PMNS}$, which contains three mixing angles, one Dirac phase, and two Majorana phases; and the Casas--Ibarra matrix $R$, parameterized by three complex angles $z_i$. We estimate the distributions of the combinations of $K_{ij}$ entering the decay widths by scanning these parameters with flat priors in the linear variables. We vary the lightest neutrino mass in the range $0\leq m_1\leq 0.1~{\rm eV}$ and the right-handed-neutrino Yukawa couplings in the range $10^{-3} \leq y^{(\mathcal{N})}_I\leq 10^{-1}$. This range is chosen as a representative perturbative range motivated by the heavy-seesaw regime discussed in Section~\ref{subsec: isocurvature}. For the PMNS matrix and the active-neutrino mass splittings, we vary the Majorana phases over $0\leq \alpha_i\leq 2\pi$ $(i=1,2)$, while fixing the oscillation parameters to the normal-ordering best-fit values of Ref.~\cite{Esteban:2024eli}: $\Delta m_{21}^2=7.49\times10^{-5}~{\rm eV}^2$, $\Delta m_{31}^2=2.513\times10^{-3}~{\rm eV}^2$, $\theta_{12}=33.68^\circ$, $\theta_{23}=43.3^\circ$, $\theta_{13}=8.56^\circ$, and $\delta_{\rm CP}=212^\circ$. For the Casas--Ibarra parameters, we scan $0\leq {\rm Re}\,z_i\leq 2\pi$ and set ${\rm Im}\,z_i=0$. This choice provides a well-motivated minimal baseline for the distributions of the relevant coefficients. Nonzero imaginary parts of $z_i$ generate hyperbolic factors in the Casas--Ibarra matrix and can therefore enhance the neutrino Yukawa couplings $y^{(\nu)}_{iI}$ exponentially; sufficiently large imaginary parts may drive the theory into a non-perturbative regime. Such regions are not representative of the generic parameter space considered here. Moderate imaginary parts may modify the tails of the distribution, but we expect the real-$R$ scan to provide a minimal baseline. A more general analysis including nonzero imaginary parts and the leptogenesis condition is left for future work. Moreover, high-energy CP phases associated with complex Casas--Ibarra angles are not strictly required for leptogenesis, since a viable lepton asymmetry can be generated from the low-energy PMNS phases in flavored leptogenesis~\cite{Pascoli:2006ci,Moffat:2018smo}.

\begin{figure}[t]
    \centering
    \includegraphics[width=0.55\linewidth]{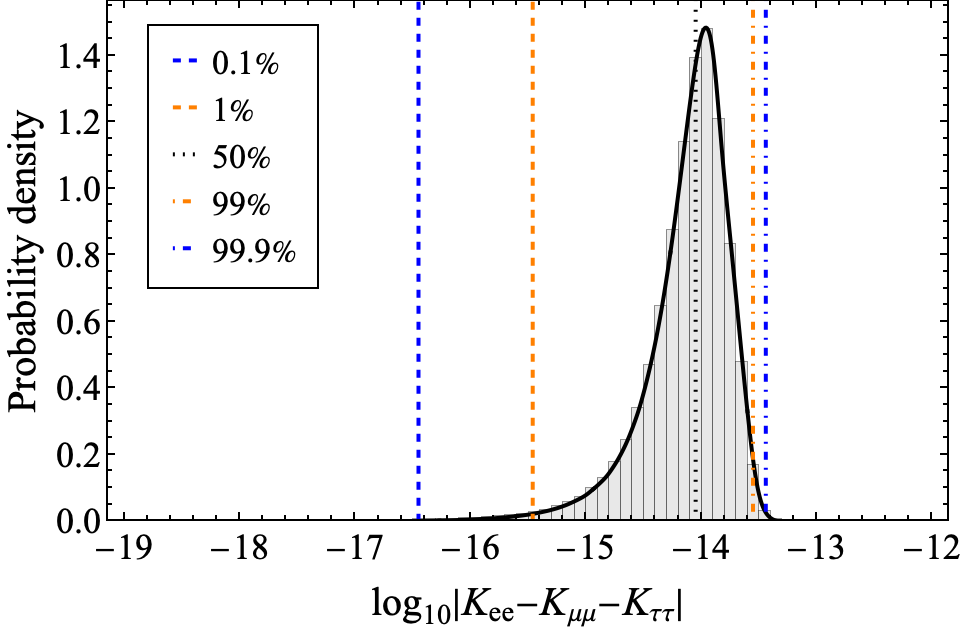}
    \caption{\small\sl 
    Scan
    distribution of the combination $|K_{11}-K_{22}-K_{33}|$, which controls the decay widths into $e^-e^+$ and $\gamma\gamma$. The distribution is obtained from $10^7$ points in the minimal majoron model, using the parameter ranges and assumptions described in the text. The central 99.8\% interval is indicated by the vertical dashed lines and corresponds to $3.5\times10^{-17}\leq |K_{11}-K_{22}-K_{33}|\leq 3.6\times10^{-14}$.
    }
    \label{fig: probability distribution}
\end{figure}

As shown in Eqs.~(\ref{eq: J to ee}) and (\ref{eq: J to ee kai}), the decay width into an $e^-e^+$ pair is proportional to the combination $|K_{11}-K_{22}-K_{33}|^2$. In the electron-loop-dominated approximation adopted in Sec.~\ref{subsubsec: constraint on J to gamma gamma}, the decay width into $\gamma\gamma$ is also proportional to $|K_{11}-K_{22}-K_{33}|^2$. We therefore consider the distribution of $|K_{11}-K_{22}-K_{33}|$. Using the scanning procedure described above, we generate $10^7$ sample points in the parameter space and obtain the distribution shown in Fig.~\ref{fig: probability distribution}. From this distribution, we find that the central 99.8\,\% interval is $3.5\times10^{-17}\leq |K_{11}-K_{22}-K_{33}|\leq 3.6\times10^{-14}$. We checked that imposing the cosmologically motivated upper bound on the lightest active SM neutrino mass, namely, $m_1<0.03\,\mathrm{eV}$~\cite{Planck:2018vyg, Esteban:2024eli}, shifts both edges of this interval inward by several tens of percent, thereby narrowing the range. This narrowing does not significantly impact our discussion in the main text. We also perform a separate parameter scan over the right-handed-neutrino Yukawa coupling range $10^{-3} \leq y_I^{(\mathcal{N})} \leq 10^{-2}$. We find that 99.9\% of the resulting $|K_{11}-K_{22}-K_{33}|$ values satisfy $|K_{11}-K_{22}-K_{33}| \leq 3.5 \times 10^{-15}$. In Figs.~\ref{fig: ee gammagamma} and~\ref{fig: ee gammagamma F}, we highlight in dark green the part of the green hatched region that also satisfies $|K_{11}-K_{22}-K_{33}| \leq 3.5 \times 10^{-15}$. Note that these intervals should not be interpreted as statistical predictions of the model, since they depend on the adopted parameter ranges and priors. They are intended only as representative estimates of the theoretically plausible range in the absence of special cancellations.

\bibliographystyle{unsrt}
\bibliography{references}

@article{Dai:1994kq,
    author = "Dai, Xian-zhe and Freed, Daniel S.",
    title = "{eta invariants and determinant lines}",
    eprint = "hep-th/9405012",
    archivePrefix = "arXiv",
    doi = "10.1063/1.530747",
    journal = "J. Math. Phys.",
    volume = "35",
    pages = "5155--5194",
    year = "1994",
    note = "[Erratum: J.Math.Phys. 42, 2343--2344 (2001)]"
}

@article{Planck:2018jri,
    author = "Akrami, Y. and others",
    collaboration = "Planck",
    title = "{Planck 2018 results. X. Constraints on inflation}",
    eprint = "1807.06211",
    archivePrefix = "arXiv",
    primaryClass = "astro-ph.CO",
    doi = "10.1051/0004-6361/201833887",
    journal = "Astron. Astrophys.",
    volume = "641",
    pages = "A10",
    year = "2020"
}

@article{Nelson:2018via,
    author = "Nelson, Ann E. and Xiao, Huangyu",
    title = "{Axion Cosmology with Early Matter Domination}",
    eprint = "1807.07176",
    archivePrefix = "arXiv",
    primaryClass = "astro-ph.CO",
    doi = "10.1103/PhysRevD.98.063516",
    journal = "Phys. Rev. D",
    volume = "98",
    number = "6",
    pages = "063516",
    year = "2018"
}

@article{Kawasaki:2018qwp,
    author = "Kawasaki, Masahiro and Sonomoto, Eisuke and Yanagida, Tsutomu T.",
    title = "{Cosmologically allowed regions for the axion decay constant $F_a$}",
    eprint = "1801.07409",
    archivePrefix = "arXiv",
    primaryClass = "hep-ph",
    reportNumber = "IPMU18-0020",
    doi = "10.1016/j.physletb.2018.05.014",
    journal = "Phys. Lett. B",
    volume = "782",
    pages = "181--184",
    year = "2018"
}

@article{Linde:1991km,
    author = "Linde, Andrei D.",
    title = "{Axions in inflationary cosmology}",
    reportNumber = "SU-ITP-883",
    doi = "10.1016/0370-2693(91)90130-I",
    journal = "Phys. Lett. B",
    volume = "259",
    pages = "38--47",
    year = "1991"
}

@article{OHare:2024nmr,
    author = "O'Hare, Ciaran A. J.",
    title = "{Cosmology of axion dark matter}",
    eprint = "2403.17697",
    archivePrefix = "arXiv",
    primaryClass = "hep-ph",
    doi = "10.22323/1.454.0040",
    journal = "PoS",
    volume = "COSMICWISPers",
    pages = "040",
    year = "2024"
}

@article{Hertzberg:2008wr,
    author = "Hertzberg, Mark P and Tegmark, Max and Wilczek, Frank",
    title = "{Axion Cosmology and the Energy Scale of Inflation}",
    eprint = "0807.1726",
    archivePrefix = "arXiv",
    primaryClass = "astro-ph",
    reportNumber = "MIT-CTP-3950",
    doi = "10.1103/PhysRevD.78.083507",
    journal = "Phys. Rev. D",
    volume = "78",
    pages = "083507",
    year = "2008"
}

@article{Kobayashi:2013nva,
    author = "Kobayashi, Takeshi and Kurematsu, Ryosuke and Takahashi, Fuminobu",
    title = "{Isocurvature Constraints and Anharmonic Effects on QCD Axion Dark Matter}",
    eprint = "1304.0922",
    archivePrefix = "arXiv",
    primaryClass = "hep-ph",
    reportNumber = "TU-933",
    doi = "10.1088/1475-7516/2013/09/032",
    journal = "JCAP",
    volume = "09",
    pages = "032",
    year = "2013"
}

@article{Blinov:2019rhb,
    author = "Blinov, Nikita and Dolan, Matthew J and Draper, Patrick and Kozaczuk, Jonathan",
    title = "{Dark matter targets for axionlike particle searches}",
    eprint = "1905.06952",
    archivePrefix = "arXiv",
    primaryClass = "hep-ph",
    reportNumber = "FERMILAB-PUB-19-197-A-T",
    doi = "10.1103/PhysRevD.100.015049",
    journal = "Phys. Rev. D",
    volume = "100",
    number = "1",
    pages = "015049",
    year = "2019"
}

@article{Yanagida:1979as,
    author = "Yanagida, Tsutomu",
    editor = "Sawada, Osamu and Sugamoto, Akio",
    title = "{Horizontal gauge symmetry and masses of neutrinos}",
    reportNumber = "KEK-79-18-95",
    journal = "Conf. Proc. C",
    volume = "7902131",
    pages = "95--99",
    year = "1979"
}

@article{Yanagida:1979gs,
    author = "Yanagida, Tsutomu",
    title = "{Horizontal Symmetry and Mass of the Top Quark}",
    reportNumber = "TU/79/196",
    doi = "10.1103/PhysRevD.20.2986",
    journal = "Phys. Rev. D",
    volume = "20",
    pages = "2986",
    year = "1979"
}

@article{Gell-Mann:1979vob,
    author = "Gell-Mann, Murray and Ramond, Pierre and Slansky, Richard",
    title = "{Complex Spinors and Unified Theories}",
    eprint = "1306.4669",
    archivePrefix = "arXiv",
    primaryClass = "hep-th",
    reportNumber = "PRINT-80-0576",
    journal = "Conf. Proc. C",
    volume = "790927",
    pages = "315--321",
    year = "1979"
}

@article{Minkowski:1977sc,
    author = "Minkowski, Peter",
    title = "{$\mu \to e\gamma$ at a Rate of One Out of $10^{9}$ Muon Decays?}",
    reportNumber = "Print-77-0182 (BERN)",
    doi = "10.1016/0370-2693(77)90435-X",
    journal = "Phys. Lett. B",
    volume = "67",
    pages = "421--428",
    year = "1977"
}

@article{Pontecorvo:1957qd,
    author = "Pontecorvo, B.",
    title = "{Inverse Beta Processes and Nonconservation of Lepton Charge}",
    journal = "Sov. Phys. JETP",
    volume = "7",
    pages = "172--173",
    year = "1958"
}

@article{Maki:1962mu,
    author = "Maki, Ziro and Nakagawa, Masami and Sakata, Shoichi",
    title = "{Remarks on the unified model of elementary particles}",
    doi = "10.1143/PTP.28.870",
    journal = "Prog. Theor. Phys.",
    volume = "28",
    pages = "870--880",
    year = "1962"
}

@article{Chikashige:1980ui,
    author = "Chikashige, Y. and Mohapatra, Rabindra N. and Peccei, R. D.",
    title = "{Are There Real Goldstone Bosons Associated with Broken Lepton Number?}",
    reportNumber = "MPI-PAE-PTH-36-80",
    doi = "10.1016/0370-2693(81)90011-3",
    journal = "Phys. Lett. B",
    volume = "98",
    pages = "265--268",
    year = "1981"
}

@article{Pilaftsis:1993af,
    author = "Pilaftsis, Apostolos",
    title = "{Astrophysical and terrestrial constraints on singlet Majoron models}",
    eprint = "hep-ph/9308258",
    archivePrefix = "arXiv",
    reportNumber = "MZ-TH-93-18",
    doi = "10.1103/PhysRevD.49.2398",
    journal = "Phys. Rev. D",
    volume = "49",
    pages = "2398--2404",
    year = "1994"
}

@article{Garcia-Cely:2017oco,
    author = "Garcia-Cely, Camilo and Heeck, Julian",
    title = "{Neutrino Lines from Majoron Dark Matter}",
    eprint = "1701.07209",
    archivePrefix = "arXiv",
    primaryClass = "hep-ph",
    reportNumber = "ULB-TH-17-01",
    doi = "10.1007/JHEP05(2017)102",
    journal = "JHEP",
    volume = "05",
    pages = "102",
    year = "2017"
}

@article{Heeck:2019guh,
    author = "Heeck, Julian and Patel, Hiren H.",
    title = "{Majoron at two loops}",
    eprint = "1909.02029",
    archivePrefix = "arXiv",
    primaryClass = "hep-ph",
    reportNumber = "UCI-TR-2019-23",
    doi = "10.1103/PhysRevD.100.095015",
    journal = "Phys. Rev. D",
    volume = "100",
    number = "9",
    pages = "095015",
    year = "2019"
}

@article{Xu:2024vdn,
    author = "Xu, Clara and Qin, Wenzer and Slatyer, Tracy R.",
    title = "{CMB limits on decaying dark matter beyond the ionization threshold}",
    eprint = "2408.13305",
    archivePrefix = "arXiv",
    primaryClass = "astro-ph.CO",
    reportNumber = "MIT-CTP/5747",
    doi = "10.1103/PhysRevD.110.123529",
    journal = "Phys. Rev. D",
    volume = "110",
    number = "12",
    pages = "123529",
    year = "2024"
}

@article{Liu:2020wqz,
    author = "Liu, Hongwan and Qin, Wenzer and Ridgway, Gregory W. and Slatyer, Tracy R.",
    title = "{Lyman-{\ensuremath{\alpha}} constraints on cosmic heating from dark matter annihilation and decay}",
    eprint = "2008.01084",
    archivePrefix = "arXiv",
    primaryClass = "astro-ph.CO",
    doi = "10.1103/PhysRevD.104.043514",
    journal = "Phys. Rev. D",
    volume = "104",
    number = "4",
    pages = "043514",
    year = "2021"
}

@article{Calore:2022pks,
    author = "Calore, Francesca and Dekker, Ariane and Serpico, Pasquale Dario and Siegert, Thomas",
    title = "{Constraints on light decaying dark matter candidates from 16~yr of INTEGRAL/SPI observations}",
    eprint = "2209.06299",
    archivePrefix = "arXiv",
    primaryClass = "hep-ph",
    doi = "10.1093/mnras/stad457",
    journal = "Mon. Not. Roy. Astron. Soc.",
    volume = "520",
    number = "3",
    pages = "4167--4172",
    year = "2023",
    note = "[Erratum: Mon.Not.Roy.Astron.Soc. 538, 132 (2025)]"
}

@article{Essig:2013goa,
    author = "Essig, Rouven and Kuflik, Eric and McDermott, Samuel D. and Volansky, Tomer and Zurek, Kathryn M.",
    title = "{Constraining Light Dark Matter with Diffuse X-Ray and Gamma-Ray Observations}",
    eprint = "1309.4091",
    archivePrefix = "arXiv",
    primaryClass = "hep-ph",
    reportNumber = "YITP-SB-29-13, FERMILAB-PUB-13-377-A-T, MCTP-13-27",
    doi = "10.1007/JHEP11(2013)193",
    journal = "JHEP",
    volume = "11",
    pages = "193",
    year = "2013"
}

@article{Esteban:2024eli,
    author = "Esteban, Ivan and Gonzalez-Garcia, M. C. and Maltoni, Michele and Martinez-Soler, Ivan and Pinheiro, Jo{\~a}o Paulo and Schwetz, Thomas",
    title = "{NuFit-6.0: updated global analysis of three-flavor neutrino oscillations}",
    eprint = "2410.05380",
    archivePrefix = "arXiv",
    primaryClass = "hep-ph",
    reportNumber = "IFT-UAM/CSIC-24-140, YITP-SB-2024-24, IPPP/24/64, IPPP/24/64, IFT-UAM/CSIC-24-140, YITP-SB-2024-24",
    doi = "10.1007/JHEP12(2024)216",
    journal = "JHEP",
    volume = "12",
    pages = "216",
    year = "2024"
}

@article{Fukugita:1986hr,
    author = "Fukugita, M. and Yanagida, T.",
    title = "{Baryogenesis Without Grand Unification}",
    reportNumber = "RIFP-641",
    doi = "10.1016/0370-2693(86)91126-3",
    journal = "Phys. Lett. B",
    volume = "174",
    pages = "45--47",
    year = "1986"
}

@article{Sheng:2025sou,
    author = "Sheng, Jie and Yanagida, Tsutomu T.",
    title = "{High quality QCD axion in the Standard Model}",
    eprint = "2510.17370",
    archivePrefix = "arXiv",
    primaryClass = "hep-ph",
    doi = "10.1103/h9ws-xgst",
    journal = "Phys. Rev. D",
    volume = "113",
    number = "5",
    pages = "055010",
    year = "2026"
}

@article{Preskill:1982cy,
    author = "Preskill, John and Wise, Mark B. and Wilczek, Frank",
    editor = "Srednicki, M. A.",
    title = "{Cosmology of the Invisible Axion}",
    reportNumber = "HUTP-82-A048, NSF-ITP-82-103",
    doi = "10.1016/0370-2693(83)90637-8",
    journal = "Phys. Lett. B",
    volume = "120",
    pages = "127--132",
    year = "1983"
}

@article{Abbott:1982af,
    author = "Abbott, L. F. and Sikivie, P.",
    editor = "Srednicki, M. A.",
    title = "{A Cosmological Bound on the Invisible Axion}",
    reportNumber = "PRINT-82-0695 (BRANDEIS)",
    doi = "10.1016/0370-2693(83)90638-X",
    journal = "Phys. Lett. B",
    volume = "120",
    pages = "133--136",
    year = "1983"
}

@article{Dine:1982ah,
    author = "Dine, Michael and Fischler, Willy",
    editor = "Srednicki, M. A.",
    title = "{The Not So Harmless Axion}",
    reportNumber = "UPR-0201T",
    doi = "10.1016/0370-2693(83)90639-1",
    journal = "Phys. Lett. B",
    volume = "120",
    pages = "137--141",
    year = "1983"
}

@article{Turner:1983he,
    author = "Turner, Michael S.",
    title = "{Coherent Scalar Field Oscillations in an Expanding Universe}",
    reportNumber = "EFI-83-29-CHICAGO",
    doi = "10.1103/PhysRevD.28.1243",
    journal = "Phys. Rev. D",
    volume = "28",
    pages = "1243",
    year = "1983"
}

@article{Planck:2018vyg,
    author = "Aghanim, N. and others",
    collaboration = "Planck",
    title = "{Planck 2018 results. VI. Cosmological parameters}",
    eprint = "1807.06209",
    archivePrefix = "arXiv",
    primaryClass = "astro-ph.CO",
    doi = "10.1051/0004-6361/201833910",
    journal = "Astron. Astrophys.",
    volume = "641",
    pages = "A6",
    year = "2020",
    note = "[Erratum: Astron.Astrophys. 652, C4 (2021)]"
}

@article{Borzumati:2000fe,
    author = "Borzumati, Francesca and Hamaguchi, Koichi and Yanagida, T.",
    title = "{Supersymmetric seesaw model for the (1+3) scheme of neutrino masses}",
    eprint = "hep-ph/0011141",
    archivePrefix = "arXiv",
    reportNumber = "KEK-TH-724, UT-916",
    doi = "10.1016/S0370-2693(00)01351-4",
    journal = "Phys. Lett. B",
    volume = "497",
    pages = "259--264",
    year = "2001"
}

@article{deSalas:2015glj,
    author = "de Salas, P. F. and Lattanzi, M. and Mangano, G. and Miele, G. and Pastor, S. and Pisanti, O.",
    title = "{Bounds on very low reheating scenarios after Planck}",
    eprint = "1511.00672",
    archivePrefix = "arXiv",
    primaryClass = "astro-ph.CO",
    reportNumber = "IFIC-15-70",
    doi = "10.1103/PhysRevD.92.123534",
    journal = "Phys. Rev. D",
    volume = "92",
    number = "12",
    pages = "123534",
    year = "2015"
}

@article{Barbieri:2025moq,
    author = "Barbieri, Nicola and Brinckmann, Thejs and Gariazzo, Stefano and Lattanzi, Massimiliano and Pastor, Sergio and Pisanti, Ofelia",
    title = "{Current Constraints on Cosmological Scenarios with Very Low Reheating Temperatures}",
    eprint = "2501.01369",
    archivePrefix = "arXiv",
    primaryClass = "astro-ph.CO",
    doi = "10.1103/j5rj-dz1k",
    journal = "Phys. Rev. Lett.",
    volume = "135",
    number = "18",
    pages = "181003",
    year = "2025"
}

@article{Davidson:2002qv,
    author = "Davidson, Sacha and Ibarra, Alejandro",
    title = "{A Lower bound on the right-handed neutrino mass from leptogenesis}",
    eprint = "hep-ph/0202239",
    archivePrefix = "arXiv",
    reportNumber = "OUTP-02-10P, IPPP-02-16, DCPT-02-32",
    doi = "10.1016/S0370-2693(02)01735-5",
    journal = "Phys. Lett. B",
    volume = "535",
    pages = "25--32",
    year = "2002"
}

@article{Buchmuller:2004nz,
    author = "Buchmuller, W. and Di Bari, P. and Plumacher, M.",
    title = "{Leptogenesis for pedestrians}",
    eprint = "hep-ph/0401240",
    archivePrefix = "arXiv",
    reportNumber = "DESY-03-100, UAB-FT-551, CERN-TH-2003-199",
    doi = "10.1016/j.aop.2004.02.003",
    journal = "Annals Phys.",
    volume = "315",
    pages = "305--351",
    year = "2005"
}

@article{Nardi:2006fx,
    author = "Nardi, Enrico and Nir, Yosef and Roulet, Esteban and Racker, Juan",
    title = "{The importance of flavor in leptogenesis}",
    eprint = "hep-ph/0601084",
    archivePrefix = "arXiv",
    doi = "10.1088/1126-6708/2006/01/164",
    journal = "JHEP",
    volume = "01",
    pages = "164",
    year = "2006"
}

@article{Hambye:2003rt,
    author = "Hambye, Thomas and Lin, Yin and Notari, Alessio and Papucci, Michele and Strumia, Alessandro",
    title = "{Constraints on neutrino masses from leptogenesis models}",
    eprint = "hep-ph/0312203",
    archivePrefix = "arXiv",
    reportNumber = "IFUP-TH-2003-48",
    doi = "10.1016/j.nuclphysb.2004.06.027",
    journal = "Nucl. Phys. B",
    volume = "695",
    pages = "169--191",
    year = "2004"
}

@article{Pilaftsis:2003gt,
    author = "Pilaftsis, Apostolos and Underwood, Thomas E. J.",
    title = "{Resonant leptogenesis}",
    eprint = "hep-ph/0309342",
    archivePrefix = "arXiv",
    reportNumber = "MC-TH-2003-09",
    doi = "10.1016/j.nuclphysb.2004.05.029",
    journal = "Nucl. Phys. B",
    volume = "692",
    pages = "303--345",
    year = "2004"
}

@article{DOnofrio:2014rug,
    author = "D'Onofrio, Michela and Rummukainen, Kari and Tranberg, Anders",
    title = "{Sphaleron Rate in the Minimal Standard Model}",
    eprint = "1404.3565",
    archivePrefix = "arXiv",
    primaryClass = "hep-ph",
    doi = "10.1103/PhysRevLett.113.141602",
    journal = "Phys. Rev. Lett.",
    volume = "113",
    number = "14",
    pages = "141602",
    year = "2014"
}

@article{Alvi:2022aam,
    author = "Alvi, S. and Brinckmann, T. and Gerbino, M. and Lattanzi, M. and Pagano, L.",
    title = "{Do you smell something decaying? Updated linear constraints on decaying dark matter scenarios}",
    eprint = "2205.05636",
    archivePrefix = "arXiv",
    primaryClass = "astro-ph.CO",
    doi = "10.1088/1475-7516/2022/11/015",
    journal = "JCAP",
    volume = "11",
    pages = "015",
    year = "2022"
}

@article{Arguelles:2022nbl,
    author = {Arg{\"u}elles, Carlos A. and Delgado, Diyaselis and Friedlander, Avi and Kheirandish, Ali and Safa, Ibrahim and Vincent, Aaron C. and White, Henry},
    title = "{Dark matter decay to neutrinos}",
    eprint = "2210.01303",
    archivePrefix = "arXiv",
    primaryClass = "hep-ph",
    doi = "10.1103/PhysRevD.108.123021",
    journal = "Phys. Rev. D",
    volume = "108",
    number = "12",
    pages = "123021",
    year = "2023"
}

@article{Akita:2023qiz,
    author = "Akita, Kensuke and Niibo, Michiru",
    title = "{Updated constraints and future prospects on majoron dark matter}",
    eprint = "2304.04430",
    archivePrefix = "arXiv",
    primaryClass = "hep-ph",
    reportNumber = "CTPU-PTC-23-10",
    doi = "10.1007/JHEP07(2023)132",
    journal = "JHEP",
    volume = "07",
    pages = "132",
    year = "2023"
}

@article{Tomsick:2023aue,
    author = "Tomsick, John A. and others",
    title = "{The Compton Spectrometer and Imager}",
    eprint = "2308.12362",
    archivePrefix = "arXiv",
    primaryClass = "astro-ph.HE",
    doi = "10.22323/1.444.0745",
    journal = "PoS",
    volume = "ICRC2023",
    pages = "745",
    year = "2023"
}

@article{Pineda:1998kn,
    author = "Pineda, A. and Soto, J.",
    title = "{Potential NRQED: The Positronium case}",
    eprint = "hep-ph/9805424",
    archivePrefix = "arXiv",
    reportNumber = "UB-ECM-PF-98-11, KFA-IKP-TH-98-9",
    doi = "10.1103/PhysRevD.59.016005",
    journal = "Phys. Rev. D",
    volume = "59",
    pages = "016005",
    year = "1999"
}

@article{Brambilla:1999xf,
    author = "Brambilla, Nora and Pineda, Antonio and Soto, Joan and Vairo, Antonio",
    title = "{Potential NRQCD: An Effective theory for heavy quarkonium}",
    eprint = "hep-ph/9907240",
    archivePrefix = "arXiv",
    reportNumber = "CERN-TH-99-199, HEPHY-PUB-716-99, UB-ECM-PF-99-06, UWTHPH-1999-34, UB-ECM-PF-99-13",
    doi = "10.1016/S0550-3213(99)00693-8",
    journal = "Nucl. Phys. B",
    volume = "566",
    pages = "275",
    year = "2000"
}

@article{Hayashi:2024not,
    author = "Hayashi, Tatsuya and Matsumoto, Shigeki and Watanabe, Yuki and Yanagida, Tsutomu T.",
    title = "{Gauge U(1)B-L dark matter above the e-e+ threshold}",
    eprint = "2408.12155",
    archivePrefix = "arXiv",
    primaryClass = "hep-ph",
    doi = "10.1103/PhysRevD.111.055012",
    journal = "Phys. Rev. D",
    volume = "111",
    number = "5",
    pages = "055012",
    year = "2025"
}

@article{Boudaud:2016mos,
    author = "Boudaud, Mathieu and Lavalle, Julien and Salati, Pierre",
    title = "{Novel cosmic-ray electron and positron constraints on MeV dark matter particles}",
    eprint = "1612.07698",
    archivePrefix = "arXiv",
    primaryClass = "astro-ph.HE",
    doi = "10.1103/PhysRevLett.119.021103",
    journal = "Phys. Rev. Lett.",
    volume = "119",
    number = "2",
    pages = "021103",
    year = "2017"
}

@article{DelaTorreLuque:2023olp,
    author = "De la Torre Luque, Pedro and Balaji, Shyam and Koechler, Jordan",
    title = "{Importance of Cosmic-Ray Propagation on Sub-GeV Dark Matter Constraints}",
    eprint = "2311.04979",
    archivePrefix = "arXiv",
    primaryClass = "hep-ph",
    doi = "10.3847/1538-4357/ad41e0",
    journal = "Astrophys. J.",
    volume = "968",
    number = "1",
    pages = "46",
    year = "2024"
}

@article{Siegert:2015knp,
    author = "Siegert, Thomas and Diehl, Roland and Khachatryan, Gerasim and Krause, Martin G. H. and Guglielmetti, Fabrizia and Greiner, Jochen and Strong, Andrew W. and Zhang, Xiaoling",
    title = "{Gamma-ray spectroscopy of Positron Annihilation in the Milky Way}",
    eprint = "1512.00325",
    archivePrefix = "arXiv",
    primaryClass = "astro-ph.HE",
    doi = "10.1051/0004-6361/201527510",
    journal = "Astron. Astrophys.",
    volume = "586",
    pages = "A84",
    year = "2016"
}

@phdthesis{Siegert:2017thesis,
    author = "Siegert, Thomas",
    title = "{Positron-Annihilation Spectroscopy throughout the Milky Way}",
    school = "Technische Universit{\"a}t M{\"u}nchen",
    year = "2017",
    address = "M{\"u}nchen, Germany",
    url = "https://mediatum.ub.tum.de/node?id=1340342"
}

@inproceedings{Siegert:2023wus,
    author = "Siegert, Thomas",
    title = "{The Positron Puzzle}",
    eprint = "2303.15582",
    archivePrefix = "arXiv",
    primaryClass = "astro-ph.HE",
    doi = "10.1007/s10509-023-04184-4",
    month = "3",
    year = "2023"
}

@article{DelaTorreLuque:2023cef,
    author = "De la Torre Luque, Pedro and Balaji, Shyam and Silk, Joseph",
    title = "{New 511 keV Line Data Provide Strongest sub-GeV Dark Matter Constraints}",
    eprint = "2312.04907",
    archivePrefix = "arXiv",
    primaryClass = "hep-ph",
    doi = "10.3847/2041-8213/ad72f4",
    journal = "Astrophys. J. Lett.",
    volume = "973",
    number = "1",
    pages = "L6",
    year = "2024",
    note = "[Erratum: Astrophys.J.Lett. 991, L29 (2025), Erratum: Astrophys.J. 991, L29 (2025)]"
}

@article{Nguyen:2025tkl,
    author = "Nguyen, Thong T. Q. and De la Torre Luque, Pedro and John, Isabelle and Balaji, Shyam and Carenza, Pierluca and Linden, Tim",
    title = "{INTEGRAL, eROSITA and Voyager constraints on light bosonic dark matter: ALPs, dark photons, scalars, B-L and Li-Lj vectors}",
    eprint = "2507.13432",
    archivePrefix = "arXiv",
    primaryClass = "hep-ph",
    reportNumber = "KCL-2025-30",
    doi = "10.1103/fn6k-1nlc",
    journal = "Phys. Rev. D",
    volume = "113",
    number = "10",
    pages = "103010",
    year = "2026"
}

@article{Tomsick:2021wed,
    author = "Tomsick, John A.",
    collaboration = "COSI",
    title = "{The Compton Spectrometer and Imager Project for MeV Astronomy}",
    eprint = "2109.10403",
    archivePrefix = "arXiv",
    primaryClass = "astro-ph.IM",
    doi = "10.22323/1.395.0652",
    journal = "PoS",
    volume = "ICRC2021",
    pages = "652",
    year = "2021"
}

@article{Aramaki:2022zpw,
    author = "Aramaki, Tsuguo and Boezio, Mirko and Buckley, James and Bulbul, Esra and von Doetinchem, Philip and Donato, Fiorenza and Harding, J. Patrick and Karwin, Chris and Kumar, Jason and Leane, Rebecca K. and Matsumoto, Shigeki and McEnry, Julie and Melia, Tom and Perez, Kerstin and Profumo, Stefano and Salazar-Gallegos, Daniel and Strong, Andrew W. and Roach, Brandon and Sanchez-Conde, Miguel A. and Shutt, Tom and Takada, Atsushi and Tanimori, Toru and Tomsick, John and Watanabe, Yu and Williams, David A.",
    title = "{Snowmass2021 Cosmic Frontier: The landscape of cosmic-ray and high-energy photon probes of particle dark matter}",
    eprint = "2203.06894",
    archivePrefix = "arXiv",
    primaryClass = "hep-ex",
    year = "2022"
}

@article{Jean:2009zj,
    author = "Jean, P. and Gillard, W. and Marcowith, A. and Ferriere, K.",
    title = "{Positron transport in the interstellar medium}",
    eprint = "0909.4022",
    archivePrefix = "arXiv",
    primaryClass = "astro-ph.HE",
    doi = "10.1051/0004-6361/200809830",
    journal = "Astron. Astrophys.",
    volume = "508",
    pages = "1099",
    year = "2009"
}

@article{Prantzos:2010wi,
    author = "Prantzos, N. and Boehm, C. and Bykov, A. M. and Diehl, R. and Ferriere, K. and Guessoum, N. and Jean, P. and Knoedlseder, J. and Marcowith, A. and Moskalenko, I. V. and Strong, A. and Weidenspointner, G.",
    title = "{The 511 keV emission from positron annihilation in the Galaxy}",
    eprint = "1009.4620",
    archivePrefix = "arXiv",
    primaryClass = "astro-ph.HE",
    doi = "10.1103/RevModPhys.83.1001",
    journal = "Rev. Mod. Phys.",
    volume = "83",
    pages = "1001--1056",
    year = "2011"
}

@article{Siegert:2021upv,
    author = "Siegert, Thomas and Berteaud, Julien and Calore, Francesca and Serpico, Pasquale D. and Weinberger, Christoph",
    title = "{Measuring the smearing of the Galactic 511-keV signal}",
    eprint = "2109.03791",
    archivePrefix = "arXiv",
    primaryClass = "astro-ph.HE",
    doi = "10.1093/mnrasl/slab113",
    journal = "Mon. Not. Roy. Astron. Soc.",
    volume = "509",
    number = "1",
    pages = "L11--L16",
    year = "2022"
}

@inproceedings{Skinner:2015,
    author = "Skinner, Gerald",
    title = "{The Galactic distribution of the 511 keV electron positron annihilation radiation}",
    booktitle = "{10th INTEGRAL Workshop: A Synergistic View of the High-Energy Sky}",
    journal = "PoS",
    volume = "INTEGRAL2014",
    pages = "054",
    year = "2015",
    doi = "10.22323/1.228.0054"
}

@article{Pascoli:2006ci,
    author = "Pascoli, S. and Petcov, S. T. and Riotto, Antonio",
    title = "{Leptogenesis and Low Energy CP Violation in Neutrino Physics}",
    eprint = "hep-ph/0611338",
    archivePrefix = "arXiv",
    reportNumber = "DCPT-06-166, CERN-PH-TH-2006-213, IPPP-06-83, SISSA-71-2006-EP",
    doi = "10.1016/j.nuclphysb.2007.02.019",
    journal = "Nucl. Phys. B",
    volume = "774",
    pages = "1--52",
    year = "2007"
}

@article{Moffat:2018smo,
    author = "Moffat, K. and Pascoli, S. and Petcov, S. T. and Turner, J.",
    title = "{Leptogenesis from Low Energy $CP$ Violation}",
    eprint = "1809.08251",
    archivePrefix = "arXiv",
    primaryClass = "hep-ph",
    reportNumber = "IPPP/18/79, SISSA 38/2018/FISI, IPMU18-0151, FERMILAB-PUB-18-382-T",
    doi = "10.1007/JHEP03(2019)034",
    journal = "JHEP",
    volume = "03",
    pages = "034",
    year = "2019"
}

@article{Laha:2020ivk,
    author = "Laha, Ranjan and Mu{\~n}oz, Julian B. and Slatyer, Tracy R.",
    title = "{INTEGRAL constraints on primordial black holes and particle dark matter}",
    eprint = "2004.00627",
    archivePrefix = "arXiv",
    primaryClass = "astro-ph.CO",
    doi = "10.1103/PhysRevD.101.123514",
    journal = "Phys. Rev. D",
    volume = "101",
    number = "12",
    pages = "123514",
    year = "2020"
}

@article{Fischer:2022pse,
    author = "Fischer, S. and Malyshev, D. and Ducci, L. and Santangelo, A.",
    title = "{New constraints on decaying dark matter from INTEGRAL/SPI}",
    eprint = "2211.06200",
    archivePrefix = "arXiv",
    primaryClass = "astro-ph.HE",
    doi = "10.1093/mnras/stad304",
    journal = "Mon. Not. Roy. Astron. Soc.",
    volume = "520",
    number = "4",
    pages = "6322--6334",
    year = "2023"
}

@article{Caputo:2022dkz,
    author = "Caputo, Andrea and Negro, Michela and Regis, Marco and Taoso, Marco",
    title = "{Dark matter prospects with COSI: ALPs, PBHs and sub-GeV dark matter}",
    eprint = "2210.09310",
    archivePrefix = "arXiv",
    primaryClass = "hep-ph",
    doi = "10.1088/1475-7516/2023/02/006",
    journal = "JCAP",
    volume = "02",
    pages = "006",
    year = "2023"
}

@article{Matsumoto:2022ojl,
    author = "Matsumoto, Shigeki and Watanabe, Yu and Watanabe, Yuki and White, Graham",
    title = "{Decay of the Mediator Particle at Threshold}",
    eprint = "2212.10739",
    archivePrefix = "arXiv",
    primaryClass = "hep-ph",
    doi = "10.1007/JHEP09(2023)015",
    journal = "JHEP",
    volume = "09",
    pages = "015",
    year = "2023"
}

@proceedings{INS:1981qlp,
  title        = {{INS Symposium on Quark and Lepton Physics}},
  organization = {{Institute for Nuclear Study (INS)}},
  address      = {Tokyo, Japan},
  year         = {1981},
  note         = {June 25--27}
}

@proceedings{Pfeil:1981vb,
  editor    = "Pfeil, W.",
  title     = "{Proceedings of Lepton--Photon}",
  publisher = "Bonn U., Inst. Phys.",
  year      = "1981",
  note      = "10th International Symposium on Lepton and Photon Interactions at High Energy, Bonn, Germany, August 24--29"
}

@article{Weinrich:2020ftb,
    author = "Weinrich, N. and Boudaud, M. and Derome, L. and Genolini, Y. and Lavalle, J. and Maurin, D. and Salati, P. and Serpico, P. and Weymann-Despres, G.",
    title = "{Galactic halo size in the light of recent AMS-02 data}",
    eprint = "2004.00441",
    archivePrefix = "arXiv",
    primaryClass = "astro-ph.HE",
    reportNumber = "LAPTH-009/20, LUPM:20-019",
    doi = "10.1051/0004-6361/202038064",
    journal = "Astron. Astrophys.",
    volume = "639",
    pages = "A74",
    year = "2020"
}

@article{Maurin:2022gfm,
    author = "Maurin, D. and Ferronato Bueno, E. and Derome, L.",
    title = "{A simple determination of the halo size from 10Be/9Be data}",
    eprint = "2203.07265",
    archivePrefix = "arXiv",
    primaryClass = "astro-ph.HE",
    doi = "10.1051/0004-6361/202243546",
    journal = "Astron. Astrophys.",
    volume = "667",
    pages = "A25",
    year = "2022"
}

@article{Lavalle:2014kca,
    author = "Lavalle, Julien and Maurin, David and Putze, Antje",
    title = "{Direct constraints on diffusion models from cosmic-ray positron data: Excluding the minimal model for dark matter searches}",
    eprint = "1407.2540",
    archivePrefix = "arXiv",
    primaryClass = "astro-ph.HE",
    reportNumber = "LUPM-14-019, LUPM:14-019",
    doi = "10.1103/PhysRevD.90.081301",
    journal = "Phys. Rev. D",
    volume = "90",
    pages = "081301",
    year = "2014"
}

@article{Yonekura:2016wuc,
    author = "Yonekura, Kazuya",
    title = "{Dai-Freed theorem and topological phases of matter}",
    eprint = "1607.01873",
    archivePrefix = "arXiv",
    primaryClass = "hep-th",
    reportNumber = "IPMU-16-0094",
    doi = "10.1007/JHEP09(2016)022",
    journal = "JHEP",
    volume = "09",
    pages = "022",
    year = "2016"
}

@article{Casas:2001sr,
    author = "Casas, J. A. and Ibarra, A.",
    title = "{Oscillating neutrinos and $\mu \to e, \gamma$}",
    eprint = "hep-ph/0103065",
    archivePrefix = "arXiv",
    reportNumber = "IEM-FT-211-01, OUTP-01-11P, IFT-UAM-CSIC-01-08",
    doi = "10.1016/S0550-3213(01)00475-8",
    journal = "Nucl. Phys. B",
    volume = "618",
    pages = "171--204",
    year = "2001"
}

@article{McMillan:2016jtx,
    author = "McMillan, Paul J.",
    title = "{The mass distribution and gravitational potential of the Milky Way}",
    eprint = "1608.00971",
    archivePrefix = "arXiv",
    primaryClass = "astro-ph.GA",
    doi = "10.1093/mnras/stw2759",
    journal = "Mon. Not. Roy. Astron. Soc.",
    volume = "465",
    number = "1",
    pages = "76--94",
    year = "2016"
}

@article{Portail:2016vei,
    author = "Portail, Matthieu and Gerhard, Ortwin and Wegg, Christopher and Ness, Melissa",
    title = "{Dynamical modelling of the galactic bulge and bar: the Milky Way's pattern speed, stellar and dark matter mass distribution}",
    eprint = "1608.07954",
    archivePrefix = "arXiv",
    primaryClass = "astro-ph.GA",
    doi = "10.1093/mnras/stw2819",
    journal = "Mon. Not. Roy. Astron. Soc.",
    volume = "465",
    number = "2",
    pages = "1621--1644",
    year = "2017"
}

@article{Cautun:2019eaf,
    author = "Cautun, Marius and Benitez-Llambay, Alejandro and Deason, Alis J. and Frenk, Carlos S. and Fattahi, Azadeh and G{\'o}mez, Facundo A. and Grand, Robert J. J. and Oman, Kyle A. and Navarro, Julio F. and Simpson, Christine M.",
    title = "{The Milky Way total mass profile as inferred from Gaia DR2}",
    eprint = "1911.04557",
    archivePrefix = "arXiv",
    primaryClass = "astro-ph.GA",
    doi = "10.1093/mnras/staa1017",
    journal = "Mon. Not. Roy. Astron. Soc.",
    volume = "494",
    number = "3",
    pages = "4291--4313",
    year = "2020"
}

@article{Bertone:2004pz,
    author = "Bertone, Gianfranco and Hooper, Dan and Silk, Joseph",
    title = "{Particle dark matter: Evidence, candidates and constraints}",
    eprint = "hep-ph/0404175",
    archivePrefix = "arXiv",
    reportNumber = "FERMILAB-PUB-04-047-A",
    doi = "10.1016/j.physrep.2004.08.031",
    journal = "Phys. Rept.",
    volume = "405",
    pages = "279--390",
    year = "2005"
}

@article{Jungman:1995df,
    author = "Jungman, Gerard and Kamionkowski, Marc and Griest, Kim",
    title = "{Supersymmetric dark matter}",
    eprint = "hep-ph/9506380",
    archivePrefix = "arXiv",
    reportNumber = "SU-4240-605, UCSD-PTH-95-02, IASSNS-HEP-95-14, CU-TP-677",
    doi = "10.1016/0370-1573(95)00058-5",
    journal = "Phys. Rept.",
    volume = "267",
    pages = "195--373",
    year = "1996"
}

@article{Peccei:1977hh,
    author = "Peccei, R. D. and Quinn, Helen R.",
    title = "{CP Conservation in the Presence of Instantons}",
    reportNumber = "ITP-568-STANFORD",
    doi = "10.1103/PhysRevLett.38.1440",
    journal = "Phys. Rev. Lett.",
    volume = "38",
    pages = "1440--1443",
    year = "1977"
}

@article{Peccei:1977ur,
    author = "Peccei, R. D. and Quinn, Helen R.",
    title = "{Constraints Imposed by CP Conservation in the Presence of Instantons}",
    reportNumber = "ITP-572-STANFORD",
    doi = "10.1103/PhysRevD.16.1791",
    journal = "Phys. Rev. D",
    volume = "16",
    pages = "1791--1797",
    year = "1977"
}

@article{Weinberg:1977ma,
    author = "Weinberg, Steven",
    title = "{A New Light Boson?}",
    reportNumber = "HUTP-77/A074",
    doi = "10.1103/PhysRevLett.40.223",
    journal = "Phys. Rev. Lett.",
    volume = "40",
    pages = "223--226",
    year = "1978"
}

@article{Wilczek:1977pj,
    author = "Wilczek, Frank",
    title = "{Problem of Strong  $P$  and  $T$  Invariance in the Presence of Instantons}",
    reportNumber = "Print-77-0939 (COLUMBIA)",
    doi = "10.1103/PhysRevLett.40.279",
    journal = "Phys. Rev. Lett.",
    volume = "40",
    pages = "279--282",
    year = "1978"
}

@article{Kim:1979if,
    author = "Kim, Jihn E.",
    title = "{Weak Interaction Singlet and Strong CP Invariance}",
    reportNumber = "UPR-0120T",
    doi = "10.1103/PhysRevLett.43.103",
    journal = "Phys. Rev. Lett.",
    volume = "43",
    pages = "103",
    year = "1979"
}

@article{Shifman:1979if,
    author = "Shifman, Mikhail A. and Vainshtein, A. I. and Zakharov, Valentin I.",
    title = "{Can Confinement Ensure Natural CP Invariance of Strong Interactions?}",
    reportNumber = "ITEP-64-1979",
    doi = "10.1016/0550-3213(80)90209-6",
    journal = "Nucl. Phys. B",
    volume = "166",
    pages = "493--506",
    year = "1980"
}

@article{Zhitnitsky:1980tq,
    author = "Zhitnitsky, A. R.",
    title = "{On Possible Suppression of the Axion Hadron Interactions. (In Russian)}",
    journal = "Sov. J. Nucl. Phys.",
    volume = "31",
    pages = "260",
    year = "1980"
}

@article{Dine:1981rt,
    author = "Dine, Michael and Fischler, Willy and Srednicki, Mark",
    title = "{A Simple Solution to the Strong CP Problem with a Harmless Axion}",
    reportNumber = "Print-81-0320 (IAS,PRINCETON)",
    doi = "10.1016/0370-2693(81)90590-6",
    journal = "Phys. Lett. B",
    volume = "104",
    pages = "199--202",
    year = "1981"
}

@article{Frampton:2002qc,
    author = "Frampton, P. H. and Glashow, S. L. and Yanagida, T.",
    title = "{Cosmological sign of neutrino CP violation}",
    eprint = "hep-ph/0208157",
    archivePrefix = "arXiv",
    reportNumber = "CERN-TH-2002-193",
    doi = "10.1016/S0370-2693(02)02853-8",
    journal = "Phys. Lett. B",
    volume = "548",
    pages = "119--121",
    year = "2002"
}

@article{Ibarra:2003up,
    author = "Ibarra, A. and Ross, Graham G.",
    title = "{Neutrino phenomenology: The Case of two right-handed neutrinos}",
    eprint = "hep-ph/0312138",
    archivePrefix = "arXiv",
    reportNumber = "CERN-TH-2003-294, OUTP-0333P",
    doi = "10.1016/j.physletb.2004.04.037",
    journal = "Phys. Lett. B",
    volume = "591",
    pages = "285--296",
    year = "2004"
}

@article{Antusch:2011nz,
    author = "Antusch, S. and Di Bari, P. and Jones, D. A. and King, S. F.",
    title = "{Leptogenesis in the Two Right-Handed Neutrino Model Revisited}",
    eprint = "1107.6002",
    archivePrefix = "arXiv",
    primaryClass = "hep-ph",
    doi = "10.1103/PhysRevD.86.023516",
    journal = "Phys. Rev. D",
    volume = "86",
    pages = "023516",
    year = "2012"
}

@article{Dodelson:1993je,
    author = "Dodelson, Scott and Widrow, Lawrence M.",
    title = "{Sterile-neutrinos as dark matter}",
    eprint = "hep-ph/9303287",
    archivePrefix = "arXiv",
    reportNumber = "FERMILAB-PUB-93-057-A",
    doi = "10.1103/PhysRevLett.72.17",
    journal = "Phys. Rev. Lett.",
    volume = "72",
    pages = "17--20",
    year = "1994"
}

@article{Shi:1998km,
    author = "Shi, Xiang-Dong and Fuller, George M.",
    title = "{A New dark matter candidate: Nonthermal sterile neutrinos}",
    eprint = "astro-ph/9810076",
    archivePrefix = "arXiv",
    doi = "10.1103/PhysRevLett.82.2832",
    journal = "Phys. Rev. Lett.",
    volume = "82",
    pages = "2832--2835",
    year = "1999"
}

@article{Asaka:2005an,
    author = "Asaka, Takehiko and Blanchet, Steve and Shaposhnikov, Mikhail",
    title = "{The nuMSM, dark matter and neutrino masses}",
    eprint = "hep-ph/0503065",
    archivePrefix = "arXiv",
    doi = "10.1016/j.physletb.2005.09.070",
    journal = "Phys. Lett. B",
    volume = "631",
    pages = "151--156",
    year = "2005"
}

@article{Asaka:2005pn,
    author = "Asaka, Takehiko and Shaposhnikov, Mikhail",
    title = "{The $\nu$MSM, dark matter and baryon asymmetry of the universe}",
    eprint = "hep-ph/0505013",
    archivePrefix = "arXiv",
    doi = "10.1016/j.physletb.2005.06.020",
    journal = "Phys. Lett. B",
    volume = "620",
    pages = "17--26",
    year = "2005"
}

@article{Laine:2008pg,
    author = "Laine, M. and Shaposhnikov, M.",
    title = "{Sterile neutrino dark matter as a consequence of nuMSM-induced lepton asymmetry}",
    eprint = "0804.4543",
    archivePrefix = "arXiv",
    primaryClass = "hep-ph",
    reportNumber = "BI-TP-2008-02, NSF-KITP-08-39",
    doi = "10.1088/1475-7516/2008/06/031",
    journal = "JCAP",
    volume = "06",
    pages = "031",
    year = "2008"
}

@article{Kusenko:2010ik,
    author = "Kusenko, Alexander and Takahashi, Fuminobu and Yanagida, Tsutomu T.",
    title = "{Dark Matter from Split Seesaw}",
    eprint = "1006.1731",
    archivePrefix = "arXiv",
    primaryClass = "hep-ph",
    reportNumber = "IPMU-10-0095",
    doi = "10.1016/j.physletb.2010.08.031",
    journal = "Phys. Lett. B",
    volume = "693",
    pages = "144--148",
    year = "2010"
}

@article{Boyarsky:2008xj,
    author = "Boyarsky, Alexey and Lesgourgues, Julien and Ruchayskiy, Oleg and Viel, Matteo",
    title = "{Lyman-alpha constraints on warm and on warm-plus-cold dark matter models}",
    eprint = "0812.0010",
    archivePrefix = "arXiv",
    primaryClass = "astro-ph",
    reportNumber = "CERN-PH-TH-2008-234, LAPTH-1290-08",
    doi = "10.1088/1475-7516/2009/05/012",
    journal = "JCAP",
    volume = "05",
    pages = "012",
    year = "2009"
}

@article{Boyarsky:2009ix,
    author = "Boyarsky, Alexey and Ruchayskiy, Oleg and Shaposhnikov, Mikhail",
    title = "{The Role of sterile neutrinos in cosmology and astrophysics}",
    eprint = "0901.0011",
    archivePrefix = "arXiv",
    primaryClass = "hep-ph",
    doi = "10.1146/annurev.nucl.010909.083654",
    journal = "Ann. Rev. Nucl. Part. Sci.",
    volume = "59",
    pages = "191--214",
    year = "2009"
}

@article{Boyarsky:2018tvu,
    author = "Boyarsky, A. and Drewes, M. and Lasserre, T. and Mertens, S. and Ruchayskiy, O.",
    title = "{Sterile neutrino Dark Matter}",
    eprint = "1807.07938",
    archivePrefix = "arXiv",
    primaryClass = "hep-ph",
    doi = "10.1016/j.ppnp.2018.07.004",
    journal = "Prog. Part. Nucl. Phys.",
    volume = "104",
    pages = "1--45",
    year = "2019"
}

@article{Kasai:2025xaw,
    author = "Kasai, Kentaro and Kawasaki, Masahiro and Murai, Kai",
    title = "{Resonant production of sterile neutrino dark matter with a refined numerical scheme}",
    eprint = "2510.01907",
    archivePrefix = "arXiv",
    primaryClass = "hep-ph",
    doi = "10.1088/1475-7516/2026/02/048",
    journal = "JCAP",
    volume = "02",
    pages = "048",
    year = "2026"
}

@article{Kaneta:2016vkq,
    author = "Kaneta, Kunio and Kang, Zhaofeng and Lee, Hye-Sung",
    title = "{Right-handed neutrino dark matter under the $B - L$ gauge interaction}",
    eprint = "1606.09317",
    archivePrefix = "arXiv",
    primaryClass = "hep-ph",
    reportNumber = "CTPU-16-17",
    doi = "10.1007/JHEP02(2017)031",
    journal = "JHEP",
    volume = "02",
    pages = "031",
    year = "2017"
}

@article{Okada:2016gsh,
    author = "Okada, Nobuchika and Okada, Satomi",
    title = "{$Z^\prime_{BL}$ portal dark matter and LHC Run-2 results}",
    eprint = "1601.07526",
    archivePrefix = "arXiv",
    primaryClass = "hep-ph",
    reportNumber = "YGHP16-03",
    doi = "10.1103/PhysRevD.93.075003",
    journal = "Phys. Rev. D",
    volume = "93",
    number = "7",
    pages = "075003",
    year = "2016"
}

@article{Sheng:2023dix,
    author = "Sheng, Jie and Cheng, Yu and Yanagida, Tsutomu T.",
    title = "{Thermal relic right-handed neutrino dark matter}",
    eprint = "2312.15637",
    archivePrefix = "arXiv",
    primaryClass = "hep-ph",
    doi = "10.1016/j.physletb.2024.138735",
    journal = "Phys. Lett. B",
    volume = "854",
    pages = "138735",
    year = "2024"
}

@article{Fujisawa:2025yqi,
    author = "Fujisawa, Subaru and Hayashi, Tatsuya and Matsumoto, Shigeki and Watanabe, Yuki",
    title = "{Detecting sterile neutrino dark matter at MeV gamma-ray observatories}",
    eprint = "2508.08695",
    archivePrefix = "arXiv",
    primaryClass = "hep-ph",
    doi = "10.1007/JHEP01(2026)097",
    journal = "JHEP",
    volume = "01",
    pages = "097",
    year = "2026"
}

@article{Lin:2022xbu,
    author = "Lin, Weikang and Visinelli, Luca and Xu, Donglian and Yanagida, Tsutomu T.",
    title = "{Neutrino astronomy as a probe of physics beyond the Standard Model: Decay of sub-MeV B-L gauge boson dark matter}",
    eprint = "2202.04496",
    archivePrefix = "arXiv",
    primaryClass = "hep-ph",
    doi = "10.1103/PhysRevD.106.075011",
    journal = "Phys. Rev. D",
    volume = "106",
    number = "7",
    pages = "075011",
    year = "2022"
}

@article{Lin:2022mqe,
    author = "Lin, Weikang and Yanagida, Tsutomu T.",
    title = "{Confronting the Galactic 511~keV emission with B-L gauge boson dark matter}",
    eprint = "2205.08171",
    archivePrefix = "arXiv",
    primaryClass = "hep-ph",
    doi = "10.1103/PhysRevD.106.075012",
    journal = "Phys. Rev. D",
    volume = "106",
    number = "7",
    pages = "075012",
    year = "2022"
}

@article{Sheng:2023iup,
    author = "Sheng, Jie and Cheng, Yu and Lin, Weikang and Yanagida, Tsutomu T.",
    title = "{F{\'e}eton (B-L gauge boson) dark matter for the 511-keV gamma-ray excess and the prediction of low-energy neutrino flux}",
    eprint = "2310.05420",
    archivePrefix = "arXiv",
    primaryClass = "hep-ph",
    doi = "10.1088/1674-1137/ad4af3",
    journal = "Chin. Phys. C",
    volume = "48",
    number = "8",
    pages = "083104",
    year = "2024"
}

@article{Cheng:2024vqb,
    author = "Cheng, Yu and Sheng, Jie and Yanagida, Tsutomu T.",
    title = "{F{\'e}eton (B {\ensuremath{-}} L gauge boson) dark matter testable in future direct detection experiments}",
    eprint = "2410.12554",
    archivePrefix = "arXiv",
    primaryClass = "hep-ph",
    doi = "10.1007/JHEP12(2024)078",
    journal = "JHEP",
    volume = "12",
    pages = "078",
    year = "2024"
}

@article{Gelmini:1980re,
    author = "Gelmini, G. B. and Roncadelli, M.",
    title = "{Left-Handed Neutrino Mass Scale and Spontaneously Broken Lepton Number}",
    reportNumber = "MPI-PAE-PTH-50-80",
    doi = "10.1016/0370-2693(81)90559-1",
    journal = "Phys. Lett. B",
    volume = "99",
    pages = "411--415",
    year = "1981"
}

@article{Obata:2026qwx,
    author = "Obata, Ippei and Yanagida, Tsutomu T.",
    title = "{Probing Majoron Dark Matter with Gravitational Wave Detectors}",
    eprint = "2604.08193",
    archivePrefix = "arXiv",
    primaryClass = "hep-ph",
    reportNumber = "KEK-TH-2823, KEK-Cosmo-0416",
    month = "4",
    year = "2026"
}

@article{Akita:2026gzk,
    author = "Akita, Kensuke and Hamaguchi, Koichi and Kitagawa, Haruto and Yokoyama, Tatsuya",
    title = "{Minimal Majoron Dark Matter}",
    eprint = "2605.12946",
    archivePrefix = "arXiv",
    primaryClass = "hep-ph",
    month = "5",
    year = "2026"
}

@article{deGiorgi:2026jqn,
    author = "de Giorgi, Arturo and Naredo-Tuero, Daniel and Ponce D{\'\i}az, Xavier",
    title = "{The Majoron Cosmological Window: Dark Matter and Thermal Leptogenesis}",
    eprint = "2605.18944",
    archivePrefix = "arXiv",
    primaryClass = "hep-ph",
    reportNumber = "IPPP/26/39",
    month = "5",
    year = "2026"
}

@article{Batell:2026avi,
    author = "Batell, Brian and Dasgupta, Arnab and Dutta, Swapnil and Ghalsasi, Akshay",
    title = "{Majoron Dark Matter, High-Scale Seesaw, and Leptogenesis}",
    eprint = "2606.02706",
    archivePrefix = "arXiv",
    primaryClass = "hep-ph",
    reportNumber = "PITT-PACC-2606",
    month = "6",
    year = "2026"
}

@article{Berezinsky:1993fm,
    author = "Berezinsky, V. and Valle, J. W. F.",
    title = "{The KeV majoron as a dark matter particle}",
    eprint = "hep-ph/9309214",
    archivePrefix = "arXiv",
    reportNumber = "FTUV-93-35, LNGS-93-79",
    doi = "10.1016/0370-2693(93)90140-D",
    journal = "Phys. Lett. B",
    volume = "318",
    pages = "360--366",
    year = "1993"
}

@article{Lattanzi:2007ux,
    author = "Lattanzi, M. and Valle, J. W. F.",
    title = "{Decaying warm dark matter and neutrino masses}",
    eprint = "0705.2406",
    archivePrefix = "arXiv",
    primaryClass = "astro-ph",
    reportNumber = "IFIC-06-39",
    doi = "10.1103/PhysRevLett.99.121301",
    journal = "Phys. Rev. Lett.",
    volume = "99",
    pages = "121301",
    year = "2007"
}

@article{Bazzocchi:2008fh,
    author = "Bazzocchi, Federica and Lattanzi, Massimiliano and Riemer-S{\o}rensen, Signe and Valle, Jose W. F.",
    title = "{X-ray photons from late-decaying majoron dark matter}",
    eprint = "0805.2372",
    archivePrefix = "arXiv",
    primaryClass = "astro-ph",
    reportNumber = "IFIC-08-25",
    doi = "10.1088/1475-7516/2008/08/013",
    journal = "JCAP",
    volume = "08",
    pages = "013",
    year = "2008"
}

@article{Kallosh:1995hi,
    author = "Kallosh, Renata and Linde, Andrei D. and Linde, Dmitri A. and Susskind, Leonard",
    title = "{Gravity and global symmetries}",
    eprint = "hep-th/9502069",
    archivePrefix = "arXiv",
    reportNumber = "SU-ITP-95-2",
    doi = "10.1103/PhysRevD.52.912",
    journal = "Phys. Rev. D",
    volume = "52",
    pages = "912--935",
    year = "1995"
}

@article{Harlow:2018jwu,
    author = "Harlow, Daniel and Ooguri, Hirosi",
    title = "{Constraints on Symmetries from Holography}",
    eprint = "1810.05337",
    archivePrefix = "arXiv",
    primaryClass = "hep-th",
    doi = "10.1103/PhysRevLett.122.191601",
    journal = "Phys. Rev. Lett.",
    volume = "122",
    number = "19",
    pages = "191601",
    year = "2019"
}

@article{Krauss:1988zc,
    author = "Krauss, Lawrence M. and Wilczek, Frank",
    title = "{Discrete Gauge Symmetry in Continuum Theories}",
    reportNumber = "YCTP-P26-88, NSF-ITP-88-187",
    doi = "10.1103/PhysRevLett.62.1221",
    journal = "Phys. Rev. Lett.",
    volume = "62",
    pages = "1221",
    year = "1989"
}

@article{Ibanez:1991pr,
    author = "Ibanez, Luis E. and Ross, Graham G.",
    title = "{Discrete gauge symmetries and the origin of baryon and lepton number conservation in supersymmetric versions of the standard model}",
    reportNumber = "CERN-TH-6111-91",
    doi = "10.1016/0550-3213(92)90195-H",
    journal = "Nucl. Phys. B",
    volume = "368",
    pages = "3--37",
    year = "1992"
}

@article{Garcia-Etxebarria:2018ajm,
    author = "Garc{\'\i}a-Etxebarria, I{\~n}aki and Montero, Miguel",
    title = "{Dai-Freed anomalies in particle physics}",
    eprint = "1808.00009",
    archivePrefix = "arXiv",
    primaryClass = "hep-th",
    reportNumber = "MPP-2018-188",
    doi = "10.1007/JHEP08(2019)003",
    journal = "JHEP",
    volume = "08",
    pages = "003",
    year = "2019"
}

@article{Kawasaki:2023mjm,
    author = "Kawasaki, Masahiro and Yanagida, Tsutomu T.",
    title = "{Dai-Freed anomaly in the standard model and topological inflation}",
    eprint = "2304.10100",
    archivePrefix = "arXiv",
    primaryClass = "hep-ph",
    doi = "10.1007/JHEP11(2023)106",
    journal = "JHEP",
    volume = "11",
    pages = "106",
    year = "2023"
}

@article{Forestell:2018txr,
    author = "Forestell, Lindsay and Morrissey, David E. and White, Graham",
    title = "{Limits from BBN on Light Electromagnetic Decays}",
    eprint = "1809.01179",
    archivePrefix = "arXiv",
    primaryClass = "hep-ph",
    doi = "10.1007/JHEP01(2019)074",
    journal = "JHEP",
    volume = "01",
    pages = "074",
    year = "2019"
}

@article{Depta:2020zbh,
    author = "Depta, Paul Frederik and Hufnagel, Marco and Schmidt-Hoberg, Kai",
    title = "{Updated BBN constraints on electromagnetic decays of MeV-scale particles}",
    eprint = "2011.06519",
    archivePrefix = "arXiv",
    primaryClass = "hep-ph",
    reportNumber = "DESY-20-160, DESY 20-160, ULB-TH/20-15",
    doi = "10.1088/1475-7516/2021/04/011",
    journal = "JCAP",
    volume = "04",
    pages = "011",
    year = "2021"
}

@article{Suzuki:2026xvf,
    author = "Suzuki, Motoo and Yokokura, Ryo",
    title = "{Lazarides-Shafi axion models as Dijkgraaf-Witten theories}",
    eprint = "2602.12345",
    archivePrefix = "arXiv",
    primaryClass = "hep-th",
    doi = "10.1103/hl5s-73lj",
    journal = "Phys. Rev. D",
    volume = "113",
    number = "9",
    pages = "L091902",
    year = "2026"
}

@article{Akita:2025txo,
    author = "Akita, Kensuke and Hamaguchi, Koichi and Ovchynnikov, Maksym",
    title = "{Maximal Parameter Space of Sterile Neutrino Dark Matter with Lepton Asymmetries}",
    eprint = "2507.20659",
    archivePrefix = "arXiv",
    primaryClass = "hep-ph",
    reportNumber = "CERN-TH-2025-149",
    doi = "10.1103/d2qg-clrz",
    journal = "Phys. Rev. Lett.",
    volume = "136",
    number = "11",
    pages = "111001",
    year = "2026"
}

@article{Ghosh:2025cxp,
    author = "Ghosh, Tathagata and Loho, Kousik and Manna, Sudip",
    title = "{From U(1){\texttimes}U(1) symmetry breaking to Majoron cosmology: Insights from NANOGrav 15-year data}",
    eprint = "2507.04342",
    archivePrefix = "arXiv",
    primaryClass = "hep-ph",
    reportNumber = "HRI-RECAPP-2025-07",
    doi = "10.1103/gd2q-qdhg",
    journal = "Phys. Rev. D",
    volume = "113",
    number = "4",
    pages = "043036",
    year = "2026"
}

@article{Manna:2022gwn,
    author = "Manna, Soumen Kumar and Sil, Arunansu",
    title = "{Majorons revisited: Light dark matter as a FIMP}",
    eprint = "2212.08404",
    archivePrefix = "arXiv",
    primaryClass = "hep-ph",
    doi = "10.1103/PhysRevD.108.075026",
    journal = "Phys. Rev. D",
    volume = "108",
    number = "7",
    pages = "075026",
    year = "2023"
}

@article{King:2024idj,
    author = "King, Stephen F. and Manna, Soumen Kumar and Roshan, Rishav and Sil, Arunansu",
    title = "{Leptogenesis with Majoron dark matter}",
    eprint = "2412.14121",
    archivePrefix = "arXiv",
    primaryClass = "hep-ph",
    doi = "10.1103/PhysRevD.111.095008",
    journal = "Phys. Rev. D",
    volume = "111",
    number = "9",
    pages = "095008",
    year = "2025"
}

@article{Buchmuller:2003gz,
    author = "Buchmuller, W. and Di Bari, P. and Plumacher, M.",
    title = "{The Neutrino mass window for baryogenesis}",
    eprint = "hep-ph/0302092",
    archivePrefix = "arXiv",
    reportNumber = "DESY-03-001, UAB-FT-540, CERN-TH-2003-016",
    doi = "10.1016/S0550-3213(03)00449-8",
    journal = "Nucl. Phys. B",
    volume = "665",
    pages = "445--468",
    year = "2003"
}

@article{Engelhard:2007kf,
    author = "Engelhard, Guy and Grossman, Yuval and Nir, Yosef",
    title = "{Relating leptogenesis parameters to light neutrino masses}",
    eprint = "hep-ph/0702151",
    archivePrefix = "arXiv",
    doi = "10.1088/1126-6708/2007/07/029",
    journal = "JHEP",
    volume = "07",
    pages = "029",
    year = "2007"
}

@article{Asaka:1999yd,
    author = "Asaka, T. and Hamaguchi, Koichi and Kawasaki, M. and Yanagida, T.",
    title = "{Leptogenesis in inflaton decay}",
    eprint = "hep-ph/9906366",
    archivePrefix = "arXiv",
    reportNumber = "UT-853, RESCEU-15-99",
    doi = "10.1016/S0370-2693(99)01020-5",
    journal = "Phys. Lett. B",
    volume = "464",
    pages = "12--18",
    year = "1999"
}

@article{Asaka:1999jb,
    author = "Asaka, T. and Hamaguchi, Koichi and Kawasaki, M. and Yanagida, T.",
    title = "{Leptogenesis in inflationary universe}",
    eprint = "hep-ph/9907559",
    archivePrefix = "arXiv",
    reportNumber = "UT-855, RESCEU-28-99",
    doi = "10.1103/PhysRevD.61.083512",
    journal = "Phys. Rev. D",
    volume = "61",
    pages = "083512",
    year = "2000"
}

@article{Blanchet:2008pw,
    author = "Blanchet, Steve and Di Bari, Pasquale",
    title = "{New aspects of leptogenesis bounds}",
    eprint = "0807.0743",
    archivePrefix = "arXiv",
    primaryClass = "hep-ph",
    doi = "10.1016/j.nuclphysb.2008.08.026",
    journal = "Nucl. Phys. B",
    volume = "807",
    pages = "155--187",
    year = "2009"
}

@article{Gallego:2025xxx,
    author = "Gallego, Savitri and Oberlack, Uwe and Lommler, Jan and Karwin, Christopher M. and Fenu, Francesco and Fioretti, Valentina and Zoglauer, Andreas and Follega, F. M. and Perinelli, A. and Battiston, Roberto and Iuppa, Roberto and Boggs, E. Steven and Mittal, Saurabh and Jean, Pierre and Kierans, Carolyn and Hartmann, H. Dieter and Tomsick, A. John",
    title = "{Preflight Background Estimates for COSI}",
    eprint = "2510.25304",
    archivePrefix = "arXiv",
    primaryClass = "astro-ph.IM",
    doi = "10.3847/1538-4357/ae32f4",
    journal = "Astrophys. J.",
    volume = "997",
    number = "2",
    pages = "284",
    year = "2026"
}

\end{document}